\documentclass[11pt,a4paper,oneside]{article}
\pdfoutput=1
\usepackage{jheppub}

\usepackage{verbatim}
\usepackage{mathrsfs}
\usepackage{appendix}
\usepackage{caption}
\usepackage{float}
\usepackage{subfig}
\usepackage[vcentermath]{youngtab}
\usepackage{alltt}
\usepackage{array,multirow}
\usepackage{arydshln}
\usepackage{lscape}

\usepackage{amssymb,amsmath,amsfonts,mathrsfs,enumerate,wrapfig,latexsym,wasysym}
\usepackage{graphicx}
\usepackage{adjustbox}
\usepackage{xcolor}  
\usepackage{slashed}    
\usepackage{url}
\usepackage{hyperref}
\usepackage{framed}
\usepackage[numbers,sort&compress]{natbib}
\usepackage{fancyvrb}
\allowdisplaybreaks
\usepackage{multirow}
\usepackage{framed}
\RequirePackage{flushend}
\usepackage{makecell}
\usepackage{color}
\usepackage{pifont}
\usepackage{cancel}
\usepackage{import}
\usepackage{subfiles}
\usepackage{filecontents}
\usepackage{threeparttable,tablefootnote}\usepackage{mathtools}
\usepackage{pbox}
\usepackage{setspace}
\usepackage{lscape}

\title{Effective Operator Bases for Beyond Standard Model Scenarios: An EFT compendium for discoveries}

\author[1]{Upalaparna Banerjee, Joydeep Chakrabortty, Suraj Prakash, Shakeel Ur Rahaman}

\author[2]{ and Michael Spannowsky}

\emailAdd{upalab, joydeep, surajprk, shakel@iitk.ac.in, michael.spannowsky@durham.ac.uk}

\affiliation[1]{Indian Institute of Technology Kanpur, Kalyanpur, Kanpur 208016, Uttar Pradesh, INDIA}

\affiliation[2]{Institute for Particle Physics Phenomenology, Department of Physics, Durham University, Durham DH1 3LE, U.K}

\preprint{IPPP/20/37}

\abstract{It is not only conceivable but likely that the spectrum of physics beyond the Standard Model (SM) is non-degenerate. The lightest non-SM particle may reside close enough to the electroweak scale that it can be kinematically probed at high-energy experiments and on account of this, it must be included as an infrared (IR) degree of freedom (DOF) along with the SM ones. The rest of the non-SM  particles are heavy enough to be directly experimentally inaccessible and can be integrated out. Now, to capture the effects of the complete theory, one must take into account the higher dimensional operators constituted of the SM DOFs and the minimal extension. This construction, BSMEFT, is in the same spirit as SMEFT but now with extra IR DOFs. Constructing a BSMEFT is in general the first step after establishing experimental evidence for a new particle. We have investigated three different scenarios where the SM is extended by additional (i) uncolored, (ii) colored particles, and (iii) abelian gauge symmetries. For each such scenario, we have included the most-anticipated and phenomenologically motivated models to demonstrate the concept of BSMEFT. 	In this paper, we have provided the full EFT Lagrangian for each such model up to mass dimension 6.  We have also identified the $CP$, baryon ($B$), and lepton ($L$) number violating  effective operators. }

\begin{document}

\maketitle

\section{Introduction}

The Standard Model (SM) of particle physics has been the most successful theory to describe the dynamics and interactions of sub-atomic particles. Every prediction that could be made based on the SM Lagrangian has been substantiated by different experiments, while the converse is not true. Several observations can not be satisfactorily explained within the Standard Model framework. Thus, the SM appears to be a theory of fundamental particles - but not the complete one, i.e. its validity does not extend to arbitrarily high energy scales. There have been several efforts to extend the SM by extending its gauge groups and (or) by adding new particles. It is believed that at very high energies, near the Planck scale, there is a unified gauge group from where all the low energy physics, including the SM, have descended. The region between the unified and electroweak scales is potentially populated with many particles of different mass scales. However, to this day, we are not confident about the exact nature of the theories beyond the SM (BSM), as we do not have sufficient experimental data that can help us to isolate a specific BSM scenario. A plethora of BSM proposals \cite{Golling:2016gvc,Blondel:2019vdq} exist and each of them has its own merits.  

For future and ongoing searches of new physics, for example at the LHC, an important question is whether it is possible to capture the essence of the new unknown physics using our knowledge about the symmetries and the particle content. Indeed, this is the underlying idea of an Effective Field theory (EFT) where the complete Lagrangian is written as: 
\begin{eqnarray}
\mathcal{L} = \mathcal{L}_{\text{renorm}} + \sum_{i=5}^{n}\sum_{j=1}^{N_i}\frac{\mathcal{C}_j^{(i)}}{\Lambda^{i-4}}\mathcal{O}_j^{(i)}.
\end{eqnarray}
Here, $(\sum_j)$ denotes the sum over effective operators ($N_i$) each having mass dimension $i$.   $\Lambda$ is the scale of new physics and thus possesses mass dimension 1.  The dimensionless coefficients  $\mathcal{C}_j^{(i)}$ are the so-called Wilson coefficients. The second term in the above equation is the effective Lagrangian ($\mathcal{L}_{EFT}$) \cite{Georgi:1994qn,Manohar:1996cq,Kaplan:1995uv,Kaplan:2005es,Burgess:2007pt,Rothstein:2003mp,Cohen:2019wxr}.  The origin of these effective operators can be understood through two possible mechanisms. First, if we have prior knowledge about the new physics Lagrangian then we can suitably integrate out the heavy modes from the UV theory while retaining the light ones, i.e. infrared (IR) degrees of freedom (DOF). The impact of heavy DOFs is captured by the effective interactions and their respective WCs. Second, to capture their effects, we can add the gauge invariant effective operators in a consistent way. In this case we need to rely only on the on-shell DOFs and the associated symmetries. It is interesting to note that even when the exact nature of the UV theory is unknown, this formalism can be very useful in sensing the integrated out new physics.
In this work, we will focus on this aspect of EFTs \cite{Manohar:1996cq,Burgess:2007pt,Georgi:1994qn,Kaplan:1995uv,Kaplan:2005es,Rothstein:2003mp,Cohen:2019wxr}. \\

\noindent
Recognising that the SM may only be valid up to a certain high energy scale beyond which the effects of new physics may become noticeable, the last decade has seen tremendous progress towards the study of SM physics as an EFT (or SMEFT) \cite{Henning:2014wua,Jegerlehner:2013cta,Contino:2016jqw,Ellis:2014jta,Berthier:2015gja,Brivio:2017vri}. More precisely, the study of higher dimensional operators (of mass dimension $\geq$ 5) has attracted a lot of attention. And these operators have been found to introduce many novel and interesting predictions. For instance, the only dimension 5 operator shows lepton number violation and generates a Majorana mass term for the neutrino. Going to even higher dimensions we even come across predictions of processes as rare as proton decay \cite{Hambye:2017qix,Fonseca:2018ehk}. SMEFT also encompasses the two paradigms of EFT - the first being the {\it top-down} approach which actually comes about through an interplay of a particular minimal extension of the SM and through a subset of higher dimension SMEFT operators. These assume the existence of the minimal extension at some high energy scale and after integrating out the heavy degree of freedom yields SMEFT effective operators \cite{Henning:2016lyp,CriadoAlamo:2019kbt,Huo:2015exa,Chiang:2015ura,Ellis:2016enq,Fuentes-Martin:2016uol}. A number of computational tools such as \texttt{CoDeX} \cite{Bakshi:2018ics}, \texttt{Wilson} \cite{Aebischer:2018bkb}, \texttt{DsixTools} \cite{Celis:2017hod}, \texttt{WCxf} \cite{Aebischer:2017ugx}, \texttt{MatchingTools} \cite{Criado:2017khh} have been developed to automatise this procedure. The second one, i.e., the {\it bottom-up} approach is concerned with the construction of complete and independent operator sets at various mass dimensions based on group-theoretic ideas \cite{Henning:2015alf,Henning:2017fpj,Lehman:2015via,Lehman:2015coa,Fonseca:2019yya}. Complete and independent sets of SMEFT operators have been constructed for mass dimensions 6 \cite{Grzadkowski:2010es}, 7 \cite{Lehman:2014jma}, 8 \cite{Murphy:2020rsh,Li:2020gnx}, and 9 \cite{Li:2020xlh,Liao:2020jmn}. Several ingeniously built modern tools such as \texttt{GrIP} \cite{Banerjee:2020bym}, \texttt{BasisGen} \cite{Criado:2019ugp}, \texttt{Sym2Int} \cite{Fonseca:2017lem}, \texttt{ECO} \cite{Marinissen:2020jmb}, and \texttt{DEFT} \cite{Gripaios:2018zrz} have made the construction of higher dimensional operators straightforward and convenient.  The operators obtained by integrating out the heavy fields from several different SM extensions turn out to be overlapping subsets of this complete set. Thus, SMEFT provides a common ground to encode the predictions of different new physics models. At the same time, some of the higher dimensional operators also provide non-leading order contributions to the predictions of the SM itself, thus enhancing the precision of theoretical calculations \cite{Banerjee:2013apa,Hays:2018zze,Dedes:2018seb}.\\

\noindent
It is worth noting that the SMEFT construction assumes that new physics appears at a particular scale and all the non-SM particles are degenerate. A completely degenerate spectrum in the UV regime of a new theory, however, is very unlikely. Instead, for a non-degenerate spectrum, there will be a non-SM degree of freedom with a small mass. If such a BSM particle is light enough to be kinematically accessible, and couples strongly to the SM DOFs, it may be counted as an IR DOF along with the SM ones, while the rest of the new particles would be heavy enough to be integrated out. SMEFT is not designed to capture such a scenario. As the on-shell IR DOFs are now extended, one has to compute the new set of effective operators in addition to the SMEFT ones, thus leading to a new effective operator basis which can be referred to as BSMEFT. This is the underlying principle behind the Effective Field Theoretic reformulation of several popular scenarios. Higher mass dimension operators have been constructed for diverse scenarios such as the extension of SM by a doubly charged scalar \cite{King:2014uha},  the Two Higgs Doublet Model \cite{Anisha:2019nzx,Trautner:2018ipq,Crivellin:2016ihg,Karmakar:2017yek,Birch-Sykes:2020btk}, and the Minimal Left Right Symmetric Model \cite{Anisha:2019nzx}. Neutrino mass models are now being studied under the framework of $\nu$SMEFT and operators of mass dimensions 6 \cite{Elgaard-Clausen:2017xkq,Chala:2020vqp} and 7 \cite{Bhattacharya:2015vja,Liao:2016qyd} have been constructed for the same. The same ideas have also been applied to low energy (below electroweak scale) models within the framework of LEFT \cite{Jenkins:2017jig,Jenkins:2017dyc,Dekens:2019ept} where operators up to mass dimension 7 have been constructed \cite{Liao:2020zyx}. These find great utility in B-physics \cite{Aebischer:2017gaw} and dark matter studies \cite{Brod:2017bsw}.\\

\noindent
To conduct a procedural analysis we must start by investigating possible minimal extensions of the SM, which are mostly phenomenologically motivated. To capture the interplay of the SM electroweak sector with the new physics models, one must address a variety of scenarios starting from SM-singlet real scalar fields \cite{Adhikari:2020vqo,OConnell:2006rsp,Barger_2008} to higher dimensional color singlet multiplets. One must also consider the extensions of the strong sector using colored scalars and fermions \cite{Buchmuller:1986zs}. These minimal extensions have been introduced in an attempt to rationalize very specific observations. It is worth mentioning that there exist multiple UV complete theories that may end up leading to the same set of IR DOFs after suitably and  partially integrating out heavy DOFs. So, looking into these minimal extensions, it is indeed difficult to identify the unique parent theory. For example, if the SM spectrum is extended by a doubly charged scalar then its UV root will be difficult to ascertain.
It can appear either as an $SU(2)_L$ singlet but non-zero hyper-charged complex scalar field or as a part of higher dimensional representations of the electroweak gauge group $SU(2)_{L}\otimes U(1)_Y$.  In such cases, the natural possibility is that there exists a hierarchy of masses between the doubly charged scalar and the other components of the multiplet.  It is also possible that the whole multiplet is lighter than the other non-SM fields. Then that should be counted as the IR DOF while constructing the effective operators. \\

\noindent
BSMEFT can be considered to be the first stride in the step by step process of unraveling a full BSM model. Collision experiments are expected to detect few non-SM particles first, rather than unveiling the complete spectrum of an extension to the SM at once. The first reaction after observing a new resonance will be to build a BSMEFT theory around this particle - as evidenced in previous occasions of eventually unconfirmed experimental excesses (see e.g. \cite{Franceschini:2015kwy,Backovic:2015fnp,Strumia:2016wys}). Thus, the BSMEFT models we provide can serve as a compendium for complete operator bases after a new resonance is observed.\\

\noindent
To promulgate the idea of BSMEFT our study must encompass several varieties of models, which is precisely the purpose of this work. We have organized the paper as follows. First,  we have meticulously described a general procedure to construct invariant operators in section \ref{sec:op-constr-procedure}. We have highlighted the various subtleties associated with it by giving suitable example operators. In this work, we have carefully selected the BSM scenarios to capture the possible impact of the effective operators on the electroweak and strong sectors. Thus we have worked with models where SM is extended by additional color singlet complex scalars and fermions that transform as different $SU(2)_L$ representations and also phenomenologically motivated Lepto-Quark scenarios. We have further adopted an abelian extension of the gauge sector of the SM, motivated by a gauge-boson dark matter scenario. In section \ref{sec:op-bases} we have  enlisted the complete and independent sets of operators of mass dimensions 5 and 6 for all these models. We have arranged the operators on the basis of their constituents and we have specifically highlighted the operators that violate baryon and lepton numbers. This will help to analyze and pin down which of the rare processes are more likely to occur for a given BSM scenario. We have showcased the flavour structures of each class of operators for each such model.

\section{Roadmap of   invariant operator construction}\label{sec:op-constr-procedure}
In calculating the invariant operators, underlying symmetries play a crucial role. The quantum fields transform under these symmetries according to their assigned charges. The goal is to find all invariants under these symmetries, i.e., singlet configurations containing any number of those quantum fields. The Lagrangian consists of all such configurations. We classify the symmetries as follows: (i) space-time and (ii) gauge symmetries. In addition to that we can have certain kinds of imposed and (or) accidental global symmetries. The requirement of their violation or conservation driven by phenomenological needs determines the presence or absence of {\it rare} operators.  In principle, the Lagrangian ($\mathcal{L}$) can contain an infinite number of such singlet terms. But not all of them are phenomenologically important. Thus it is preferred to write down $\mathcal{L}$ as a polynomial of the invariant operators and the mass dimension is chosen to be the order of that polynomial. This allows one to keep the terms up to a mass dimension based on the experimental precision possibly achieved in the ongoing and (or) future experiments.  In the following subsections we will demonstrate the role of individual symmetries and the issues related to the dynamical nature of these fields, e.g., equation of motions and integration by parts.

\subsection{Tackling space-time symmetry: Lorentz Invariance}\label{subsec:Lorentz-symm}

The quantum fields under consideration have different spins, which are determined by their transformation properties under the (3+1)-dimensional space-time symmetry, dictated here  by the Lorentz group $SO(3,1)$. In this work, our primary focus is on the scalar, vector and spinorial representations of the Lorentz group. The scalars, spin-0 fields,  transform trivially, i.e. they are singlets under the Lorentz group. While the vectors, i.e. spin-1 and spinors, i.e. spin-1/2 are non-singlet representations under $SO(3,1)$. We must recall, here,  that our prescription for computing the invariant operators deals with finite-dimensional unitary representations. The Lorentz group being non-compact does not have finite-dimensional unitary representations. Hence, we will realize the representations of $SO(3,1)$ in terms of unitary finite-dimensional representations of its compact form $SU(2)_L\times SU(2)_R$, and we will work within the Weyl basis where the gamma matrices take the following forms:

\vspace{-0.5cm}
{\small\begin{eqnarray}\label{eq:gamma-matrices}
\gamma^\mu=
\begin{pmatrix}
0& \sigma^\mu_{\alpha\dot{\beta}} \\
\overline{\sigma}^{\mu\dot{\alpha}\beta} & 0
\end{pmatrix},\,
\gamma^5=
\begin{pmatrix}
\text{-}\mathbb{I} & & 0\\
0 & & \mathbb{I}
\end{pmatrix}.
\end{eqnarray}}
\noindent
Here, $\sigma^\mu=(\mathbb{I},\sigma^i)$, $\overline{\sigma}^\mu=(\mathbb{I},-\sigma^i) $, with $\sigma^i$ being the Pauli spin-matrices and $\mathbb{I}$ is a 2$\times$2 identity matrix.
In this  basis, the non-zero spin fields possess definite chirality.  In the case of fermions, we will work with Weyl spinors $\Psi_{L}$ and $\Psi_{R}$ instead of the Dirac spinors $\Psi$ and $\overline{\Psi}$ which are defined as \cite{Dreiner:2008tw}:

\vspace{-0.5cm}
{\small\begin{eqnarray}\label{eq:spinor-4comp-2comp}
\Psi =
\begin{pmatrix}
\chi_\alpha \\
\xi^{\dagger\dot{\alpha}}
\end{pmatrix},\;\;
\overline{\Psi} =\Psi^\dagger\gamma^0=
\begin{pmatrix}
\xi^\alpha & \chi^{\dagger}_{\dot{\alpha}}
\end{pmatrix}.
\end{eqnarray}}
\noindent
We can define the two component Weyl spinors $\Psi_L$ and $\Psi_R$ as four component ones in the following manner\footnote{$\Psi_{L,R}$ are obtained from $\Psi$ using the projection operators $\frac{1\mp\gamma^5}{2}$, i.e., 
$\Psi_{L} = \frac{1-\gamma^5}{2} \Psi$, and $\Psi_{R} = \frac{1+\gamma^5}{2} \Psi$.}:

\vspace{-0.5cm}
{\small\begin{eqnarray}\label{eq:chiral-spinors}
\Psi_L =
\begin{pmatrix}
\chi_\alpha \\
0
\end{pmatrix},\;\;
\overline{\Psi}_L =\Psi_L^\dagger\gamma^0=
\begin{pmatrix}
0 & \chi^{\dagger}_{\dot{\alpha}}
\end{pmatrix}, \;\;
\Psi_R =
\begin{pmatrix}
0 \\
\xi^{\dagger\dot{\alpha}}
\end{pmatrix},\;\;
\overline{\Psi}_R =\Psi_R^\dagger\gamma^0=
\begin{pmatrix}
\xi^\alpha & 0
\end{pmatrix}.
\end{eqnarray}}
Following a similar principle,  the field strength tensor $X^{\mu\nu}$ and its dual $\tilde{X}_{\mu\nu}=\frac{1}{2}\epsilon_{\mu\nu\rho\sigma}X^{\rho\sigma}$,  transforming under $SO(3,1)$, must be written in terms of representations of $SU(2)_L\times SU(2)_R$, i.e., $X_{L,\mu\nu}$ and $X_{R,\mu\nu}$ as:

\vspace{-0.5cm}
{\small\begin{eqnarray}\label{eq:fsL-fsR-def}
X_{L,\mu\nu} &=& \frac{1}{2}\left(X_{\mu\nu} - i\tilde{X}_{\mu\nu}\right), \hspace{0.5cm} (X_L)_{\alpha\beta}=\sigma^{\mu}_{\alpha\dot{\beta}}\,\overline{\sigma}^{\nu\dot{\beta}\kappa}\,\epsilon_{\kappa\beta}\,X_{L,\mu\nu} ,\nonumber \\
X_{R,\mu\nu} &=& \frac{1}{2}\left(X_{\mu\nu} + i\tilde{X}_{\mu\nu}\right), \hspace{0.5cm} (X_R)^{\dot{\alpha}\dot{\beta}}=\overline{\sigma}^{\mu\dot{\alpha}\kappa}\,\sigma^{\nu}_{\kappa\dot{\kappa}}\,\,\epsilon^{\dot{\kappa}\dot{\beta}}\,X_{R,\mu\nu} .
\end{eqnarray}}
\noindent
To proceed further, we have identified the quantum fields\footnote{In our analysis, we have put the covariant derivative ($\mathcal{D}$) on an equal footing as the quantum fields.} as the representations of $SU(2)_L\times SU(2)_R$ and demarcated them by their respective spin values $(j_L, j_R)$ as:

\vspace{-0.5cm}
{\small\begin{eqnarray}\label{eq:su2xsu2-irrep}
\hspace{-0.7cm} \Phi &\equiv& \left(0,0\right), \hspace{0.1cm} \Psi_L \equiv \left(\frac{1}{2}, 0\right), \hspace{0.1cm} \Psi_R \equiv \left(0, \frac{1}{2}\right),  \hspace{0.1cm}
\mathcal{D} \equiv \left(\frac{1}{2}, \frac{1}{2}\right), \hspace{0.1cm} X_L \equiv \left(1, 0\right), \hspace{0.1cm} X_R \equiv \left(0, 1\right). 
\end{eqnarray}}
Here, $\Phi$ refers to a scalar, and $\Psi_{L,R}$, $X_{L,R}$ are defined in Eqns.~\eqref{eq:chiral-spinors} and \eqref{eq:fsL-fsR-def} respectively. 

As mentioned earlier, our primary aim is to construct a set of  Lorentz invariant operators $(\mathcal{O} )$ using these fields and that can be mathematically framed as follows:

\vspace{-0.5cm}
{\small\begin{eqnarray}\label{eq:op-class-determination-1}
\mathcal{O} &\equiv& \Phi^{p} \times \Psi_L^{q_1}  \times \Psi_R^{q_2} \times \mathcal{D}^{r} \times X_L^{s_1}  \times X_R^{s_2}, \\
\label{eq:op-class-determination-2}
\implies (0,0) &\equiv& (0,0)^{p} \times \left(\frac{1}{2},0\right)^{q_1} \times \left(0,\frac{1}{2}\right)^{q_2} \times \left(\frac{1}{2},\frac{1}{2}\right)^{r} \times \left(1,0\right)^{s_1} \times \left(0,1\right)^{s_2}.
\end{eqnarray}}
\noindent
Here $p, q_1, q_2, r, s_1, s_2$ are the number of times the different fields appear in the operator. All these are non-negative integers. 
The equivalent relation in terms of mass dimension can be written as:

\vspace{-0.7cm}
{\small
\begin{eqnarray}
 [M]^{d} &\equiv& [M]^{p} \times [M]^{3q_1/2} \times [M]^{3q_2/2} \times [M]^{r} \times [M]^{2s_1} \times [M]^{2s_2}, 
\end{eqnarray}
}
and equating mass dimensions on both sides we find
{\small
\begin{eqnarray}\label{eq:mass-dim}
d &=& p + \frac{3}{2}(q_1+q_2) + r + 2(s_1 + s_2).
\end{eqnarray}
}
Here, $d$ is the mass dimension of the Lorentz invariant  operator and that for fermionic and bosonic fields, and field strength tensors are $3/2$, $1$, and $2$ respectively\footnote{The covariant derivative $\mathcal{D}$ has mass dimension 1.}. Similarly, the relation derived from Eqn.~\eqref{eq:op-class-determination-2} can be expressed in terms of the spin ($j$) as: 
{\small
	\begin{eqnarray}\label{eq:op-spin}
	0 &\equiv& [0]^{p} \oplus [1/2]^{q_1} \oplus [0]^{q_2} \oplus [1/2]^{r} \oplus [1]^{s_1} \oplus [0]^{s_2},  \nonumber\\
		0 &\equiv& [0]^{p} \oplus [0]^{q_1} \oplus [1/2]^{q_2} \oplus [1/2]^{r} \oplus [0]^{s_1} \oplus [1]^{s_2},  
	\end{eqnarray}
}
or equivalently in terms of $SU(2)$ representations ($2j+1$) as:
{\small
	\begin{eqnarray}\label{eq:op-SU2-rep}
	 1 &\equiv& [1]^{p} \otimes [2]^{q_1} \otimes [0]^{q_2} \otimes [2]^{r} \otimes [3]^{s_1} \otimes [1]^{s_2},  \nonumber\\
	1 &\equiv& [1]^{p} \otimes [1]^{q_1} \otimes [2]^{q_2} \otimes [2]^{r} \otimes [1]^{s_1} \otimes [3]^{s_2}.
	\end{eqnarray}
}
\noindent 
Here, $[1/2]^{q}$ in Eqn.~\eqref{eq:op-spin} and $[2]^{q}$ in Eqn.~\eqref{eq:op-SU2-rep} imply $\underbrace{\vec{\frac{1}{2}}+\cdots+\vec{\frac{1}{2}}}_{q}$
 and $\underbrace{2\otimes\cdots\otimes 2}_{q}$ respectively.
Now simultaneously solving Eqns.~\eqref{eq:mass-dim}, \eqref{eq:op-spin}, and \eqref{eq:op-SU2-rep} we can find Lorentz invariant operators\footnote{These are also the operator classes at a given mass dimension.}. The number of possible operator classes keeps on increasing as the mass dimension increases.   In Table \ref{table:op-class-1to6} we have listed all possible operator classes up to dimension 6 consisting of $\Phi$, $\Psi_L$, $\Psi_R$, $X_L$, $X_R$, and $\mathcal{D}$. But they are not written in covariant forms which are necessary for further analysis.  Below, we have explicitly shown how  the Lorentz indices must be assigned to the constituent fields to write down the invariant operator in a covariant form. 
%\newpage
\begin{table}[h]
	\centering
	\renewcommand{\arraystretch}{1.7}
	{\small\begin{tabular}{||c|c|cc|c|cc|c||c|c|cc|c|cc|c||}
			\hline
			dim - $(d)$&
			$p$&
			$q_1$&
			$q_2$&
			$r$&
			$s_1$&
			$s_2$&
			\textbf{Class}&
			dim - $(d)$&
			$p$&
			$q_1$&
			$q_2$&
			$r$&
			$s_1$&
			$s_2$&
			\textbf{Class}\\
			\hline
			
			{\Large \bf 1}&
			1&
			0&
			0&
			0&
			0&
			0&
			$\Phi$&
			\multirow{4}{*}{{\Large \bf 3}}&
			3&
			0&
			0&
			0&
			0&
			0&
			$\Phi^3$\\
			\cline{1-8}
			
			\multirow{3}{*}{{\Large \bf 2}}&
			2&
			0&
			0&
			0&
			0&
			0&
			$\Phi^2$&
			&
			0&
			2&
			0&
			0&
			0&
			0&
			$\Psi_L^2$\\
			
			&
			0&
			0&
			0&
			2&
			0&
			0&
			\textcolor{purple}{$\mathcal{D}^2$}&			
			&
			0&
			0&
			2&
			0&
			0&
			0&
			$\Psi_R^2$\\
			
			&
			&
			&
			&
			&
			&
			&
			&
			&1
			&0
			&0
			&2
			&0
			&0
			&\textcolor{purple}{$\Phi \,\mathcal{D}^2$}\\
			\hline
			
			\multirow{5}{*}{{\Large \bf 4}}&
			4&
			0&
			0&
			0&
			0&
			0&
			$\Phi^4$&
			\multirow{5}{*}{{\Large \bf 4}}&
			0&
			0&
			0&
			4&
			0&
			0&
			\textcolor{purple}{$\mathcal{D}^4$}\\
			
			&
			1&
			2&
			0&
			0&
			0&
			0&
			$\Psi_L^2\,\Phi$&
			&
			1&
			0&
			2&
			0&
			0&
			0&
			$\Psi_R^2\,\Phi$\\
			
			&
			0&
			1&
			1&
			1&
			0&
			0&
			$\Psi_L\,\Psi_R\,\mathcal{D}$&
			&
			2&
			0&
			0&
			2&
			0&
			0&
			$\Phi^2\,\mathcal{D}^2$\\
			
			&
			0&
			0&
			0&
			0&
			2&
			0&
			$X_L^2$&
			&
			0&
			0&
			0&
			0&
			0&
			2&
			$X_R^2$\\
			
			&
			0&
			0&
			0&
			2&
			1&
			0&
			\textcolor{purple}{$\mathcal{D}^2\, X_L$}&
			&
			0&
			0&
			0&
			2&
			0&
			1&
			\textcolor{purple}{$\mathcal{D}^2\, X_R$}\\
			\hline
			
			\multirow{7}{*}{{\Large \bf 5}}&
			2&
			2&
			0&
			0&
			0&
			0&
			$\Psi_L^2\,\Phi^2$&
			\multirow{7}{*}{{\Large \bf 5}}&
			2&
			0&
			2&
			0&
			0&
			0&
			$\Psi_R^2\,\Phi^2$\\
			\cdashline{2-8}\cdashline{10-16}
			
			&
			5&
			0&
			0&
			0&
			0&
			0&
			$\Phi^5$&
			&
			1&
			1&
			1&
			1&
			0&
			0&
			$\Psi_L\,\Psi_R\,\Phi\,\mathcal{D}$\\

			&
			0&
			2&
			0&
			2&
			0&
			0&
			$\Psi_L^2\,\mathcal{D}^2$&
			&
			0&
			0&
			2&
			2&
			0&
			0&
			$\Psi_R^2\,\mathcal{D}^2$\\
			
			&
			1&
			0&
			0&
			0&
			2&
			0&
			$\Phi \,X_L^2$&
			&
			1&
			0&
			0&
			0&
			0&
			2&
			$\Phi \,X_R^2$\\
			
			&
			0&
			2&
			0&
			0&
			1&
			0&
			$\Psi_L^2 \,X_L$&
			&
			0&
			0&
			2&
			0&
			0&
			1&
			$\Psi_R^2 \,X_R$\\
			
			&
			1&
			0&
			0&
			4&
			0&
			0&
			\textcolor{purple}{$\Phi\,\mathcal{D}^4$}&
			&
			3&
			0&
			0&
			2&
			0&
			0&
			$\Phi^3\mathcal{D}^2$\\
			
			&
			1&
			0&
			0&
			2&
			1&
			0&
			$\Phi \,X_L\, \mathcal{D}^2$&
			&
			1&
			0&
			0&
			2&
			0&
			1&
			$\Phi\, X_R\, \mathcal{D}^2$\\
			\hline
			
			\multirow{11}{*}{{\Large \bf 6}}&
			6&
			0&
			0&
			0&
			0&
			0&
			$\Phi^6$&
			\multirow{11}{*}{{\Large \bf 6}}&
			4&
			0&
			0&
			2&
			0&
			0&
			$\Phi^4\,\mathcal{D}^2$\\
			
			&
			2&
			0&
			0&
			0&
			2&
			0&
			$\Phi^2\,X_L^2$&
			&
			2&
			0&
			0&
			0&
			0&
			2&
			$\Phi^2\,X_R^2$\\
			
			&
			1&
			2&
			0&
			0&
			1&
			0&
			$\Psi_L^2\, \Phi X_L$&
			&
			1&
			0&
			2&
			0&
			0&
			1&
			$\Psi_R^2\, \Phi X_R$\\
			
			&
			0&
			0&
			0&
			0&
			3&
			0&
			$X_L^3$&
			&
			0&
			0&
			0&
			0&
			0&
			3&
			$X_R^3$\\
			
			&
			3&
			2&
			0&
			0&
			0&
			0&
			$\Psi_L^2\, \Phi^3$&
			&
			3&
			0&
			2&
			0&
			0&
			0&
			$\Psi_R^2\, \Phi^3$\\

			&
			0&
			4&
			0&
			0&
			0&
			0&
			$\Psi_L^4$&
			&
			0&
			0&
			4&
			0&
			0&
			0&
			$\Psi_R^4$\\
			
			&
			0&
			2&
			2&
			0&
			0&
			0&
			$\Psi_L^2\, \Psi_R^2$&
			&
			2&
			1&
			1&
			1&
			0&
			0&
			$\Psi_L\, \Psi_R\, \Phi^2\, \mathcal{D}$\\
			\cdashline{2-8}\cdashline{10-16}
			
			&
			0&
			0&
			0&
			2&
			2&
			0&
			$\mathcal{D}^2\, X_L^2$&
			&
			0&
			0&
			0&
			2&
			0&
			2&
			$\mathcal{D}^2\, X_R^2$\\
			
			&
			2&
			0&
			0&
			2&
			1&
			0&
			$\Phi^2\, X_L\, \mathcal{D}^2$&
			&
			2&
			0&
			0&
			2&
			0&
			1&
			$\Phi^2\, X_R\, \mathcal{D}^2$\\
			
			&
			1&
			2&
			0&
			2&
			0&
			0&
			$\Psi_L^2\,\Phi\,\mathcal{D}^2$&
			&
			1&
			0&
			2&
			2&
			0&
			0&
			$\Psi_R^2\,\Phi\,\mathcal{D}^2$\\
			
			&
			2&
			0&
			0&
			4&
			0&
			0&
			\textcolor{purple}{$\Phi^2\,\mathcal{D}^4$}&
			&
			0&
			0&
			0&
			2&
			1&
			1&
			$\mathcal{D}^2 \,X_L \,X_R$
			\\

			&
			0&
			1&
			1&
			1&
			1&
			0&
			$\Psi_L\,\Psi_R\,X_L\,\mathcal{D}$&
			&
			0&
			1&
			1&
			1&
			0&
			1&
			$\Psi_L\,\Psi_R\,X_R\,\mathcal{D}$
			\\
			\hline
	\end{tabular}}
	\caption{Lorentz Invariant Operator classes (in Weyl representation) upto mass dimension 6. For the case of dimensions 5 and 6, only the operator classes above the dashed line appear in SMEFT. The terms in red are total derivative terms and are therefore excluded from the Lagrangian. These operator classes are not all independent.}
	\label{table:op-class-1to6}
\end{table}
\clearpage 

\begin{itemize}
	\item \underline{Total derivative terms}: The Lorentz invariant total derivative operators that appear up to dimension 6 are of the following forms:
	
	\vspace{-0.8cm}
	{\small\begin{eqnarray}
		\mathcal{D}^{2} &\rightarrow& \mathcal{D}_{\mu}\mathcal{D}^{\mu}, \,\,\,
		\Phi\mathcal{D}^{2} \rightarrow \mathcal{D}_{\mu}(\mathcal{D}^{\mu}\Phi), \,\,\,
		\mathcal{D}^{4} \rightarrow (\mathcal{D}_{\mu}\mathcal{D}^{\mu})^2, 
		\nonumber\\
		(\mathcal{D}^{2}X_L + \mathcal{D}^{2}X_R) &\rightarrow& (\mathcal{D}_{\mu}\mathcal{D}_{\nu}X^{\mu\nu} + \mathcal{D}_{\mu}\mathcal{D}_{\nu}\tilde{X}^{\mu\nu}), \nonumber\\
		\Phi\mathcal{D}^{4} &\rightarrow& \mathcal{D}_{\mu}\mathcal{D}_{\nu}\mathcal{D}^{\nu}(\mathcal{D}^{\mu}\Phi), \,\,\,\, 
		\Phi^2\mathcal{D}^{4} \rightarrow \mathcal{D}_{\mu}\mathcal{D}^{\mu}(\mathcal{D}^{\nu}\Phi)(\mathcal{D}_{\nu}\Phi).
		\end{eqnarray}}
	 But being total derivatives, they do not leave any impact. Thus they are suitably removed from the Lagrangian density. 
\item \underline{Operators containing only scalar fields}: 
Spin-0 field ($\Phi$) is a Lorentz scalar. Therefore the operators

\vspace{-1.1cm} 
{\small\begin{eqnarray}
	\Phi, \,\,\, \Phi^2, \,\,\, \Phi^3, \,\,\, \Phi^4, \,\,\, \Phi^5, \,\,\, \Phi^6,\cdots
	\end{eqnarray}}
which consist of $\Phi$ only  are  Lorentz invariant.  	

\item \underline{Operators containing fermion bi-linears}: 
In the Weyl basis, there exist three different fermion bi-linears: $\Psi_L^2$, $\Psi_R^2$, $\Psi_L\Psi_R$. The first two terms can form  Lorentz invariant operators of mass dimension three. The last one  appears only as a constituent of higher dimensional operators, since it transforms as the $\left(\frac{1}{2}, \frac{1}{2}\right)$ representation of $SU(2)_L\times SU(2)_R$. These fermion bi-linears can be written in multiple covariant forms:

\vspace{-0.8cm}
{\small\begin{eqnarray}
\Psi_{L,R}^2 &\rightarrow& \underline{\Psi_{L,R}^T\,C\,\Psi_{L,R}}, \,\, \underline{\overline{\Psi}_{R,L}\,\Psi_{L,R}}, \,\, \Psi^T_{L,R}\,C\,\sigma^{\mu\nu}\,\Psi_{L,R}, \,\, \overline{\Psi}_{R,L}\,\sigma^{\mu\nu}\,\Psi_{L,R}, \nonumber\\
\Psi_L\Psi_R &\rightarrow& \overline{\Psi}_L\,\gamma^{\mu}\,\Psi_L, \,\, \overline{\Psi}_R\,\gamma^{\mu}\,\Psi_R.
\end{eqnarray}}
Here, $\sigma^{\mu\nu} = \frac{i}{4}[\gamma^{\mu}, \gamma^{\nu}]$ and $C$ is the charge conjugation operator. In the above equation only underlined terms are Lorentz invariant. The remaining structures combine with other Lorentz non-singlet terms to form higher dimensional operators. Some of those invariant structures have been listed below:

\vspace{-0.7cm}
{\small\begin{eqnarray}
	\overline{\Psi}_L\,\gamma^{\mu}\,\Psi_L\,\,\times\,\,\mathcal{D}_{\mu}\,&\equiv&\, \Psi_L\Psi_R\, \mathcal{D}, \hspace{0.5cm} \overline{\Psi}_L\,\gamma^{\mu}\,\Psi_L\,\,\times\,\,\Phi\,\mathcal{D}_{\mu}\,\Phi\,\equiv\, \Psi_L\Psi_R\,\Phi^2\,\mathcal{D},\nonumber \\
	\overline{\Psi}_L\,\gamma^{\mu}\,\Psi_L\,\,\times\,\,\overline{\Psi}_R\,\gamma^{\mu}\,\Psi_R\,&\equiv&\, \Psi_L^2\Psi_R^2, \hspace{1.1cm} 
	\overline{\Psi}_R\,\sigma^{\mu\nu}\,\Psi_L\,\,\times\,\,X_{\mu\nu}\,\equiv\, \Psi_L^2 X_L, \nonumber\\
	\overline{\Psi}_R\,\Psi_L\,\,\times\,\,\Phi\,&\equiv&\, \Psi_L^2 \Phi, \hspace{1.6cm}
	\overline{\Psi}_R\,\Psi_L\,\,\times\,\,\overline{\Psi}_R\,\Psi_L\,\equiv\, \Psi_L^4.
	\end{eqnarray}}
\item \underline{Operators containing  Field strength tensors}: 
As $X_L,X_R$ transform as $(1,0),\, (0,1)$ respectively under $SU(2)_L\times SU(2)_R$, they form the following  Lorentz scalars: $X_L^2$, $X_R^2,\, X_L^3,\, X_R^3$ up to dimension 6. They can be expressed in terms of  $X_{\mu\nu},\,  \tilde{X}_{\mu\nu} \in SO(3,1)$ as:

\vspace{-1cm}
{\small\begin{eqnarray}
	X_L^2\,+\,X_R^2\,\,&\rightarrow&\,\, X_{\mu\nu}X^{\mu\nu} + \tilde{X}_{\mu\nu}X^{\mu\nu},\nonumber\\
	X_L^3\,+\,X_R^3\,\,&\rightarrow&\,\, \underline{X^{\mu}{}_{\nu}X^{\nu}{}_{\kappa}X^{\kappa}{}_{\mu}} + \underline{\tilde{X}^{\mu}{}_{\nu}X^{\nu}{}_{\kappa}X^{\kappa}{}_{\mu}}.
	\end{eqnarray}}
The tri-linear terms being overall traces of the combination of three antisymmetric tensors vanish. This method can be adopted to construct higher dimensional operators, e.g., at dimension 8 we will have:

\vspace{-0.8cm}
{\small\begin{eqnarray}
	X_L^4+X_L^2X_R^2+X_R^4\,\rightarrow\, (X_{\mu\nu}X^{\mu\nu})\,(X_{\kappa\lambda}X^{\kappa\lambda}) + (\tilde{X}_{\mu\nu}X^{\mu\nu})\,(X_{\kappa\lambda}X^{\kappa\lambda}) + (\tilde{X}_{\mu\nu}X^{\mu\nu})\,(\tilde{X}_{\kappa\lambda}X^{\kappa\lambda}).\nonumber
	\end{eqnarray}}
	The field strength tensor $X_{L/R}$ may combine with other Lorentz non-singlet objects to form an invariant operator class, e.g., 
	
	\vspace{-0.7cm}
	{\small\begin{eqnarray}
\mathcal{D}^2X^2_{L,R} &\equiv& 	(\mathcal{D}_{\mu}X^{\mu\nu})^2,\,\,\,\,\,
X_{L,R}\Phi^2\mathcal{D}^2	 \equiv (\mathcal{D}_{\mu}X^{\mu\nu})(\Phi\mathcal{D}_{\nu}\Phi),\,\,\,\,\,
\mathcal{D}^2 X_L X_R \equiv (\mathcal{D}_{\nu}\mathcal{D}_{\mu} X^{\mu\kappa}	 \tilde{X}_{\kappa}^{\nu}). \nonumber
	\end{eqnarray}} 
\end{itemize}
\noindent

\subsection{Role of  Gauge Symmetry}\label{subsec:internal-symm}
So far we have discussed the possible structures of the operators which are constituted of quantum fields with spin $0,\, 1/2,\, 1$ only and taking only the space-time symmetry into account. In a realistic particle physics model, there are additional local and (or) global internal symmetries. As a result of this, there could be particles of different internal quantum numbers but possessing the same spin. Such fields will be equivalent to each other with respect to the Lorentz symmetry. But based on their internal charges, these fields will combine in a variety of ways leading to different sub-categories of operators within the same class. Thus, while Lorentz invariance provides us a list of possible operator classes, it is the internal symmetry which ultimately decides which combinations are permitted and which ones are not. This can be elucidated through the most popular example: the Standard Model  gauge symmetry. Looking into its particle content and their quantum numbers in Table~\ref{table:SM-fields}, it is evident that many of the operator classes in Table \ref{table:op-class-1to6}  do not respect the SM gauge symmetry. Thus they are excluded from the operator basis. Here, we have systematically explained the impact of internal gauge symmetries.
\begin{itemize}
	\item \underline{$\Phi^n$ operator class with integer $n$:} The SM  Higgs transforms under $SU(3)_C\otimes SU(2)_L\otimes U(1)_Y$ as $(1,2,1/2)$. Thus, the operators containing an odd number of  $H$ fields violate both $SU(2)$ and $U(1)$ symmetries. If $n$ is an even integer then all the operators of the forms $H^n$, $(H^{\dagger})^n$, and $H^{\frac{n}{2}}(H^{\dagger})^{\frac{n}{2}}$ are $SU(2)$ invariant. But only the $(H^\dagger H)$ and its powers are SM singlets. In BSM scenarios that contain multiple scalars, we may end up with more intricate structures. For example, if we add an $SU(2)$ triplet scalar $\Delta$ with hypercharge of +1, there will be an invariant operator $H^{T}\,\Delta^{\dagger}\,H \in \Phi^3$-class.	
	
	\item \underline{Operators involving field strength tensors:} Lorentz invariance allows us to construct terms containing an even number of field strength tensors. But the internal symmetry prevents their mixing, e.g., in SM there are no cross-terms between $B_{\mu\nu}$, $W_{\mu\nu}^{I}$ and $G_{\mu\nu}^{A}$. But this need not be true for certain BSM scenarios. For example, if there are multiple abelian symmetries, then we can expect some mixing in the gauge kinetic sector. Looking into the Lorentz symmetry only, the term involving tri-linear field strengths vanishes due to its anti-symmetric structure. But internal non-abelian gauge symmetries allow such terms at the dimension 6 level. Within the SM, $B^{\mu}{}_{\nu}B^{\nu}{}_{\kappa}B^{\kappa}{}_{\mu}$ is absent but $f^{ABC}\,G^{A\mu}_{\nu}\,G^{B\nu}_{\kappa}\,G^{C\kappa}_{\mu}$ and $\epsilon^{IJK}\,W^{I\mu}_{\nu}\,W^{J\nu}_{\kappa}\,W^{K\kappa}_{\mu}$ possess non-vanishing contributions. Here, the anti-symmetric tensors $f^{ABC}$ and $\epsilon^{IJK}$ are $SU(3)$ and $SU(2)$ structure constants respectively.

	\item \underline{Operators containing of  bi-linear fermion fields:}  Lorentz invariance allows fermion mass terms of the forms $\Psi_L^2$ (Majorana) and $\overline{\Psi}_L\,\Psi_R$ (Dirac). But in the SM, left and right chiral fermions are on a different footing. Hence, these terms are forbidden by the internal symmetries. Further, the quantum numbers of the fields allow the couplings of fermion bi-linears with the Higgs scalar in the form of the Yukawa interactions - $\overline{L}\,e\,H$, $\overline{Q}\,d\,H$ and $\overline{Q}\,u\,i\tau_2\,H^{*}$. In addition, the $SU(3)$ symmetry prevents the appearance of terms like $\overline{L}\,u$, $\overline{L}\,d$ and $\overline{Q}\,e$ \footnote{These terms appear as constituents of certain dimension 9 operators.}.
	Also, the operator class $(\overline{\Psi}_L\,\sigma_{\mu\nu}\,\Psi_R)\,\Phi\,X^{\mu\nu}$ appears at mass dimension 6. The choice of $X^{\mu\nu}$ and $\Psi$'s is fixed by the internal symmetries. There are fermion bi-linears which are not Lorentz scalar but may appear in higher mass dimensional operator class $(\overline{\Psi}\,\gamma_{\mu}\,\Psi)\,(\overline{\Psi^{\prime}}\,\gamma^{\mu}\,\Psi^{\prime})$ \footnote{The fermion fields $\Psi$ and $\Psi^{\prime}$ need not be same always.}.
	
\end{itemize}     

\subsection{Removal of redundancies and forming Operator basis}

So far we have learnt how to compute the invariant operators of any mass dimension based on the space-time and internal symmetries. But we must keep in mind the fact that these operators need to satisfy another criteria to be  phenomenologically relevant. The operators at each mass dimension must form a basis, i.e., they must be mutually independent. Thus it is necessary to remove all the redundancies, if any,  to compute the operator basis. In this construction, we have noted three different ways in which the operators can be interrelated: (i) integration by parts (IBP), (ii) equation of motion (EOM),  and (iii)  identities of symmetry generators.  Here, we have discussed these sources of redundancies briefly with examples based on SMEFT and beyond.

\subsubsection*{\underline{Integration by Parts (IBP)}}\label{subsec:IBP}

In our prescription, the covariant derivative ($\mathcal{D}_{\mu}$) participates in the operator construction in a similar way as the quantum fields. Due to the distributive property of  $\mathcal{D}_{\mu}$ and incorporating integration by parts (IBPs), two or more invariant operators may be related to each other by a total derivative. As we know such a term in the Lagrangian has no role to play, thus it can be removed. Therefore the multiple operators can not be treated independently and only one of them should be included in the operator basis. This duplication due to IBP occurs among different operators belonging to the same operator class. For example, at mass dimension 6, the operator  $\Psi_L\Psi_R\,\Phi^2\mathcal{D}$ can be recast in the following form:

\vspace{-0.5cm} 
{\small\begin{eqnarray}\label{eq:IBP}
i\mathcal{D}_{\mu}\,(\overline{\Psi}_{L,R}\,\gamma^{\mu}\,\Psi_{L,R}\,\Phi^{\dagger}\Phi) &=& \overline{\Psi}_{L,R}\,\gamma^{\mu}i\mathcal{D}_{\mu}\,\Psi_{L,R}\,\Phi^{\dagger}\Phi - \overline{\Psi}_{L,R}\,\gamma^{\mu}i\overleftarrow{\mathcal{D}_{\mu}}\,\Psi_{L,R}\,\Phi^{\dagger}\Phi \nonumber\\ 
& & + \overline{\Psi}_{L,R}\,\gamma^{\mu}\,\Psi_{L,R}\,\Phi^{\dagger}i\mathcal{D}_{\mu}\Phi - \overline{\Psi}_{L,R}\,\gamma^{\mu}\,\Psi_{L,R}\,\Phi^{\dagger}i\overleftarrow{\mathcal{D}_{\mu}}\Phi \nonumber\\
&=& (\overline{\Psi}_{L,R}\,\gamma^{\mu}i \overleftrightarrow{\mathcal{D}}_{\mu}\,\Psi_{L,R})\,\Phi^{\dagger}\Phi + \overline{\Psi}_{L,R}\,\gamma^{\mu}\,\Psi_{L,R}\,(\Phi^{\dagger}i\overleftrightarrow{\mathcal{D}}_{\mu}\,\Phi). 
\end{eqnarray}}
Here, $i\overleftrightarrow{\mathcal{D}}_{\mu} \equiv i \mathcal{D}_{\mu} - i \overleftarrow{\mathcal{D}}_{\mu}$ has been introduced to combine the first two and the last two operators to form  $(\overline{\Psi}_{L,R}\,\gamma^{\mu}i \overleftrightarrow{\mathcal{D}}_{\mu}\,\Psi_{L,R})\,\Phi^{\dagger}\Phi$ and $\overline{\Psi}_{L,R}\,\gamma^{\mu}\,\Psi_{L,R}\,(\Phi^{\dagger}i\overleftrightarrow{\mathcal{D}}_{\mu}\,\Phi)$ which are self-hermitian. It is evident from Eqn.~\eqref{eq:IBP}, that these operators are related to each other by a total derivative term $\mathcal{D}_{\mu}\,(\overline{\Psi}_{L,R}\,\gamma^{\mu}\,\Psi_{L,R}\,\Phi^{\dagger}\Phi)$. 
So, in the operator basis we will include only one of them. Here, our choice of the independent operator will be the one where the derivative acts on the scalar field. This is because the latter structure where the derivative acts on the fermions is related to other operators through equations of motion. We will justify this choice in the following section.
 
\subsubsection*{\underline{Equation of Motion (EOM)}}\label{subsec:eom}
The quantum fields representing the particles are dynamical in nature and each of them satisfies their respective equation of motion.  It has been noted that two or more operators may be related to each other through the EOMs of the involved fields along with the IBPs \cite{Henning:2017fpj,Grzadkowski:2010es}. Unlike the previous case, the EOMs can relate operators belonging to different classes. We have explained how EOM leads to redundancy using a few examples:

\begin{itemize}
	\item $\boxed{\Psi_L\Psi_R\,\Phi^2\mathcal{D}}$ : In the Weyl basis we can have two possible covariant structures for this operator: $(\overline{\Psi}_{L,R}\,\gamma^{\mu}i \overleftrightarrow{\mathcal{D}}_{\mu}\,\Psi_{L,R})\,\Phi^{\dagger}\Phi$ and $\overline{\Psi}_{L,R}\,\gamma^{\mu}\,\Psi_{L,R}\,(\Phi^{\dagger}i\overleftrightarrow{\mathcal{D}}_{\mu}\,\Phi)$. We have already noted that these two operators differ from each other by a total derivative.  There we have further mentioned that we have selected the operator where the derivative is acting on the scalars. The reason behind that choice is that after incorporating the EOMs of  $\Psi$ or its conjugate $\overline{\Psi}$, this operator reduces to an operator belonging to $\Psi_{L,R}^2 \Phi^3$ class:
	
	\vspace{-0.7cm}
	{\small\begin{eqnarray}
	(\overline{\Psi}_{L,R}\,\gamma^{\mu}i \overleftrightarrow{\mathcal{D}}_{\mu}\,\Psi_{L,R})\,\Phi^{\dagger}\Phi \,\, \propto \,\, \overline{\Psi}_{L,R}\,\Psi_{R,L}\,\Phi\,(\Phi^{\dagger}\Phi) \,\, \equiv \Psi_{L,R}^2\,\Phi^3.
	\end{eqnarray}}
	\item $\boxed{\Psi_{L,R}^2\,\Phi\,D^2}$ : The unique  covariant form of this operator is  $(\overline{\Psi}_{L}\,\Psi_R)\,\mathcal{D}^2\,\Phi$. After implementing  the EOM of the scalar field: 
	$	\mathcal{D}^2\,\Phi = c_1\,\Phi + c_2\,\Phi\,(\Phi^{\dagger}\Phi) + c_3\,\overline{\Psi}_{R}\,\Psi_L$, this operator can be reduced in the following form:
	
	\vspace{-0.7cm}
	{\small\begin{eqnarray}
	(\overline{\Psi}_{L}\Psi_R)\,\mathcal{D}^2\,\Phi &=& c_1\,\underbrace{(\overline{\Psi}_{L}\Psi_R\,\Phi)}_{\text{dim-4 term}} + c_2\,(\overline{\Psi}_{L}\Psi_R\,\Phi)\,(\Phi^{\dagger}\Phi) + c_3\,(\overline{\Psi}_{L}\Psi_R)\,(\overline{\Psi}_{R}\Psi_L),
	\end{eqnarray}} 
	with $c_1, c_2$ and $c_3$ being complex numbers. Thus, the operator class $\Psi^2_{L,R}\,\Phi\,\mathcal{D}^2$ can be expressed as a linear combination of two other dimension 6 classes $\Psi_{L,R}^2 \Phi^3$ and $\Psi_{L}^2 \Psi_{R}^2$, and therefore is excluded from the set of independent operators.  
	
	\item $\boxed{\mathcal{D}^2\,X^2_{L,R}, \, \mathcal{D}^2\,X_L\,X_R}$ : The possible covariant form of the operators are (i) $(\mathcal{D}_{\mu}X^{\mu\nu})^2$, (ii) $(\mathcal{D}_{\mu}X^{\mu\nu})(\mathcal{D}_{\mu}\tilde{X}^{\mu\nu})$, and (iii) $(\mathcal{D}_{\mu}\tilde{X}^{\mu\nu})^2$. It is interesting to note that after implementing the EOM of field strength tensors:
	
	\vspace{-0.7cm}
	{\small\begin{eqnarray}\label{eq:eom-gauge-boson}
	\mathcal{D}_{\mu}\tilde{X}^{\mu\nu} = 0, \hspace{0.7cm} \mathcal{D}_{\mu}X^{\mu\nu} = \overline{\Psi}_{L,R}\gamma^{\nu}\,\Psi_{L,R} + \Phi^{\dagger}i\overleftrightarrow{\mathcal{D}}^{\nu}\Phi,
	\end{eqnarray}}
	the last two structures (ii) and (iii)  identically vanish.  The very  first operator can be rewritten either as:
	
	\vspace{-0.7cm}
	{\small\begin{eqnarray}
	(\mathcal{D}_{\mu}X^{\mu\nu})^2 &=& a_1 (\overline{\Psi}_{L,R}\,\gamma_{\nu}\,\Psi_{L,R})(\mathcal{D}_{\mu}\,X^{\mu\nu}) + a_2(\Phi^{\dagger}i\overleftrightarrow{\mathcal{D}}_{\nu}\Phi)(\mathcal{D}_{\mu}\,X^{\mu\nu}),
	\end{eqnarray}} 
	or as:
	{\small\begin{eqnarray}
	(\mathcal{D}_{\mu}X^{\mu\nu})^2 &=& b_1 \,(\overline{\Psi}_{L,R}\,\gamma^{\nu}\,\Psi_{L,R})^2 + b_2\,(\Phi^{\dagger}i\overleftrightarrow{\mathcal{D}}^{\nu}\Phi)^2 + b_3(\Phi^{\dagger}i\overleftrightarrow{\mathcal{D}}_{\nu}\Phi)(\overline{\Psi}_{L,R}\,\gamma^{\nu}\,\Psi_{L,R}).
	\end{eqnarray}} 
	Here, $a_i, b_i$ are complex numbers. Thus we can generate operators belonging to $\Phi^4 \mathcal{D}^2,\, \Psi^4,\, \Phi^2 \Psi^2 \mathcal{D}$ starting from  $\mathcal{D}^{2}X^2$ class of operators and thus it is redundant and  can not be a part of the operator basis. 
	
	Alternatively, using the notion of integration by parts (IBP) we have the following relation:
	
	\vspace{-0.5cm}
	{\small\begin{eqnarray}
	(\mathcal{D}_{\mu}\,X^{\mu\nu})^2, \hspace{0.1cm} (\mathcal{D}_{\mu}\,X^{\mu\nu})(\mathcal{D}_{\mu}\,\tilde{X}^{\mu\nu})\,\, &\xRightarrow{\text{IBP}}& [\mathcal{D}_{\mu},\,\mathcal{D}_{\nu}] X^{[\mu\kappa} \,X_{\kappa}^{\nu]}, \hspace{0.1cm} [\mathcal{D}_{\mu},\,\mathcal{D}_{\nu}] X^{[\mu\kappa} \,\tilde{X}_{\kappa}^{\nu]}\nonumber\\
	&\equiv& X_{\mu\nu} X^{\mu\kappa} X_{\kappa}^{\nu}, \hspace{0.1cm} X_{\mu\nu} X^{\mu\kappa} \tilde{X}_{\kappa}^{\nu}.
	\end{eqnarray}}
	Here,  $[\mathcal{D}_{\mu},\,\mathcal{D}_{\nu}]$ is suitably replaced by $X_{\mu\nu}$ and we have obtained  $X^3$ class of operators. So, we conclude that with the help of EOMs and IBPs, the operators belonging to $\mathcal{D}^2\,X^2_{L,R}$ and $\mathcal{D}^2\,X_{L}\,X_{R}$ classes can always be recast into operators of other classes. Thus these two are excluded from the operator basis. 
	 
	\item $\boxed{\Phi^2\,X_{L,R}\,\mathcal{D}^2}$ : The covariant form of this operator  $(\Phi^{\dagger}i\overleftrightarrow{\mathcal{D}}_{\nu}\,\Phi)\,\mathcal{D}_{\mu}\,X^{\mu\nu}$ can be rewritten using  Eqn.~\eqref{eq:eom-gauge-boson} as:
	{\small\begin{eqnarray}
	(\Phi^{\dagger}i\overleftrightarrow{\mathcal{D}}_{\nu}\,\Phi)\,\mathcal{D}_{\mu}\,X^{\mu\nu} = a^{\prime}\, (\overline{\Psi}_{L,R}\,\gamma_{\nu}\,\Psi_{L,R})(\Phi^{\dagger}\,i\overleftrightarrow{\mathcal{D}}^{\nu}\,\Phi) + b^{\prime}\,(\Phi^{\dagger}\,i\overleftrightarrow{\mathcal{D}}_{\nu}\,\Phi)(\Phi^{\dagger}\,i\overleftrightarrow{\mathcal{D}}^{\nu}\,\Phi),
	\end{eqnarray}}
	where $a^{\prime}, b^{\prime}$ are complex numbers. Similar to the previous case,  $\Phi^2 X\,\mathcal{D}^2$ can be rewritten in terms of  operator classes $\Psi_L\Psi_R\,\Phi^2\mathcal{D}$ and $\Phi^4\mathcal{D}^2$.  This justifies the absence of  $\Phi^2\,X_{L,R}\,\mathcal{D}^2$ class from the independent operator set.  
	
	\item $\boxed{\Psi_L\,\Psi_R\,X_{L,R}\,\mathcal{D}}$ : We find two different covariant forms $X^{\mu\nu}\,(\overline{\Psi}_{L,R}\,\gamma_{\mu}\,\mathcal{D}_{\nu}\,\Psi_{L,R})$ and $(\mathcal{D}_{\mu}\,X^{\mu\nu})(\overline{\Psi}_{L,R}\,\gamma_{\nu}\,\Psi_{L,R})$. These operators can be further reduced with the help of suitable EOMs as:
	
	\vspace{-0.7cm}
	{\small\begin{eqnarray}
	X^{\mu\nu}\,(\overline{\Psi}_{L,R}\,\gamma_{\mu}\,\mathcal{D}_{\nu}\,\Psi_{L,R}) &=& X^{\mu\nu}\,(\overline{\Psi}_{L,R}\,\gamma_{\mu}\,\gamma_{\nu}\,\slashed{\mathcal{D}}\,\Psi_{L,R}) = X^{\mu\nu}\,(\overline{\Psi}_{L,R}\,\gamma_{[\mu}\,\gamma_{\nu]}\,\slashed{\mathcal{D}}\,\Psi_{L,R}) \nonumber\\
	&=& X^{\mu\nu}\,(\overline{\Psi}_{L,R}\,\sigma_{\mu\nu}\,\Psi_{R,L})\,\Phi \,\,\,\,\,\equiv \,\,\, \Psi^{2}\,\Phi\,X,\\
	(\mathcal{D}_{\mu}\,X^{\mu\nu})(\overline{\Psi}_{L,R}\,\gamma_{\nu}\,\Psi_{L,R}) &=&  c^{\prime}_1\,(\overline{\Psi}\,\gamma^{\nu}\,\Psi)\,(\overline{\Psi}_{L,R}\,\gamma_{\nu}\,\Psi_{L,R}) + c^{\prime}_2\,(\Phi^{\dagger}\,i\overleftrightarrow{\mathcal{D}}^{\nu}\,\Phi)\,(\overline{\Psi}_{L,R}\,\gamma_{\nu}\,\Psi_{L,R})\nonumber\\
	&\equiv& \Psi^4_{L,R}\,/\,\Psi^2_{L}\,\Psi^2_{R} + \Psi_{L}\,\Psi_{R}\,\Phi^2\,\mathcal{D}.
	\end{eqnarray}}
	Thus it is quite evident why this class is also counted as redundant.
\end{itemize}

\begin{figure}[h]
	\centering
	{
		\includegraphics[scale=0.7]{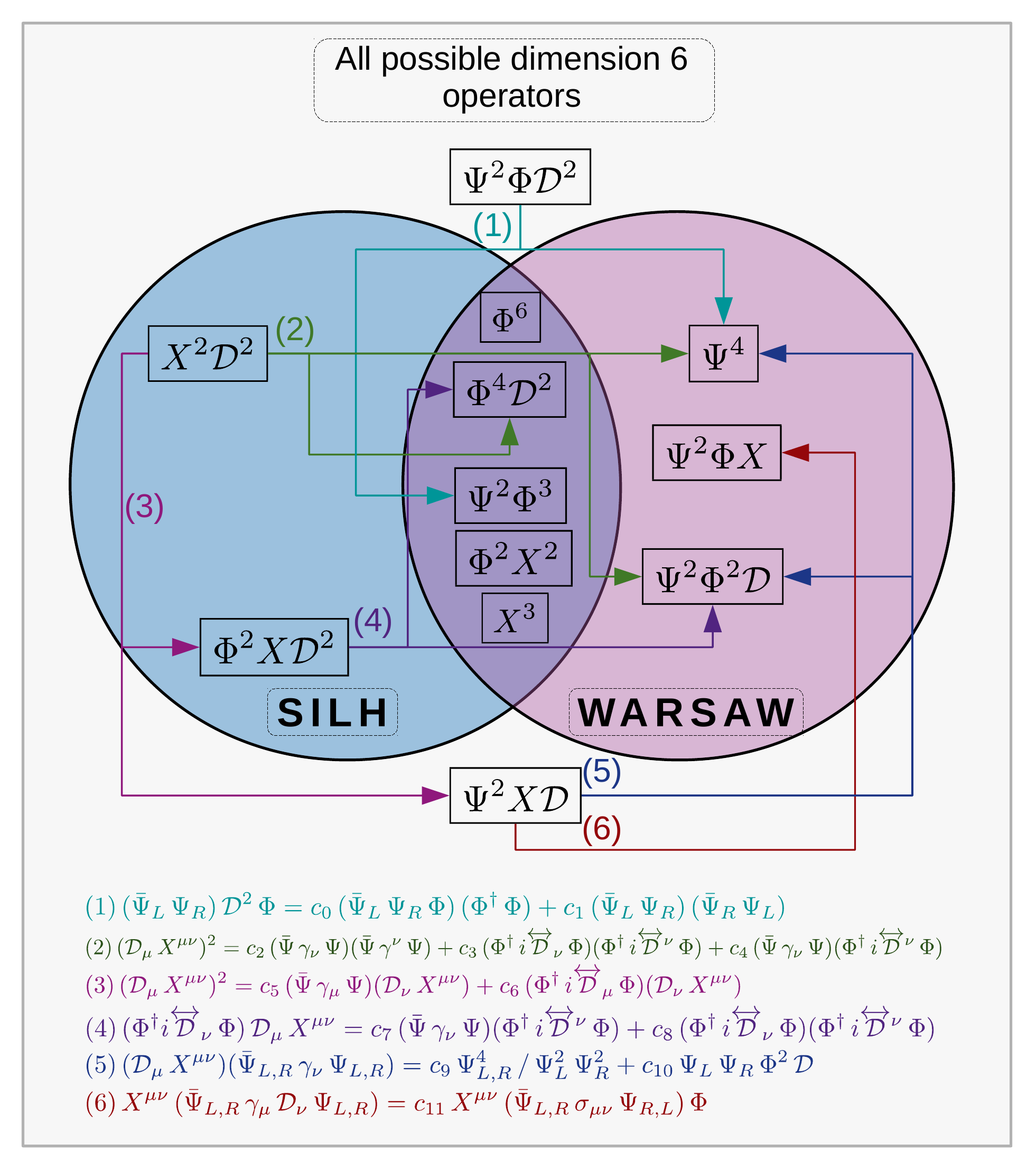}
	}
	\caption{All possible Lorentz invariant dimension 6 operator classes shown as part of the Warsaw and SILH bases for SMEFT. The arrows depict relations among the classes based on the equations of motion (EOMs) of various fields.}
	\label{fig:1}
\end{figure}

\noindent 
In summary, the symmetries of the theory play a crucial role in constructing the invariant operator set. But it is not guaranteed that all of them are independent and thus the set of operators is always over-complete. 
To be a part of the Lagrangian the operators of any mass dimension must form a basis, i.e., the operators should be independent. To ensure that we have shown through some toy examples how the EOMs and IBPs relate different operators and thus can be used as constraints in this computation. 
In the latter part of this paper, we have computed the dimension 6 operator basis for a plethora of models. As the ``Warsaw" is the only known complete operator basis, we have tabulated our results in this basis only. There is another popular choice  - the SILH (Strongly Interacting Light Higgs) basis which trades away the fermion rich operator classes $\Psi^4,\, \Psi^2\Phi X,\,  \Psi^2\Phi^2 \mathcal{D}$ from the Warsaw one and includes $\mathcal{D}^2 X^2,\, \Phi^2 X \mathcal{D}^2$, see  Fig.~\ref{fig:1}.

\subsubsection*{\underline{ Symmetry generators and their identities}}\label{subsec:fierz-identities}

The quantum fields that are the building blocks of the operators transform under the assigned space-time and internal symmetries. The symmetry generators (specifically for the non-abelian case) respect the pre-fixed algebras and satisfy a few identities. For example,  the Lorentz symmetry generators  $\sigma^{\mu}, \overline{\sigma}^{\mu}$ together form the $\sigma^{\mu\nu}, \overline{\sigma}^{\mu\nu}$ matrices, defined in Eqn.~\eqref{eq:gamma-mu-nu}: 

\vspace{-1cm}
{\small\begin{eqnarray}\label{eq:gamma-mu-nu}
		(\sigma^{\mu\nu})_{\alpha}^{\beta}=(\sigma^{\mu})_{\alpha\dot{\beta}} (\overline{\sigma}^{\nu})^{\dot{\beta}\beta}, \hspace{1cm} (\overline{\sigma}_{\mu\nu})_{\dot{\alpha}}^{\dot{\beta}}=(\overline{\sigma}_{\mu})^{\dot{\beta}\beta} (\sigma_{\nu})_{\beta\dot{\alpha}}\,\, .
\end{eqnarray}}
They also satisfy the following identities \cite{Dreiner:2008tw}:

\vspace{-0.6cm}
{\small\begin{eqnarray}
		\label{eq:sigma-idntt1} 
		(\sigma^\mu)_{\alpha\dot{\alpha}}(\sigma_\mu)_{\beta\dot{\beta}} 
	&	= & 2\epsilon_{\alpha\beta}\epsilon_{\dot{\alpha}\dot{\beta}}\,,\\ 
		\label{eq:sigma-idntt2}
		(\sigma^\mu)_{\alpha\dot{\alpha}}(\overline{\sigma}_\mu)^{\dot{\beta}\beta} 
	&	= &  2\delta^{\beta}_{\alpha} \delta^{\dot{\beta}}_{\dot{\alpha}}\,,\\
		\label{eq:sigma-idntt3}
		(\overline{\sigma}^\mu)^{\dot{\alpha}\alpha}(\overline{\sigma}_\mu)^{\dot{\beta}\beta} 
	&	= &  2\epsilon^{\alpha \beta}\epsilon^{\dot{\alpha}\dot{\beta}}\,,\\
		\label{eq:sigma-idntt4}
		[\sigma^\mu \overline{\sigma}^\nu + \sigma^\nu \overline{\sigma}^{\mu}]_{\alpha}^{\beta}
	&	= &  2 g^{\mu\nu} \delta_{\alpha}^{\beta}\,. 
	\end{eqnarray}}
The internal symmetry generators respect their algebra as well as some related identities. For example, 
the  $SU(2)$ and $SU(3)$ generators, Pauli matrices $\tau^I (I=1,2,3)$ and the Gell-Mann matrices $T^A (A=1,2,\cdots,8)$ respectively satisfy the following identities:

\vspace{-0.6cm}
{\small\begin{eqnarray}
			\label{eq:int-generator-idntt1}
	\tau^I_{ij}\tau^I_{kl}
= & 2 \delta_{il}\delta_{jk} -\delta_{ij} \delta_{kl} \,,\\
\label{eq:int-generator-idntt2}
T^A_{ij}T^A_{kl}
= & \frac{1}{2}\delta_{il}\delta_{jk} -\frac{1}{6}\delta_{ij} \delta_{kl}\,.
\end{eqnarray}}
\noindent
While constructing the covariant form of the operators we may encounter two different structures with the same field content. But they need not be two independent operators and may be related to each other through  these identities Eqns.~\eqref{eq:sigma-idntt1}-\eqref{eq:int-generator-idntt2}.
Here, we have demonstrated how the utilisation of these identities could help us to relate different covariant-structured  dimension 6 operators with a few examples\footnote{For this particular discussion we have suppressed the chiral-indices.}.

\begin{itemize}
	\item $ \boxed{\Psi^4} $: We have considered two  dimension 6 operators  $(\overline{d}\,\gamma^\mu\,T^A\, d)(\overline{Q}\,\gamma_\mu\,T^A\,Q)$ and $(\overline{d}\,\gamma^\mu\,d)(\overline{Q}\,\gamma_\mu\,Q)$ from this class. Using the identities in Eqns.~\eqref{eq:sigma-idntt1}-\eqref{eq:int-generator-idntt1}, these operators can be expressed as:
	{\small\begin{eqnarray} \label{eq:fierz-psi4-cls-1}
			& (\overline{d}\gamma^\mu T^A d)(\overline{Q} T^A \gamma_\mu Q)
			& = (\overline{d}^{\alpha}\sigma^{\mu}_{\alpha\dot{\alpha}} T^A d^{\dot{\alpha}}) (\overline{Q}_{\dot{\beta}} T^A \overline{\sigma}^{\mu\dot{\beta}\beta} Q_{\beta}) = 2 \overline{d}^\alpha T^A Q_{\beta} \overline{Q}_{\dot{\beta}} T^A d^{\dot{\alpha}}  \delta^{\beta}_{\alpha} \delta^{\dot{\beta}}_{\dot{\alpha}}  \nonumber \\
			& & = 2 (\overline{d}\,T^A\,Q)\,(\overline{Q}\, T^A\,d)  = 2 (\overline{d}_a\, [T^A]_b^a\, Q^{b})(\overline{Q}_{c}\,[T^A]_e^c\,d^{e}) \nonumber \\
			& & = (\overline{d}\,d)\,(\overline{Q}\,Q) - \frac{1}{3}(\overline{d}\,Q) (\overline{Q}\,d)\,. \\
			\label{eq:fierz-psi4-cls-2}
			& (\overline{d}\gamma^\mu d)(\overline{Q} \gamma_\mu Q)
			& = (\overline{d}^{\alpha}\sigma^{\mu}_{\alpha\dot{\alpha}} d^{\dot{\alpha}}) (\overline{Q}_{\dot{\beta}}  \overline{\sigma}^{\mu\dot{\beta}\beta} Q_{\beta}) = 2 \overline{d}^\alpha  Q_{\beta} \overline{Q}_{\dot{\beta}}  d^{\dot{\alpha}}  \delta^{\beta}_{\alpha} \delta^{\dot{\beta}}_{\dot{\alpha}}  \nonumber \\
			& & = 2 (\overline{d} Q)(\overline{Q} d). 
	\end{eqnarray}}
It is quite evident from Eqns~\eqref{eq:fierz-psi4-cls-1} and \eqref{eq:fierz-psi4-cls-2}, that with the fields $d$, $\bar{d}$, $Q$,  and $\bar{Q}$ we can only have two independent operators that should be included in SMEFT dimension 6 operator basis.
Similarly, with fields $e$, $\bar{L}$, $u$,  and $\bar{Q}$ we have following relation
	
	\vspace{-0.8cm}
	{\small\begin{eqnarray}\label{eq:fierz-psi4-cls-3}
			& (\bar{L}\sigma_{\mu\nu}e)(\bar{Q}\sigma^{\mu\nu}u)
			& = ( (\bar{L})^{\alpha} (\sigma_{\mu\nu})_{\alpha}^{\beta} (e)_{\beta}) ((\bar{Q})^{\rho}(\sigma^{\mu\nu})_{\rho}^{\theta} (u)_{\theta}) \nonumber  \\
			& & = ((\bar{L})^{\alpha}(\sigma_{\mu})_{\alpha \dot{\beta}} (\overline{\sigma}_{\nu})^{\dot{\beta} \beta} (e)_{\beta}) ((\bar{L})^{\rho}(\sigma^{\mu})_{\rho \dot{\theta}} (\overline{\sigma}^{\nu})^{\dot{\theta} \theta} (u)_{\theta}) \nonumber \\
			& & = 4(\bar{L} e) (\bar{Q} u) -8 (\bar{L} u) (\bar{Q} e) \,.
	\end{eqnarray}}%
and  thus, only $(\bar{L}\sigma_{\mu\nu}e)(\bar{Q}\sigma^{\mu\nu}u)$ and $(\bar{L}\,e)(\bar{Q}\,u)$ are included in the operator set.
	
	\item $\boxed{\Phi^6}$ : Here, we are looking into the   quartic subpart of the dimension 6 operator $(H^{\dagger}\,H)^3$. It is interesting to note using Eqn.~\eqref{eq:int-generator-idntt1} that inclusion of $SU(2)$ generators does not lead to an independent operator in the SMEFT basis \cite{Grzadkowski:2010es}:
	{\small\begin{eqnarray}\label{eq:fierz-phi6-cls}
			& (H^\dagger \tau^I H)(H^\dagger \tau^I H)
			& = (H^\dagger_i \tau^I_{ij} H_j)(H^\dagger_k \tau^I_{kl} H_l)  = H^\dagger_i H_j H^\dagger_k H_l (2 \delta_{il}\delta_{jk} -\delta_{ij} \delta_{kl}) \nonumber \\
			& & = 2 (H^\dagger H)^2 -(H^\dagger H)^2  = (H^\dagger H)^2 \,.
	\end{eqnarray}}	
	\item $ \boxed{\Phi^4\mathcal{D}^2} $ : To illustrate the redundancy in this class of operators, we have considered an operator involving a scalar Lepto-Quark ($\chi_1$) transforming as $ (3,2,1/6) $ under the SM gauge group. 
	{\small\begin{eqnarray} \label{eq:fierz-phi-D-cls}
	& (H^\dagger i\overleftrightarrow{\mathcal{D}}_{\mu}^I H) (\chi_{1}^{\dagger} i\overleftrightarrow{\mathcal{D}}^{\mu I}\chi_1)
	& = (H^\dagger (\tau^Ii\mathcal{D}_{\mu}-i\overleftarrow{\mathcal{D}}_{\mu} \tau^I) H) (\chi^{\dagger}_1 (\tau^Ii\mathcal{D}^{\mu}-i\overleftarrow{\mathcal{D}}^{\mu} \tau^I) \chi_{1}) \nonumber \\
	& & = (H^{\dagger}_{i} \tau^{I}_{ij} (i\mathcal{D}_{\mu} H)_{j} - (i\mathcal{D}_{\mu} H)^{\dagger}_{k} \tau^{I}_{kl} H_{l}) (\chi_{1a}^{\dagger} \tau^{I}_{ab} (i\mathcal{D}^{\mu} \chi_{1})_{b} - (i\mathcal{D}^{\mu} \chi_{1})^{\dagger}_{c} \tau^I_{cd} \chi_{1d} ) \nonumber \\
	& & = - (H^\dagger i\overleftrightarrow{\mathcal{D}}_{\mu} H) (\chi_{1}{^\dagger} i\overleftrightarrow{\mathcal{D}}^{\mu}\chi_{1}) + 2\,(H^\dagger_{b} i\overleftrightarrow{\mathcal{D}}_{\mu}H^{a})(\chi_{1a}^\dagger i\overleftrightarrow{\mathcal{D}}^{\mu}\chi_1^{b}). 
	\end{eqnarray}}
	As three operators are related through the above relation, only two of these can be independent and we may include $(H^\dagger i\overleftrightarrow{\mathcal{D}}_{\mu}^I H) (\chi_{1}^{\dagger} i\overleftrightarrow{\mathcal{D}}^{\mu I}\chi_1)$ and $(H^\dagger i\overleftrightarrow{\mathcal{D}}_{\mu} H) (\chi_{1}^{\dagger} i\overleftrightarrow{\mathcal{D}}^{\mu}\chi_1)$ in the operator basis for this scenario.
	
	\item $ \boxed{\Psi^2 \Phi^2 \mathcal{D}} $ : In this class we can have following three operators involving the Lepto-Quark $\chi_1$:
	
	\vspace{-0.7cm}
	{\small\begin{eqnarray}\label{eq:fierz-psi-phi-D-cls-1}
	(\overline{Q} \tau^I \gamma^{\mu} Q) \,(\chi^{\dagger}_1 i\overleftrightarrow{\mathcal{D}}_{\mu}^I \chi_1)
	&= & (\overline{Q} \tau^I \gamma^\mu Q)[\chi_{1}^{\dagger} \tau^I (i\mathcal{D}_{\mu}\chi_{1})+(i\mathcal{D}_{\mu}\chi_{1})^{\dagger} \tau^I\chi_{1}] \nonumber \\
	& = & 2(\overline{Q} \gamma^{\mu} (i\mathcal{D}_{\mu}\chi_{1})) (Q \chi_{1}^{\dagger}) - (\overline{Q} \gamma^{\mu} Q )(\chi_{1}^{\dagger}  (i\mathcal{D}_{\mu}\chi_{1})) \nonumber \\
	&  & + 2(\overline{Q} \gamma^{\mu} \chi_{1}) (Q  (i\mathcal{D}_{\mu}\chi_{1})^{\dagger}) - (\overline{Q} \gamma^{\mu} Q )((i\mathcal{D}_{\mu}\chi_{1})^{\dagger} \chi_{1}), \\
	\label{eq:fierz-psi-phi-D-cls-2}
	(\overline{Q} T^A \gamma^{\mu} \,Q) \,(\chi^{\dagger}_1 i\overleftrightarrow{\mathcal{D}}_{\mu}^A \chi_1)
			&=& (\overline{Q} T^A \gamma^\mu Q)[\chi_{1}^{\dagger} T^A  (i\mathcal{D}_{\mu}\chi_{1})+(i\mathcal{D}_{\mu}\chi_{1})^{\dagger}T^A \chi_{1}] \nonumber \\
			& = & \frac{1}{2} (\overline{Q} \gamma^{\mu} (i\mathcal{D}_{\mu}\chi_{1})) (Q \chi_{1}^{\dagger}) - \frac{1}{6}(\overline{Q} \gamma^{\mu} Q )(\chi_{1}^{\dagger}  (i\mathcal{D}_{\mu}\chi_{1})) \nonumber \\
			&  & +\frac{1}{2}(\overline{Q} \gamma^{\mu} \chi_{1}) (Q (i\mathcal{D}_{\mu}\chi_{1})^{\dagger}) - \frac{1}{6}(\overline{Q} \gamma^{\mu} Q )((i\mathcal{D}_{\mu}\chi_{1})^{\dagger} \chi_{1}), \\
			\label{eq:fierz-psi-phi-D-cls-3}
			(\overline{Q} \,T^A \,\tau^I \,\gamma^{\mu} \,Q) \,(\chi^{\dagger}_1 \,T^A \,i\overleftrightarrow{\mathcal{D}}_{\mu}^I \,\chi_1)
			&= & (\overline{Q} T^A \tau^I \gamma^\mu Q)[\chi_{1}^{\dagger} T^A \tau^I (i\mathcal{D}_{\mu}\chi_{1})+(i\mathcal{D}_{\mu}\chi_{1})^{\dagger}T^A \tau^I\chi_{1}] \nonumber \\
			& = & \frac{1}{2}[(\overline{Q} \tau^I \gamma^{\mu} (i\mathcal{D}_{\mu}\chi_{1})) (Q \tau^I \chi_{1}^{\dagger})] -\frac{1}{6}[(\overline{Q} \tau^I \gamma^{\mu} Q )(\chi_{1}^{\dagger} \tau^I (i\mathcal{D}_{\mu}\chi_{1}))] \nonumber \\
			&  & +\frac{1}{2}[(\overline{Q} \tau^I \gamma^{\mu} \chi_{1}) (Q \tau^I (i\mathcal{D}_{\mu}\chi_{1})^{\dagger})] -\frac{1}{6}[(\overline{Q} \tau^I \gamma^{\mu} Q )((i\mathcal{D}_{\mu}\chi_{1})^{\dagger} \tau^I \chi_{1})] \nonumber \\
			& = & [(\overline{Q} \gamma^{\mu} \chi_{1}^{\dagger} ) ( (i\mathcal{D}_{\mu}\chi_{1}) Q )- \frac{1}{2} ((\overline{Q} \gamma^{\mu} (i\mathcal{D}_{\mu}\chi_{1})) (Q \chi_{1}^{\dagger}) )] \nonumber \\
			& & -\frac{1}{6}[(\overline{Q} \gamma^{\mu} (i\mathcal{D}_{\mu}\chi_{1}) )(Q \chi_{1}^{\dagger}) - (\overline{Q} \gamma^{\mu} Q )(\chi_{1}^{\dagger} (i\mathcal{D}_{\mu}\chi_{1}))] \nonumber \\
			& & + [(\overline{Q} \gamma^{\mu} (i\mathcal{D}_{\mu}\chi_{1})^{\dagger}) (\chi_{1}Q) 
			- \frac{1}{2}(\overline{Q} \gamma^{\mu} \chi_{1}) (Q (i\mathcal{D}_{\mu}\chi_{1})^{\dagger}) ] \nonumber \\
			& & -\frac{1}{6}[ (\overline{Q} \gamma^{\mu} \chi_{1}) ( Q (i\mathcal{D}_{\mu}\chi_{1})^{\dagger}) - (\overline{Q} \gamma^{\mu} Q )((i\mathcal{D}_{\mu} \chi_{1})^{\dagger} \chi_{1})].
	\end{eqnarray}}%
	Thus, it is evident that the three operators in the LHS of the above equation along with $ (\overline{Q} \gamma^{\mu} Q) \,(\chi^{\dagger}_1 i\overleftrightarrow{\mathcal{D}}_{\mu} \chi_1) $, comprise a set of four independent operators and qualify to be in the operator basis.
\end{itemize}

\subsection{Additional impacts of the  Global (Accidental) Symmetries}\label{subsec:accidental-symm}

The effect of global symmetries is very similar to the gauge ones in the construction of invariant operators.  But, unlike the gauge symmetry, the global symmetry need not be strictly imposed and it may be allowed to be broken softly in specific interactions as demanded by the phenomenology.  This leads to the appearance of global charge violating effective operators that induce rare processes.

Baryon ($B$) and lepton ($L$) numbers appear as accidental global symmetries in the tree-level SM Lagrangian, see Eqn.~\eqref{eq:SM-ren-lag}. But they may be violated through higher dimensional operators. If we assign the leptons an $L$ charge of -1 unit and the quarks a $B$ charge of 1/3 units respectively, then we can generate a Majorana neutrino mass for the SM neutrinos through dimension 5 $H^2L^2$ operator. As this operator is suppressed by a high scale, the smallness of neutrino masses can be explained. Similarly within the SMEFT framework, we find operators violating $B$ and $L$ by $(0,-2)$, $(1,-1)$, $(1,1)$ units at mass dimensions 5, 6 and 7. Recently it has been noted \cite{Hambye:2013jsa} that a similar violation by  $(1,-3)$  units appears at dimension 9 and this can induce a new decay mode of the proton to three charged leptons.

In the case of BSM scenarios, there could be additional global symmetries and the amount of their breaking would be completely phenomenologically driven. This controls the appearance of certain kinds of operators at different mass dimensions.

\section{BSMEFT Operator Bases}\label{sec:op-bases}
\begin{figure}[h]
	\centering
	{
		\includegraphics[scale=0.8, trim= 30 40 0 0]{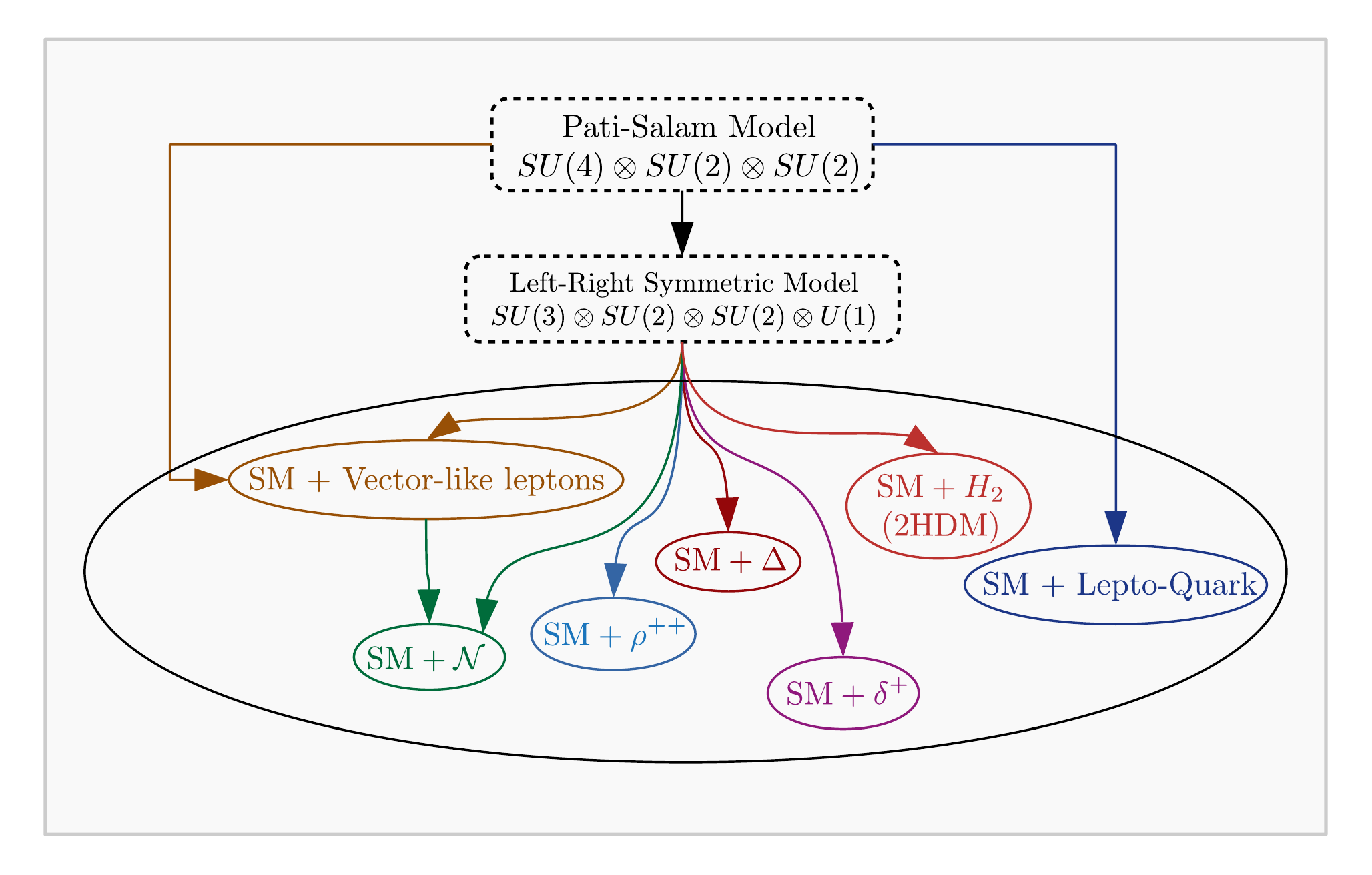}
	}
	\caption{An elucidation of the inter-connectedness of several BSM scenarios: paving the path to BSMEFT.}
	\label{fig:2}
\end{figure}

The spectrum of the UV complete theory  that is expected to explain all shortcomings of the SM is non-degenerate. This implies the existence of a multitude of scales associated with BSM fields of different masses. Thus even if all the non-SM particles are integrated out, all the higher dimensional operators will not be suppressed by a single cut-off scale ($\Lambda$). The natural scenario would be the presence of a tower of effective operators involving different $\Lambda$'s   lying between the electroweak  and the unknown UV scales. Unless the BSM spectrum is really compressed, the lightest  non-SM particle is expected to be within the reach of the ongoing experiments ($\leq \mathcal{O}$(TeV)) where the rest of the {\it new} particles are heavy enough to be successfully integrated out. In this framework, that lightest non-SM particle should be treated as an IR-DOF along with the SM ones and we must compute the effective operators involving them to capture the effects of the full UV theory. This has been the motivation of our BSMEFT construction.  The most generic choices for non-SM IR-DOFs are   real and complex scalar and fermion multiplets, vector like fermions, and Lepto-Quark bosons under the SM gauge symmetry. There may be additional gauge bosons as well.  The choice of these fields is motivated from the fact that most of the phenomenologically interesting scenarios contain these DOFs in their (non)minimal versions. We have schematically demonstrated the idea using some example scenarios in Fig.~\ref{fig:2} where it is quite evident that there could be multiple parent UV theories which may lead to the same set of lighter particles. Thus one  BSMEFT operator basis qualifies to encapsulate the features of all such UV theories treating them degenerate. To discriminate between them, we need to identify the subset of that BSMEFT operator basis corresponding to each of the UV theories. This is beyond the goal of this paper and will be discussed in our upcoming article.

\noindent
Based on the previous discussion, we have considered three different extensions of the SM and for each such scenario we have included multiple examples to encompass the most popular choices. For each example model, we have constructed the complete and independent BSMEFT operator bases up to mass dimension 6. Here, we have tabulated only the additional effective operators beyond SMEFT. For the sake of completeness the SMEFT dimension 6 operators are noted in the appendix.

\subsection{Standard Model extended by uncolored particles}

To start with, we have considered the scenarios where the SM is extended by suitable addition of extra uncolored particle(s), e.g., $SU(2)_L$ complex singlets and higher multiplets with fermionic and bosonic degrees of freedom. These particles can be part of an $SU(2)_L$ multiplet as well. The electromagnetic charges of the singlet fields are solely determined by their assigned hypercharges\footnote{Our working formula is $Q=T_3+Y$ where $Q,T_3,Y$ are electromagnetic charge, 3rd component of isospin and hypercharge respectively.}. 
We have summarized the quantum numbers of the non-SM fields in Table~\ref{table:models-quantum-no}.
   
\begin{table}[h]
	\centering
	\renewcommand{\arraystretch}{1.6}
	{\scriptsize\begin{tabular}{|c|c|c|c|c|c|}
			\hline
			\multirow{2}{*}{\textbf{Model No.}}&
			\textbf{Non-SM IR DOFs}&
			\multirow{2}{*}{$SU(3)_C$}&
			\multirow{2}{*}{$SU(2)_L$}&
			\multirow{2}{*}{$U(1)_Y$}&
			\multirow{2}{*}{\textbf{Spin}}\\
			
			&
			\textbf{(Color Singlets)}&
			&
			&
			&
			\\
			\hline
			
			1&
			$\delta^{ + }$&
			1&
			1&
			1&
			0\\
			\hline
			
			2&
			$\rho^{++}$&
			1&
			1&
			2&
			0\\
			\hline
			
			3&
			$\Delta$&
			1&
			3&
			1&
			0\\
			\hline

			4&
			$\Sigma$&
			1&
			3&
			0&
			1/2\\
			\hline
			
			\multirow{3}{*}{5}&
			$V_{L,R} $ &
			1 &
			2 &
			-1/2 &
			1/2 \\
			\cline{2-6}
			
			&
			$E_{L,R} $ &
			1 &
			1 &
			-1 &
			1/2\\
			\cline{2-6}
			
			&
			$N_{L,R} $ &
			1 &
			1 &
			0 &
			1/2\\
			\hline
	\end{tabular}}
	\caption{\small Additional IR DOFs (Color Singlets) as representations of the SM gauge groups along with their spin quantum numbers.}
	\label{table:models-quantum-no}
\end{table} 

\subsubsection*{\underline{SM + Singly Charged Scalar ($\delta^{+}$)}}

We have considered the extension of SM by an $SU(2)_L$ singlet complex scalar field ($\delta^{+}$) of hypercharge 1, see Table~\ref{table:models-quantum-no}.  After the spontaneous electroweak symmetry breaking this field emerges as a singly charged physical scalar field\footnote{Although these particles are added to the unbroken SM gauge symmetry, but looking into this feature we will identify this and other fields by their electromagnetic charges.}. It is interesting to note that when the SM is embedded in an extended gauge symmetry, e.g., Left Right Symmetric Model (LRSM), then the appearance of singly charged scalar(s) is unavoidable once the additional symmetry is broken to the SM.   There are attempts to generate neutrino masses either radiatively or through higher dimensional operators where the SM is extended by mutiple $SU(2)$ singlet complex scalars, e.g., see Ref.~\cite{Cheng:1980qt,Babu:1988ki,Babu:2002uu,AristizabalSierra:2006gb}.
This motivates us to construct an effective theory with this simplest non-trivial extension of the SM. We have categorized the effective operators involving $\delta^{+}$ of dimensions 5 and 6 in Tables~\ref{table:SM+SinglyChargedScalar-dim5-ops-1} and \ref{table:SM+SinglyChargedScalar-dim6-ops-1}. The operators with distinct hermitian conjugates have been coloured blue.
\\
\\  
\noindent\textbf{Features of the additional operators:}
\begin{itemize}
	\item Here, we have noted two types of dimension 5 operators - i)  $B,L$ conserving $\tilde{\mathcal{O}}_{QdH\delta}$, $\tilde{\mathcal{O}}_{uQH\delta}$, $\tilde{\mathcal{O}}_{LeH\delta}$, and ii) $L$ violating  $\tilde{\mathcal{O}}_{e\delta}$ which are highlighted in red colour in Table~\ref{table:SM+SinglyChargedScalar-dim5-ops-1}.

	\item The additional dimension 6 operators of class $\Phi^6$ and $\Phi^4\mathcal{D}^2$ mimic their SM counterparts.
	
	\item Since, $\delta^{+}$ is an $SU(2)_L$ singlet there is no mixing between $B_{\mu\nu}$ and $W^{I}_{\mu\nu}$ in the $\Phi^2X^2$ class nor do we obtain higher tensor products in the $\Psi^2\Phi^2\mathcal{D}$ class. 
		
	\item The operators, highlighted in red colour in Table~\ref{table:SM+SinglyChargedScalar-dim6-ops-1},   violate lepton number by two units in the $\Psi^2\Phi^2\mathcal{D}$, $\Psi^2\Phi^3$ and $\Psi^2\Phi X$ classes. 
\end{itemize}

\begin{table}[h]
	\centering
	\renewcommand{\arraystretch}{1.9}
	{\scriptsize\begin{tabular}{||c|c||c|c||}
			\hline
			\hline
			\multicolumn{4}{||c||}{$\Psi^2\Phi^2$}\\
			\hline
			$\tilde{\mathcal{O}}_{QdH\delta} $&
			$\color{blue}{(N^2_f) \,\epsilon_{ij} \,(\overline{Q}_{p\alpha i} \,d_{	q}^{\alpha}) \,(\tilde{H}_j \,\delta)}$&
			$\tilde{\mathcal{O}}_{uQH\delta} $&
			$\color{blue}{(N^2_f) \,(\overline{u}_{p\alpha} \,Q^{\alpha i}_{q}) \,(\tilde{H}_i \,\delta)}$\\
			
			$\tilde{\mathcal{O}}_{LeH\delta} $&
			$\color{blue}{(N^2_f) \,\epsilon_{ij} \,(\overline{L}_{p i} \,e_{q}) \,(\tilde{H}_j \,\delta)}$&
			$\tilde{\mathcal{O}}_{e\delta} $&
			$\color{purple}{\frac{1}{2} (N^2_f+N_f) \,(e_{p}^T \,C \,e_{q}) \,\delta^2}$\\
			\hline
			\end{tabular}}
			\caption{SM extended by Singly Charged Scalar ($\delta$): Additional operators of  dimension 5. $\delta$, $\delta^{\dagger}$ represent $\delta^{+}$ and $\delta^{-}$ respectively. Here $i, j$ and $\alpha$ are the $ SU(2)$ and $SU(3)$ indices respectively. $p,q=1,2,\cdots,N_f$ are the flavour indices. The operator in {\it red} violates lepton number.}
	\label{table:SM+SinglyChargedScalar-dim5-ops-1}
\end{table}

\begin{table}[h]
	\centering
	\renewcommand{\arraystretch}{1.9}
	{\scriptsize\begin{tabular}{||c|c||c|c||}
			\hline
			\hline
			\multicolumn{2}{||c||}{$\Phi^6$}&
			\multicolumn{2}{c||}{$\Phi^4\mathcal{D}^2$}
			\\
			\hline
			$\mathcal{O}_{\delta} $&
			$(\delta^{\dagger} \,\delta)^3$&
			$\mathcal{O}_{\delta\square} $&
			$(\delta^{\dagger} \,\delta) \,\square \,(\delta^{\dagger} \,\delta)$\\
			
			$\mathcal{O}_{H^2\delta^4} $&
			$(H^{\dagger} \,H) \,(\delta^{\dagger} \,\delta)^2$&
			$\mathcal{O}_{ H\delta \mathcal{D}}^{(1)} $&
			$(\delta^{\dagger}\,\delta)\,\left[(\mathcal{D}^{\mu}\,H)^{\dagger}(\mathcal{D}_{\mu}\,H)\right]$\\
			
			$\mathcal{O}_{H^4\delta^2} $&
			$(H^{\dagger} \,H)^2 \,(\delta^{\dagger} \,\delta)$&
			$\mathcal{O}_{H \delta \mathcal{D}}^{(2)}$&
			$(H^{\dagger}\,H)\,\left[(\mathcal{D}^{\mu}\,\delta)^{\dagger}(\mathcal{D}_{\mu}\,\delta)\right]$
			\\
			\hline
			\hline
			\multicolumn{2}{||c||}{$\Phi^2X^2$}&
			\multicolumn{2}{c||}{$\Psi^2\Phi^2\mathcal{D}$}\\
			\hline
			
			$\mathcal{O}_{ B \delta} $&
			$B_{\mu\nu} \,B^{\mu\nu} \,(\delta^{\dagger} \,\delta)$&
			$\mathcal{O}_{ Q \delta \mathcal{D}} $&
			$(N^2_f) (\,\overline{Q}_{p\alpha i} \,\gamma^{\mu} \,Q^{\alpha i}_{q}) \,(\delta^{\dagger} \,i\overleftrightarrow{\mathcal{D}}_{\mu} \,\delta)$\\
			
			$\mathcal{O}_{\tilde{B} \delta} $&
			$\tilde{B}_{\mu\nu} \,B^{\mu\nu} \,(\delta^{\dagger} \,\delta)$&
			$\mathcal{O}_{L \delta \mathcal{D}} $&
			$(N^2_f) (\,\overline{L}_{pi} \,\gamma^{\mu} \,L^{i}_{q}) \,(\delta^{\dagger} \,i\overleftrightarrow{\mathcal{D}}_{\mu} \,\delta)$\\
			
			$\mathcal{O}_{G \delta} $&
			$G^{A}_{\mu\nu}\,G^{A\mu\nu}\,(\delta^{\dagger} \,\delta)$&
			$\mathcal{O}_{u \delta \mathcal{D}} $&
			$(N^2_f) (\,\overline{u}_{p\alpha} \,\gamma^{\mu} \,u^{\alpha}_{q}) \,(\delta^{\dagger} \,i\overleftrightarrow{\mathcal{D}}_{\mu} \,\delta)$\\
			
			$\mathcal{O}_{\tilde{G} \delta } $&
			$\tilde{G}^{A}_{\mu\nu}\,G^{A\mu\nu}\,(\delta^{\dagger} \,\delta)$&
			$\mathcal{O}_{d \delta \mathcal{D}} $&
			$(N^2_f) (\,\overline{d}_{p\alpha} \,\gamma^{\mu} \,d^{\alpha}_{q}) \,(\delta^{\dagger} \,i\overleftrightarrow{\mathcal{D}}_{\mu} \,\delta)$\\
			
			$\mathcal{O}_{ W \delta} $&
			$W^{I}_{\mu\nu} \,W^{I\mu\nu} \,(\delta^{\dagger} \,\delta)$&
			$\mathcal{O}_{e \delta \mathcal{D}} $&
			$(N^2_f) (\,\overline{e}_{p} \,\gamma^{\mu} \,e_{q}) \,(\delta^{\dagger} \,i\overleftrightarrow{\mathcal{D}}_{\mu} \,\delta)$\\

			$\mathcal{O}_{ \tilde{W} \delta } $&
			$\tilde{W}^{I}_{\mu\nu} \,W^{I\mu\nu} \,(\delta^{\dagger} \,\delta)$&
			$\mathcal{O}_{L e H \delta \mathcal{D}} $&
			$\textcolor{purple}{(N^2_f) \,((L_{p}^i)^T  \,\gamma^{\mu}\,e_{q}) \,(\tilde{H}^\dagger_i\,i\mathcal{D}_{\mu}\delta)}$
			\\
			\hline
			\hline
			\multicolumn{4}{||c||}{$\Psi^2\Phi^3$}\\
			\hline
			
			$\mathcal{O}_{L e H \delta} $&
			$\textcolor{blue}{(N^2_f) \,(\overline{L}_{pi} \,e_{q}) \,H^i \,(\delta^{\dagger} \,\delta)}$&
			$\mathcal{O}_{L H \delta} $&
			$\textcolor{purple}{(N^2_f) \,\epsilon_{ij} \,((L_{p}^i)^T \,C \,L_{q}^k) \,\delta \,(H^{\dagger}_k \,H^j)}$\\
			
			$\mathcal{O}_{Q u H \delta} $&
			$\textcolor{blue}{(N^2_f) \,\epsilon_{ij} \,(\overline{Q}_{p\alpha i} \,u^{\alpha}_{q}) \,\tilde{H}_j \,(\delta^{\dagger} \,\delta)}$&
			$\mathcal{O}_{L \delta} $&
			$\boxed{\textcolor{purple}{\frac{1}{2} (N^2_f-N_f) \,\epsilon_{ij} \,((L_{p}^i)^T \,C \,L_{q}^j) \,\delta \,(\delta^{\dagger} \,\delta)}}$\\
			
			$\mathcal{O}_{Q d H \delta} $&
			$\textcolor{blue}{(N^2_f) \,(\overline{Q}_{p\alpha i} \,d^{\alpha}_{q}) \,H^i \,(\delta^{\dagger} \,\delta)}$&
			&
			\\
			
			\hline
			\hline
			\multicolumn{4}{||c||}{$\Psi^2\Phi X$}\\
			\hline
			
			$\mathcal{O}_{B L \delta} $&
			$\textcolor{purple}{\frac{1}{2} (N^2_f+N_f) \,\epsilon_{ij} \,B_{\mu\nu} \,((L^i_{p})^T\,C\,\sigma^{\mu\nu} \,L^j_{q}) \,\delta}$&
			$\mathcal{O}_{W L \delta} $&
			$\boxed{\textcolor{purple}{\frac{1}{2} (N^2_f-N_f) \,\epsilon_{ij} \,W^I_{\mu\nu} \,((L^i_{p})^T \,C \,\sigma^{\mu\nu} \,\tau^I \,L^j_{q}) \,\delta}}$\\
			
			\hline
	\end{tabular}}
	\caption{SM extended by Singly Charged Scalar ($\delta$): Additional operators of  dimension 6. Boxed operators vanish for single flavour. $\delta$, $\delta^{\dagger}$ represent $\delta^{+}$ and $\delta^{-}$ respectively. Here $i, j$ and $\alpha$ are the $SU(2)$ and $SU(3)$ indices respectively. $\tau^I$ is $SU(2)$ generator. $A =1,2,\cdots,8$ and $I=1,2,3.$ $p,q=1,2,\cdots,N_f$ are the flavour indices. Operators in {\it red} violate lepton number.}
	\label{table:SM+SinglyChargedScalar-dim6-ops-1}
\end{table}

\newpage
\subsubsection*{\underline{SM + Doubly Charged Scalar ($\rho^{++}$)}}

Similar to the earlier case, when the SM is emerged from LRSM gauge theory, the right handed complex triplet may lead to an additional scalar of hypercharge 2 which is further identified as a doubly charged scalar ($\rho^{++}$), see Table~\ref{table:models-quantum-no}. Also scenarios like in Ref.~\cite{Cheng:1980qt,Babu:1988ki,Rizzo:1981xx,Akeroyd:2005gt,Maalampi:2002vx,Babu:2002uu,AristizabalSierra:2006gb} contain a single doubly charged scalar.  Here, our primary concern is to construct the effective operators involving the additional doubly charged scalar and thus mimic the concept of Refs.~\cite{Chakrabortty:2015zpm,King:2014uha}. We have provided the effective operators involving $\rho^{++}$ up to dimension 6 in Table~\ref{table:SM+DoublyChargedScalar-dim6-ops-1}. Operators with distinct hermitian conjugates have been coloured blue.

\begin{table}[h]
	\centering
	\renewcommand{\arraystretch}{1.9}
	{\scriptsize\begin{tabular}{||c|c||c|c||}
			\hline
			\hline
			\multicolumn{2}{||c||}{$\Phi^6$}&
			\multicolumn{2}{c||}{$\Phi^4\mathcal{D}^2$}\\
			\hline
			$\mathcal{O}_{\rho} $&
			$(\rho^{\dagger} \,\rho)^3$&
			$\mathcal{O}_{\rho\square} $&
			$(\rho^{\dagger} \,\rho) \,\square \,(\rho^{\dagger} \,\rho)$\\
			
			$\mathcal{O}_{H^2 \rho^4} $&
			$(H^{\dagger} \,H) \,(\rho^{\dagger} \,\rho)^2$&
			$\mathcal{O}_{H\rho \mathcal{D}}^{(1)} $&
			$(\rho^{\dagger}\,\rho)\,\left[(\mathcal{D}^{\mu}\,H)^{\dagger}(\mathcal{D}_{\mu}\,H)\right]$
			\\
			
			$\mathcal{O}_{H^4 \rho^2} $&
			$(H^{\dagger} \,H)^2 \,(\rho^{\dagger} \,\rho)$&
			$\mathcal{O}_{H \rho \mathcal{D}}^{(2)}$&
			$(H^{\dagger}\,H)\,\left[(\mathcal{D}^{\mu}\,\rho)^{\dagger}(\mathcal{D}_{\mu}\,\rho)\right]$
			\\
			\hline
			\hline
			\multicolumn{2}{||c||}{$\Phi^2X^2$}&
			\multicolumn{2}{c||}{$\Psi^2\Phi^2\mathcal{D}$}\\
			\hline
			
			$\mathcal{O}_{B\rho} $&
			$B_{\mu\nu} \,B^{\mu\nu} \,(\rho^{\dagger} \,\rho)$&
			$\mathcal{O}_{Q \rho \mathcal{D}} $&
			$(N_f^2) \,(\overline{Q}_{p\alpha i} \,\gamma^{\mu} \,Q^{\alpha i}_{q}) \,(\rho^{\dagger} \,i\overleftrightarrow{\mathcal{D}}_{\mu} \,\rho)$\\
			
			$\mathcal{O}_{\tilde{B} \rho } $&
			$\tilde{B}_{\mu\nu} \,B^{\mu\nu} \,(\rho^{\dagger} \,\rho)$&
			$\mathcal{O}_{L \rho \mathcal{D}} $&
			$(N_f^2) \,(\overline{L}_{pi} \,\gamma^{\mu} \,L^{i}_{q}) \,(\rho^{\dagger} \,i\overleftrightarrow{\mathcal{D}}_{\mu} \,\rho)$\\
			
			$\mathcal{O}_{G \rho } $&
			$G^{A}_{\mu\nu}\,G^{A\mu\nu}\,(\rho^{\dagger} \,\rho)$&
			$\mathcal{O}_{u \rho \mathcal{D}} $&
			$(N_f^2) \,(\overline{u}_{p\alpha} \,\gamma^{\mu} \,u^{\alpha}_{q}) \,(\rho^{\dagger} \,i\overleftrightarrow{\mathcal{D}}_{\mu} \,\rho)$\\
			
			$\mathcal{O}_{\tilde{G} \rho } $&
			$\tilde{G}^{A}_{\mu\nu}\,G^{A\mu\nu}\,(\rho^{\dagger} \,\rho)$&
			$\mathcal{O}_{d \rho \mathcal{D}} $&
			$(N_f^2) \,(\overline{d}_{p\alpha} \,\gamma^{\mu} \,d^{\alpha}_{q}) \,(\rho^{\dagger} \,i\overleftrightarrow{\mathcal{D}}_{\mu} \,\rho)$\\
			
			$\mathcal{O}_{W \rho } $&
			$W^{I}_{\mu\nu} \,W^{I\mu\nu} \,(\rho^{\dagger} \,\rho)$&
			$\mathcal{O}_{e \rho \mathcal{D}} $&
			$(N_f^2) \,(\overline{e}_{p} \,\gamma^{\mu} \,e_{q}) \,(\rho^{\dagger} \,i\overleftrightarrow{\mathcal{D}}_{\mu} \,\rho)$\\

			$\mathcal{O}_{\tilde{W} \rho } $&
			$\tilde{W}^{I}_{\mu\nu} \,W^{I\mu\nu} \,(\rho^{\dagger} \,\rho)$&
			$\mathcal{O}_{L e H \rho \mathcal{D}}$&
			$\textcolor{purple}{(N^2_f) \,((L^i_{p})^T  \,\gamma^{\mu}\,e_{q})\,(\,H^\dagger_i \,i\mathcal{D}_{\mu} \rho)}$
			\\
			\hline
			\hline
			\multicolumn{4}{||c||}{$\Psi^2\Phi^3$}\\
			\hline
			
			$\mathcal{O}_{L e H \rho} $&
			$\textcolor{blue}{(N_f^2) \,(\overline{L}_{pi} \,e_{q}) \,H^i \,(\rho^{\dagger} \,\rho)}$&
			$\mathcal{O}_{L H \rho} $&
			$\textcolor{purple}{\frac{1}{2} (N^2_f+N_f) \,((L^i_{p})^T \,C \,L^j_{q}) \,\rho \,(\tilde{H}_i \,\tilde{H}_j)}$\\

			$\mathcal{O}_{Q u H \rho} $&
			$\textcolor{blue}{(N_f^2) \,\epsilon_{ij} \,(\overline{Q}_{p\alpha i} \,u^{\alpha}_{q}) \,\tilde{H}^j \,(\rho^{\dagger} \,\rho)}$&
            $\mathcal{O}_{e \rho} $&
			$\textcolor{purple}{\frac{1}{2} (N^2_f+N_f) \,(e_{p}^T \,C \,e_{q}) \,\rho \,(\rho^{\dagger} \,\rho)}$
			\\

			$\mathcal{O}_{Q d H \rho} $&
			$\textcolor{blue}{(N_f^2) \,(\overline{Q}_{p\alpha i} \,d^{\alpha}_{q}) \,H^i \,(\rho^{\dagger} \,\rho)}$&
	        $\mathcal{O}_{e H \rho} $&
			$\color{purple}{\frac{1}{2} (N^2_f+N_f) \,(e_{p}^T \,C \,e_{q}) \,\rho \,(H^{\dagger} \,H)}$
			\\
			\hline
			\hline
			\multicolumn{4}{||c||}{$\Psi^2\Phi X$}\\
			\hline
			
			$\mathcal{O}_{B e \rho} $&
			\multicolumn{3}{c|}{$\boxed{\textcolor{purple}{\frac{1}{2} (N^2_f-N_f) \,B_{\mu\nu} \,(e_{p}^T \,C \,\sigma^{\mu\nu} \,e_{q}) \,\rho}}$}\\
			\hline
	\end{tabular}}
	\caption{SM extended by Doubly Charged Scalar ($\rho$): Additional operators of dimension 6. Boxed operators vanish for single flavour. $\rho$, $\rho^{\dagger}$ represent $\rho^{++}$ and $\rho^{--}$ respectively.  Here $i, j$ and $\alpha$ are the $SU(2)$ and $SU(3)$ indices respectively. $A =1,2,\cdots,8$ and $I=1,2,3.$ $p,q=1,2,\cdots,N_f$ are the flavour indices. Operators in {\it red} violate lepton number.}
	\label{table:SM+DoublyChargedScalar-dim6-ops-1}
\end{table}
  
\noindent\textbf{Features of the additional operators:}
\begin{itemize}
	\item One of the differences between the operator sets containing $\delta^{+}$ and $\rho^{++}$ is the absence of dimension 5 operators in the latter case.
	
	\item Similar to the earlier case, at dimension 6 there is no mixing between $B_{\mu\nu}$ and $W_{\mu\nu}^{I}$ within the $\Phi^2X^2$ class and there are no higher tensor products in the $\Psi^2\Phi^2\mathcal{D}$ class, on account of $\rho^{++}$ being an $SU(2)_L$ singlet. 
	
	\item We have found new operators that violate lepton number by two units in the $\Psi^2\Phi^2\mathcal{D}$, $\Psi^2\Phi^3$ and $\Psi^2\Phi X$ classes. These operators have been highlighted in red colour in Table~\ref{table:SM+DoublyChargedScalar-dim6-ops-1}. 
\end{itemize}

\subsubsection*{\underline{SM + Complex Triplet Scalar ($\Delta$)}}

Here, we have explored other possible scenarios where electroweak multiplets are assumed to be the lighter DOFs.   First, we have considered a complex $SU(2)_L$ triplet scalar ($\Delta$) having hypercharge of +1, Table~\ref{table:models-quantum-no}. After the spontaneous breaking of electroweak symmetry its components can be assigned definite electromagnetic charges\footnote{An $SU(2)$ triplet has $T_3$ values (+1, 0, -1). So, using $Q = T_3 + Y$, we obtain the electromagnetic charges (+2, +1, 0) since $Y = 1$.} $(\Delta^{++},\,\,\Delta^{+},\,\,\Delta^{0})$.   The complex triplet is instrumental in mediating lepton number and flavour violating processes \cite{delAguila:2013yaa,delAguila:2013hla,Fonseca:2018aav}, interesting collider signatures \cite{Chakrabortty:2015zpm}, and also facilitates the generation of neutrino mass \cite{Cheng:1980qt,Konetschny:1977bn,Schechter:1980gr,Gu_2006,Cai:2017mow}. These observables may get affected by the interactions between the  heavier particles and this complex triplet, which can be captured through the effective operators involving $\Delta$.   A complex $SU(2)_L$ triplet can  descend from an LRSM once it is spontaneously broken to the SM, see Fig.~\ref{fig:2}. There are many phenomenological models \cite{Georgi:1985nv,Chiang:2015kka,Logan:2015xpa,Chiang:2018cgb,Chiang:2015amq,Chiang:2012cn} where the SM is extended by a complex triplet accompanied by multiple scalars and fermions.  In that case if the other particles are sufficiently heavier than the $\Delta$, then they can be integrated out leading to an effective Lagrangian with IR DOFs as SM ones and the complex triplet.

Here, We have listed the  complete set of effective operators involving $\Delta$, see Tables~\ref{table:SM+ComplexTripletScalar-dim5-ops-1} and~\ref{table:SM+ComplexTripletScalar-dim6-ops-1} for dimension 5 and 6 respectively. The operators with distinct hermitian conjugates have been coloured blue. While writing the operators, $\Delta$ has been expressed  as a $2\times 2$ matrix $\Delta^{I}\cdot\tau^{I}$ with $I$ = $1,2,3$ and $\tau^{I}$ being the Pauli matrices. 

\begin{table}[h]
	\centering
	\renewcommand{\arraystretch}{1.9}
	{\scriptsize\begin{tabular}{||c|c||c|c||}
			\hline
			\hline
			\multicolumn{2}{||c||}{$\Psi^2\Phi^2$}&
			\multicolumn{2}{c||}{$\Phi^5$}\\
			\hline
			$\tilde{\mathcal{O}}_{L e H \Delta} $&
			$\color{blue}{N_f^2 \,(\overline{L}_{pi} \,e_{q} \,\Delta \,\tilde{H}_{i})}$
			&$\tilde{\mathcal{O}}_{H^2 \Delta^3}^{(1)}$&
			$\color{blue}{(H^T \Delta^{\dagger} \,H) \,Tr[(\Delta^{\dagger} \,\Delta)]}$\\
			
			$\tilde{\mathcal{O}}_{Q d H \Delta}$&
			$\color{blue}{N_f^2 \,(\overline{Q}_{p\alpha i} \,d^{\alpha}_{q} \,\Delta \,\tilde{H}_{i})}$& $\tilde{\mathcal{O}}_{H^2 \Delta^3}^{(2)}$&
			$\color{blue}{(H^T \Delta^{\dagger} \Delta^{\dagger} \,\Delta \,H)}$
			\\
			
			$\tilde{\mathcal{O}}_{Q u H \Delta} $&
			$\color{blue}{N_f^2 \,(\overline{Q}_{p\alpha i} \,u^{\alpha}_{q} \,\Delta^{\dagger} \,\tilde{H}_{i})}$&
			$\tilde{\mathcal{O}}_{H^4 \Delta} $&
			$\color{blue}{(H^T \,\Delta^{\dagger} \,H) \,(H^{\dagger} \,H)}$\\
			
			$\tilde{\mathcal{O}}_{e \Delta} $&
			$\color{purple}{\frac{1}{2}(N_f^2+N_f) \,(e_{p}^T \,C \,e_{q})\,Tr[ \,\Delta \,\Delta]}$&
			&
			\\
			\hline
	\end{tabular}}
	\caption{SM extended by Complex Triplet Scalar ($\Delta$): Additional operators of  dimension 5. Here $i$ and $\alpha$ are $SU(2)$ and $SU(3)$ indices respectively. $p,q=1,2,\cdots,N_f$ are the flavour indices. The operator in {\it red} violates lepton number.}
	\label{table:SM+ComplexTripletScalar-dim5-ops-1}
\end{table}

\noindent\textbf{Features of the additional operators:}
\begin{itemize}
\item Contrary to $\delta^{+}$ and $\rho^{++}$, $\Delta$ transforms as an $SU(2)_L$ triplet. This offers multiple ways to contract its indices to form invariant operators, e.g., within $\mathcal{O}^{(1),(2),(3)}_{H^{2}\Delta^{4}}$ class we have noted the following partitions:

\vspace{-0.7cm}
{\small\begin{eqnarray}
	\mathcal{O}^{(1)}_{H^{2}\Delta^{4}}\,\,&\equiv&\,\, H^{\dagger} \,(\Delta^{\dagger} \,\Delta) \,(\Delta^{\dagger} \,\Delta) \,H \hspace{0.9cm}\rightarrow (2\otimes 3 \otimes 3 \otimes 3 \otimes 3 \otimes 2), \nonumber\\
	\mathcal{O}^{(2)}_{H^{2}\Delta^{4}}\,\,&\equiv&\,\, Tr[(\Delta^{\dagger} \,\Delta) \,(\Delta^{\dagger} \,\Delta)] \,(H^{\dagger} \,H) \rightarrow (3\otimes 3 \otimes 3 \otimes 3) \otimes (2 \otimes 2), \nonumber\\
	\mathcal{O}^{(3)}_{H^{2}\Delta^{4}}\,\,&\equiv&\,\, Tr[(\Delta^{\dagger} \,\Delta)] \,(H^{\dagger}\,\Delta^{\dagger} \,\Delta \, H) \hspace{0.25cm} \rightarrow (3\otimes 3) \otimes (2 \otimes 3 \otimes 3 \otimes 2). \nonumber
	\end{eqnarray}}   
Here, we pick a singlet representation from the tensor product within a parenthesis. 

\begin{table}[h]
	\centering
	\renewcommand{\arraystretch}{1.9}
	{\scriptsize\begin{tabular}{||c|c||c|c||}
			\hline
			\hline
			\multicolumn{2}{||c||}{$\Phi^6$}&
			\multicolumn{2}{c||}{$\Phi^4\mathcal{D}^2$}
			\\
			\hline 
			$\mathcal{O}_{\Delta}^{(1)} $&
			$Tr[(\Delta^{\dagger} \,\Delta)]^3$&
			$\mathcal{O}_{\Delta\square} $&
			$Tr[(\Delta^{\dagger} \,\Delta) \,\square \,(\Delta^{\dagger} \,\Delta)]$\\
			
			$\mathcal{O}_{\Delta}^{(2)}$&
			$Tr[(\Delta^{\dagger} \,\Delta) \,(\Delta^{\dagger} \,\Delta)] \,Tr[(\Delta^{\dagger} \,\Delta)]$&
			$\mathcal{O}_{\Delta\mathcal{D}}^{(1)} $&
			$Tr[(\Delta^{\dagger}\,i\overleftrightarrow{\mathcal{D}_{\mu}}\,\Delta)(\Delta^{\dagger}\,i\overleftrightarrow{\mathcal{D}}^{\mu}\,\Delta)]$
			\\
			
			$\mathcal{O}_{H^2\Delta^4}^{(1)} $&
			$H^{\dagger} \,(\Delta^{\dagger} \,\Delta) \,(\Delta^{\dagger} \,\Delta) \,H$&
			$\mathcal{O}_{\Delta \mathcal{D}}^{(2)} $&
			$Tr[(\Delta^{\dagger} \,\Delta)] \,Tr[(\mathcal{D}^{\mu} \,\Delta^{\dagger}) \,(\mathcal{D}_{\mu} \Delta)] $
			\\
			
			$\mathcal{O}_{H^2\Delta^4}^{(2)}$&
			$Tr[(\Delta^{\dagger} \,\Delta) \,(\Delta^{\dagger} \,\Delta)] \,(H^{\dagger} \,H)$&
			$\mathcal{O}_{H \Delta \mathcal{D}}^{(1)}$&
			$[H^{\dagger} \,(\mathcal{D}_{\mu} \Delta)]\,[(\mathcal{D}^{\mu} \,\Delta)^{\dagger}\,H] $\\
			
			$\mathcal{O}_{H^2\Delta^4}^{(3)}$&
			$Tr[(\Delta^{\dagger} \,\Delta)] \,(H^{\dagger}\,\Delta^{\dagger} \,\Delta \, H)$&
			$\mathcal{O}_{H \Delta \mathcal{D}}^{(2)}$&
			$[(\mathcal{D}^{\mu} \,H)^{\dagger}\,\Delta]\,[\Delta^{\dagger} \,(\mathcal{D}_{\mu} H)] $\\
			
			$\mathcal{O}_{H^4\Delta^2}^{(1)}$&
			$(H^{\dagger}\, \Delta^{\dagger} \, H)\,(H^{\dagger}  \,\Delta \,H)$&
			$\mathcal{O}_{H \Delta \mathcal{D}}^{(3)}$ &
			$Tr[(\Delta^{\dagger} \,\Delta)] \,(\mathcal{D}^{\mu} H)^{\dagger} \,(\mathcal{D}_{\mu} H)$
			\\
			
			$\mathcal{O}_{H^4\Delta^2}^{(2)}$&
			$(H^{\dagger} \,\Delta^{\dagger} \,\Delta \, \,H) \,(H^{\dagger} \,H)$&
			$\mathcal{O}_{H \Delta \mathcal{D}}^{(4)}$ &
			$(H^{\dagger} \,H) \,Tr[(\mathcal{D}^{\mu} \Delta)^{\dagger} \,(\mathcal{D}_{\mu} \Delta)]$
			\\

			$\mathcal{O}_{H^4\Delta^2}^{(3)}$&
			$Tr[(\Delta^{\dagger} \,\Delta)] \,(H^{\dagger} \,H)^2$ & &
			\\
			
			$\mathcal{O}_{H\Delta H}$&
			$\color{blue}{(H^T \,\Delta^{\dagger} \,H)^2}$& &
			\\
			
			\hline
			\hline
			\multicolumn{2}{||c||}{$\Phi^2X^2$}&
			\multicolumn{2}{c||}{$\Psi^2\Phi^2\mathcal{D}$}\\
			\hline
			
			$\mathcal{O}_{B \Delta } $&
			$B_{\mu\nu} \,B^{\mu\nu} \,Tr[(\Delta^{\dagger} \,\Delta)]$&
			$\mathcal{O}_{Q \Delta \mathcal{D}}^{(1)} $&
			$(N_f^2) \,(\overline{Q}_{p\alpha i} \,\gamma^{\mu} \,Q^{\alpha i}_{q}) \,Tr[(\Delta^{\dagger} \,i\overleftrightarrow{\mathcal{D}}_{\mu} \,\Delta)]$\\
			
			$\mathcal{O}_{\tilde{B} \Delta } $&
			$\tilde{B}_{\mu\nu} \,B^{\mu\nu} \,Tr[(\Delta^{\dagger} \,\Delta)]$&
			$\mathcal{O}_{Q \Delta \mathcal{D}}^{(2)}$&
			$(N_f^2) \,(\overline{Q}_{p\alpha i} \,\gamma^{\mu}\,\tau^{I} \,Q^{\alpha i}_{q}) \,Tr[(\Delta^{\dagger} \,i\overleftrightarrow{\mathcal{D}}^{I}_{\mu} \,\Delta)]$
			\\
			
			$\mathcal{O}_{G \Delta } $&
			$G^{A}_{\mu\nu}\,G^{A\mu\nu}\,Tr[(\Delta^{\dagger} \,\Delta)]$&
			$\mathcal{O}_{L \Delta \mathcal{D}}^{(1)}$&
			$(N_f^2) \,(\overline{L}_{pi} \,\gamma^{\mu} \,L^{i}_{q}) \,Tr[(\Delta^{\dagger} \,i\overleftrightarrow{\mathcal{D}}_{\mu} \,\Delta)]$
			\\
			
			$\mathcal{O}_{\tilde{G} \Delta } $&
			$\tilde{G}^{A}_{\mu\nu}\,G^{A\mu\nu}\,Tr[(\Delta^{\dagger} \,\Delta)]$&
			$\mathcal{O}_{L \Delta \mathcal{D}}^{(2)}$&
			$(N_f^2) \,(\overline{L}_{pi} \,\gamma^{\mu}\,\tau^{I} \,L^{i}_{q}) \,Tr[(\Delta^{\dagger} \,i\overleftrightarrow{\mathcal{D}}^{I}_{\mu} \,\Delta)]$
			\\
			
			$\mathcal{O}_{W \Delta }^{(1)} $&
			$W^{I}_{\mu\nu} \,W^{I\mu\nu} \,Tr[(\Delta^{\dagger} \,\Delta)]$&
			$\mathcal{O}_{u \Delta \mathcal{D}}$&
			$(N_f^2) \,(\overline{u}_{p\alpha} \,\gamma^{\mu} \,u^{\alpha}_{q}) \,Tr[(\Delta^{\dagger} \,i\overleftrightarrow{\mathcal{D}}_{\mu} \,\Delta)]$
			\\
			
			$\mathcal{O}_{W \Delta }^{(2)}$&
			$Tr[\Delta^{\dagger} \,W_{\mu\nu}\,\Delta \,W^{\mu\nu}]$&
			$\mathcal{O}_{d \Delta  \mathcal{D}}$&
			$(N_f^2) \,(\overline{d}_{p\alpha} \,\gamma^{\mu} \,d^{\alpha}_{q}) \,Tr[(\Delta^{\dagger} \,i\overleftrightarrow{\mathcal{D}}_{\mu} \,\Delta)]$
			\\
			
			$\mathcal{O}_{\tilde{W} \Delta}^{(1)} $&
			$\tilde{W}^{I}_{\mu\nu} \,W^{I\mu\nu} \,Tr[(\Delta^{\dagger} \,\Delta)]$&
			$\mathcal{O}_{e \Delta \mathcal{D}}$&
			$(N_f^2) \,(\overline{e}_{p} \,\gamma^{\mu} \,e_{q}) \,Tr[(\Delta^{\dagger} \,i\overleftrightarrow{\mathcal{D}}_{\mu} \,\Delta)]$
			\\

			$\mathcal{O}_{\tilde{W} \Delta }^{(2)}$&
			$Tr[\Delta^{\dagger} \,W_{\mu\nu} \,\Delta \,\tilde{W}^{\mu\nu}]$&
			$\mathcal{O}_{L e H \Delta \mathcal{D}}$
			&
			\color{purple}{$(N_f^2) \,Tr[L^T_{p}\,C\,i\tau_2  \,(\gamma^{\mu}\mathcal{D}_{\mu}\,\Delta)\,H\,e_{q} ]$}
			\\
			
			$\mathcal{O}_{B W \Delta }$&
			$Tr[\Delta^{\dagger} \,W_{\mu\nu} \,\Delta ]\,B^{\mu\nu}$&
			&
			\\
			
			$\mathcal{O}_{B\tilde{W} \Delta }$&
			$Tr[\Delta^{\dagger} \,\tilde{W}_{\mu\nu} \,\Delta ]\,B^{\mu\nu}$&
			&
			\\
			\hline
			\hline
			\multicolumn{4}{||c||}{$\Psi^2\Phi^3$}\\
			\hline
			
			$\mathcal{O}_{L e H \Delta}^{(1)}$&
			\color{blue}{$(N_f^2) \,(\overline{L}_{pi} \,e_{q}) \,H^i \,Tr[(\Delta^{\dagger} \,\Delta)]$}&
			$\mathcal{O}_{L e H \Delta}^{(2)}$&
			$\color{blue}{(N_f^2) \,(\overline{L}_{pi} \,e_{q}) \,\Delta^{\dagger} \,\Delta \,H^{i}}$
			\\
			
			$\mathcal{O}_{Q d H \Delta}^{(1)}$&
			$\color{blue}{(N_f^2) \,(\overline{Q}_{p\alpha i} \,d^{\alpha}_{q}) \,H^i \,Tr[(\Delta^{\dagger} \,\Delta)]}$& $\mathcal{O}_{Q d H \Delta}^{(2)}$&
			$\color{blue}{(N_f^2) \,(\overline{Q}_{p\alpha i} \,d^{\alpha}_{q}) \,\Delta^{\dagger} \,\Delta \,H^{i}}$
			\\
			
			$\mathcal{O}_{Q u H \Delta}^{(1)}$&
			$\color{blue}{(N_f^2) \,\epsilon_{ij} \,(\overline{Q}_{p\alpha i} \,u^{\alpha}_{q}) \,\tilde{H}_j \,Tr[(\Delta^{\dagger} \,\Delta)]}$&
			$\mathcal{O}_{Q u H \Delta}^{(2)}$&
			$\color{blue}{(N_f^2) \,(\overline{Q}_{p\alpha i} \,u^{\alpha}_{q}) \,\Delta^{\dagger} \,\Delta \,\tilde{H}_{i}}$
			\\
			
			$\mathcal{O}_{L \Delta}^{(1)}$&
			$\color{purple}{\frac{1}{2}(N_f^2+N_f) \,(L^T_{p} \,C\,i\tau_2 \,\Delta \,L_{q}) \,Tr[(\Delta^{\dagger} \,\Delta)]}$&
			$\mathcal{O}_{L \Delta}^{(2)}$&
			$\color{purple}{\frac{1}{2}(N_f^2+N_f) \,(L^T_{p} \,C\,i\tau_2 \,\Delta \,\Delta^{\dagger} \,\Delta \,L_{q})}$
			\\
			
			$\mathcal{O}_{L H \Delta}^{(1)}$&
			$\color{purple}{\frac{1}{2}(N_f^2+N_f)\,(L^T_{p} \,C\,i\tau_2 \,\Delta \,L_{q}) \,(H^{\dagger} \,H)}$&
			$\mathcal{O}_{L H \Delta}^{(2)}$&
			$\color{purple}{(N_f^2)\,(L^T_{pi} \,C\,i\tau_2 \,\Delta \,H^{i} \,H^{\dagger}_{j} \,L^{j}_{q})}$
			\\
			
			$\mathcal{O}_{e \Delta}$&
			$\color{purple}{\frac{1}{2}(N_f^2+N_f) \,(e_{p}^T \,C \,e_{q}) \,(H^T \,\Delta \,H)}$&
			&
			\\
			\hline
			\hline
			\multicolumn{4}{||c||}{$\Psi^2\Phi X$}\\
			\hline
			
			$\mathcal{O}_{W L \Delta}$&
			\color{purple}{$(N_f^2) \,Tr[(L^T_{p} \,C\,i\tau_2 \,\Delta \,\sigma_{\mu\nu}  \,L_{q})\, W^{\mu\nu}]$}&
			$\mathcal{O}_{B L \Delta}$&
			$\boxed{\color{purple}{\frac{1}{2}(N_f^2-N_f) \,(L^T_{p} \,C\,i\tau_2 \,\Delta \,\sigma_{\mu\nu}  \,L_{q})\, B^{\mu\nu}}}$\\
			
			\hline
	\end{tabular}}
	\caption{SM  extended by Complex Triplet Scalar ($\Delta$): Additional operators of  dimension 6. Here $i, j$ and $\alpha$ are the $SU(2)$ and $SU(3)$ indices respectively. $\tau^I$ are the $SU(2)$ generators. $A=1,2,\cdots,8$ and $I=1,2,3$. $p,q=1,2,\cdots,N_f$ are the flavour indices. Also, $\Delta = \Delta^{I}\cdot\tau^{I}$ and $W_{\mu\nu} = W^{I}_{\mu\nu}\cdot \tau^{I}$. Operators in {\it red} violate lepton number.}
	\label{table:SM+ComplexTripletScalar-dim6-ops-1}
\end{table}
\clearpage

	\item At dimension 5 in addition to the $\Psi^2\Phi^2$ class, there are new operators of the $\Phi^5$ class unlike the previous cases.

	\item Since $\Delta$ transforms as an $SU(2)_L$ triplet the $\Psi^2\Phi^2\mathcal{D}$ class has operators constituted of higher tensor products $\mathcal{O}^{(2)}_{L\Delta\mathcal{D}}$ and $\mathcal{O}^{(2)}_{Q\Delta\mathcal{D}}$ unlike the previous models.
	
	\item Lepton number violation too is observed within the $\Psi^2\Phi^2\mathcal{D}$, $\Psi^2\Phi^3$ and $\Psi^2\Phi X$ classes. These operators are highlighted in red colour.
	
\end{itemize}

\subsubsection*{\underline{SM + Left-Handed Triplet Fermion ($\Sigma$)}}

The extra IR DOF can be fermionic in nature instead of scalar. To demonstrate the  feature of such cases, we have considered a specific example, where the SM is extended by an  $SU(2)_L$ real triplet fermion $\Sigma=(\Sigma_{1},\,\,\Sigma_{2},\,\,\Sigma_{3})$. This additional DOF plays a central role in the generation of neutrino masses and mixing  \cite{Foot:1988aq, Chakrabortty:2008zh, Franceschini_2008,Cai:2017mow,Chen:2011de, Chakrabortty:2010rq, Chakrabortty:2010az, Gu:2011yx}, lepton flavour violating decays \cite{Abada:2008ea,Cheng:1980tp,Marciano:1977wx,Eboli:2011ia,Sirunyan:2017qkz,ATLAS:2018ghc,Goswami:2018jar},  explaining dark matter  \cite{Chaudhuri:2015pna,Choubey:2017yyn,Restrepo:2019ilz,FileviezPerez:2008sr}, and CP \& matter-antimatter asymmetry
\cite{Bajc:2007zf,Franco:2015pva,Branco:2011zb,Hambye:2012fh}. In most of these scenarios, this triplet fermion is accompanied by other particles which can be integrated out to construct an effective Lagrangian described by SM DOFs and $\Sigma$.  This motivates us to compute a complete set of effective operators which will capture all such extended BSM scenarios.  Here, we have classified the effective operators of dimensions 5 and 6 containing $\Sigma$ in Tables \ref{table:SM+LeftHandedTripletFermion-dim5-ops-1} and \ref{table:SM+LeftHandedTripletFermion-dim6-ops-1} respectively. The operators with distinct hermitian conjugates have been coloured blue.
\\
\\  
\noindent\textbf{Features of the additional operators:}
\begin{itemize}
	\item Since $\Sigma$ has zero hypercharge, in addition to the $\Psi^2\Phi^2$ operators, we also obtain operators of the class $\Psi^2X$ at dimension 5.
	
	\item On account of $\Sigma$ being an $SU(2)_L$ triplet we obtain multiple operators of similar structure whenever the other doublets $L, Q, H$ or the triplet $W^{I}_{\mu\nu}$ are involved.
	
	\item We obtain lepton and baryon number violation among operators of the class $\Psi^4$, $\Psi^2\Phi X$ and $\Psi^2\Phi^3$. These operators have been coloured red.
\end{itemize}

\begin{table}[h]
	\centering
	\renewcommand{\arraystretch}{1.9}
	{\scriptsize\begin{tabular}{||c|c||c|c||}
			\hline
			\hline
			\multicolumn{2}{||c||}{$\Psi^2\Phi^2$}&
			\multicolumn{2}{c||}{$\Psi^2 X$}\\
			\hline
			$\tilde{\mathcal{O}}_{\Sigma H}$&
			$\color{blue}{(N_f^2) \,((\Sigma^I_{p})^T \,C \,\Sigma^I_{q}) \,(H^{\dagger} \,H)}$
			&
			$\tilde{\mathcal{O}}_{B \Sigma }$&
			$\boxed{\color{blue}{\frac{1}{2}(N_f^2-N_f) \,B_{\mu\nu} \,((\Sigma^I_{p})^T \,C \,\sigma^{\mu\nu} \,\Sigma^I_{q})}}$
			\\
			
			$\tilde{\mathcal{O}}_{e \Sigma H}$&
			$\color{purple}{(N_f^2) \,\epsilon_{ij} \,(\overline{\Sigma}^I_{p} \,e_{q}) \,(H^i \,\tau^I H^j)}$&
			$\tilde{\mathcal{O}}_{W \Sigma}$&
			$\color{blue}{\frac{1}{2}(N_f^2+N_f) \,\epsilon_{IJK} \,W^I_{\mu\nu} \,((\Sigma^J_{p})^T \,C \,\sigma^{\mu\nu} \,\Sigma^K_{q})}$
			\\
			
			\hline
			\end{tabular}}
			\caption{SM extended by Left-Handed Triplet Fermion ($\Sigma$): Additional operators of dimension 5. Here $i, j$ are the $SU(2)$ indices. $\tau^I$ is the $SU(2)$ generator. $I=1,2,3$ and $p,q=1,2,\cdots,N_f$ are the flavour indices. The operator in {\it red} violates lepton number.}
	\label{table:SM+LeftHandedTripletFermion-dim5-ops-1}
\end{table}

\begin{table}[h]
	\centering
	\renewcommand{\arraystretch}{1.9}
	{\scriptsize\begin{tabular}{||c|c||c|c||}
			\hline
			\hline
			\multicolumn{2}{||c||}{$\Psi^2\Phi X$}&
			\multicolumn{2}{c||}{$\Psi^2\Phi^2\mathcal{D}$}
			\\
			\hline
			$\mathcal{O}_{BL\Sigma H}$&
			$\color{purple}{(N_f^2) \,\epsilon_{ij} \,B_{\mu\nu} \,((L^i_{p})^T\,C \,\sigma^{\mu\nu} \,\Sigma^I_{q}) \,\tau^I \,H^j}$&
			$\mathcal{O}_{\Sigma H}^{(1)}$&
			$(N_f^2) \,(\overline{\Sigma}^I_{p} \gamma^{\mu} \Sigma^I_{q}) \,(H^{\dagger} \,i\overleftrightarrow{\mathcal{D}}_{\mu} \,H)$\\
			
			$\mathcal{O}_{WL\Sigma H}^{(1)}$&
			$\color{purple}{(N_f^2) \,\epsilon_{ij} \,W^I_{\mu\nu} \,((L^i_{p})^T\, C \,\sigma^{\mu\nu}  \,\Sigma^I_{q}) \,H^j}$&
			$\mathcal{O}_{\Sigma H}^{(2)}$&
			$(N_f^2) \,\epsilon_{IJK} \,(\overline{\Sigma}^I_{p} \,\gamma^{\mu} \,\Sigma^J_{q}) \,(H^{\dagger} \,\tau^K \,i\overleftrightarrow{\mathcal{D}}_{\mu} \,H)$
			\\
			
			$\mathcal{O}_{WL\Sigma H}^{(2)}$&
			$\color{purple}{(N_f^2) \,\epsilon_{IJK} \,\epsilon_{ij} \,W^I_{\mu\nu} \,((L^i_{p})^T\,C \,\sigma^{\mu\nu}  \,\Sigma^J_{q}) \,\tau^K \,H^j}$&
			&
			\\
			
			\hline
			\hline
			\multicolumn{4}{||c||}{$\Psi^4$}\\
			\hline
			
			$\mathcal{O}_{u\Sigma}$&
			$(N_f^4) \,(\overline{u}_{p\alpha} \,\gamma_{\mu} \,u^{\alpha}_{q}) \,(\overline{\Sigma}^I_{r} \,\gamma^{\mu} \,\Sigma^I_{s})$&$\mathcal{O}_{d\Sigma}$&
			$(N_f^4) \,(\overline{d}_{p\alpha} \,\gamma_{\mu} \,d^{\alpha}_{q}) \,(\overline{\Sigma}^I_{r} \,\gamma^{\mu} \,\Sigma^I_{s})$
			\\
			
			$\mathcal{O}_{e\Sigma}$&
			$(N_f^4) \,(\overline{e}_{p} \,\gamma_{\mu} \,e_{q}) \,(\overline{\Sigma}^I_{r} \,\gamma^{\mu} \,\Sigma^I_{s})$&
			$\mathcal{O}_{\Sigma\Sigma}$&
			$(\frac{3}{4}N_f^4+\frac{1}{2}N_f^3+\frac{3}{4}N_f^2) \,(\overline{\Sigma}^I_{p} \,\gamma_{\mu} \,\Sigma^I_{q}) \,(\overline{\Sigma}^J_{r} \,\gamma^{\mu} \,\Sigma^J_{s})$
			\\
			
			$\mathcal{O}_{Q\Sigma}^{(1)}$&
			$(N_f^4) \,(\overline{Q}_{p\alpha i} \,\gamma_{\mu} \,Q^{\alpha i}_{q}) \,(\overline{\Sigma}^I_{r} \,\gamma^{\mu} \,\Sigma^I_{s})$&
			$\mathcal{O}_{Q\Sigma}^{(2)}$&
			$(N_f^4) \,\epsilon_{IJK} \,(\overline{Q}_{p\alpha i} \,\gamma_{\mu} \,\tau^I \,Q^{j\alpha}_{q}) \,(\overline{\Sigma}^J_{r} \,\gamma^{\mu} \,\Sigma^K_{s})$
			\\
			
			$\mathcal{O}_{L\Sigma}^{(1)}$&
			$(N_f^4) \,(\overline{L}_{pi} \,\gamma_{\mu} \,L^{i}_{q}) \,(\overline{\Sigma}^I_{r} \,\gamma^{\mu} \,\Sigma^I_{s})$&
			$\mathcal{O}_{L\Sigma}^{(2)}$&
			$(N_f^4) \,\epsilon_{IJK} \,(\overline{L}_{pi} \,\gamma_{\mu} \,\tau^I \,L^i_{q}) \,(\overline{\Sigma}^J_{r} \,\gamma^{\mu} \,\Sigma^K_{s})$
			\\
			
			$\mathcal{O}_{eL\Sigma}$&
			$\color{purple}{(N_f^4) \,\epsilon_{ij} \,((L^i_{p})^T \,C \,\tau^I \,L^j_{q}) \,(\overline{e}_{r} \,\Sigma^I_{s})}$&
			$\mathcal{O}_{\Sigma^2}$&
			$\color{blue}{\frac{1}{4}(N_f^4+3N_f^2) \,((\Sigma^I_p)^T \,C \,\Sigma^I_q)\,((\Sigma^I_r)^T \,C \,\Sigma^I_s)}$
			\\
			
			$\mathcal{O}_{QLd\Sigma}^{(1)}$&
			$\color{purple}{(N_f^4) \,\epsilon_{ij} \,((L^i_{p})^T\, C \,\tau^I \,Q^{\alpha j}_{q}) \,(\overline{d}_{r\alpha} \,\Sigma^I_{s})}$&
			$\mathcal{O}_{QLd\Sigma}^{(2)}$&
			$\color{purple}{(N_f^4) \,\epsilon_{ij} \,((L^i_{p})^T \,C\,\sigma_{\mu\nu} \,\tau^I \,Q^{\alpha j}_{q}) \,(\overline{d}_{r\alpha} \,\sigma^{\mu\nu} \,\Sigma^I_{s})}$
			\\
			
			$\mathcal{O}_{QLu\Sigma}$&
			$\color{purple}{(N_f^4) \,(\overline{Q}_{p\alpha i} \, u^{\alpha}_{q}) \,[\tau^I]^{i}_{j} \,((L^j_{r})^T\, C \,\Sigma^I_{s})}$&
			$\mathcal{O}_{Qd\Sigma}$&
			$\boxed{\color{purple}{\frac{1}{2}N_f^3(N_f-1) \,\epsilon_{\alpha\beta\gamma} \,\epsilon_{ij} \,(\overline{\Sigma}^I_{p} \,d^{\alpha}_{q}) \,((Q^{\beta i}_{r})^T \,C \,\tau^I \,Q^{\gamma j}_{s})}}$
			\\
			
			\hline
			\hline
			\multicolumn{4}{||c||}{$\Psi^2\Phi^3$}\\
			\hline
			
			$\mathcal{O}_{L\Sigma HH}^{(1)}$&
			$\color{purple}{(N_f^2) \,\epsilon_{ij} \,((L^i_{p})^T\, C \,\Sigma^I_{q}) \,\tau^I \,H^j \,(H^{\dagger} \,H)}$&
			$\mathcal{O}_{L\Sigma HH}^{(2)}$&
			$\color{purple}{(N_f^2) \,\epsilon_{IJK} \,\epsilon_{ij} \,((L^i_{p})^T\, C \,\Sigma^I_{q}) \,\tau^J \,H^j \,(H^{\dagger} \,\tau^K \,H)}$\\
			
			\hline
	\end{tabular}}
	\caption{SM extended by Left-Handed Triplet Fermion ($\Sigma$): Additional operators of  dimension 6. Here $i, j$ and $\alpha, \beta, \gamma$ are the $SU(2)$ and $SU(3)$ indices respectively. $\tau^I$ are the $SU(2)$ generators. $I,J,K=1,2,3$. $p,q,r,s=1,2,\cdots,N_f$ are the flavour indices. Operators in {\it red} violate lepton and baryon numbers.}
	\label{table:SM+LeftHandedTripletFermion-dim6-ops-1}
\end{table}

\newpage
%\clearpage

\subsubsection*{\underline{SM + Vector-like Leptons ($ V_{L,R},\,E_{L,R},\,N_{L,R} $)}}
It may be possible that the SM is extended by a {\it set} of lighter degrees of freedom. To discuss that kind of scenario, here, we have considered an example model where the IR DOFs are vector like leptons: 
lepton doublets ($ V_{L,R} $) with hypercharge $ \frac{1}{2} $, singlets with hypercharge -1 ($E_{L,R}$), and 0 ($N_{L,R}$). This subset of particles can be embedded in a rather complete scenario where parity is respected, e.g., Pati-Salam, LRSM etc. The vector like fermions may induce first order Electroweak Phase Transition (EWPT) and explain the origin of baryon asymmetry  \cite{Angelescu:2020yzf, Angelescu_2019,Huo_2015,Ishiwata_2015,borah2020subtev}. They also affect low energy observables \cite{Falkowski_2014,Ellis_2014,crivellin2020global}.  The effects of  parity conserving complete theories can be captured through the effective operators involving these vector like fermions. In that case, it would be important to note how the tree-level predictions get affected in the presence of the higher dimensional operators. Here, we have listed the dimension 5 operators  in Table~\ref{table:VLL-dim5-ops}. We have catalogued the set of dimension 6 operators in Tables~\ref{table:VLL-dim6-ops1}-\ref{table:VLL-dim6-psi4-ops5-contd}. The operators with distinct hermitian conjugates are depicted in blue colour.
\\
\\
\noindent\textbf{Features of the additional operators:}
\begin{itemize}
\item At dimension 5, we have both $B,L$ conserving and $L$ violating operators within the $\Psi^2\Phi^2$ and $\Psi^2X$ classes. 
	
\item At dimension 6, we have the freedom to write down multiple covariant forms corresponding to a particular operator. But not all of them are independent. For example, the operator $\mathcal{O}_{ V_L V_R N_L N_R }$ in Table~\ref{table:VLL-dim6-psi4-ops3} can be written in a covariant form as either $ (\overline{V}_{Rpi}\,  V_{Lq}^{i} )\, (\overline{N}_{Lr}\,  N_{Rq}  ) $ or $ (\overline{V}_{Rpi}\, \gamma^\mu\, N_{Rq} )\, (\overline{N}_{Lr}\, \gamma_{\mu}\,  V_{Ls}^{i} ) $. But these two structures are related to each other through the identities mentioned in subsection \ref{subsec:fierz-identities}. Therefore, we have included only one of them to avoid any redundancy in the operator set.

\item This model offers the violation of baryon and lepton numbers of different amount unlike other scenarios discussed in this paper.  We have noted the $(\Delta B, \Delta L)$ of following amounts: $(0, \pm2)$, $(0, \pm4)$. $(\pm1, \pm1)$ and $(\pm1, \mp1)$ within the $\Psi^4$, $\Psi^2\Phi^3$, $\Psi^2\Phi^2\mathcal{D}$, and $\Psi^2\Phi X$ classes. 
\end{itemize}

\begin{table}[h]
	\centering
	\renewcommand{\arraystretch}{1.9}
	{\scriptsize
\begin{tabular}{||c|c||c|c||}
			\hline
			\hline
			\multicolumn{4}{||c||}{$\Psi^2\Phi^2$}\\
			\hline

			$\tilde{\mathcal{O}}_{H E_L E_R}$&
			$ \color{blue}(N_f^2) \, (H^\dagger_i \, H^i) \, (\overline{E}_{Rp} \, E_{Lq}) $&
			$\tilde{\mathcal{O}}_{H N_L N_R}$&
			$ \color{blue}(N_f^2) \, (H^\dagger_i \, H^i) \, (\overline{N}_{Lp} \, N_{Rq})  $ \\
			
			$\tilde{\mathcal{O}}_{H L V_R} ^{(1)}$&
			$ \color{blue}(N_f^2) \, (H^\dagger_i \, H^i) \, (\overline{L}_{pj} \, V_{Rq}^j) $&
			$\tilde{\mathcal{O}}_{H V_L V_R} ^{(1)}$&
			$ \color{blue}(N_f^2) \, (H^\dagger_i \, H^i) \, (\overline{V}_{Lpj} \, V_{Rq}^j) $ \\

			$\tilde{\mathcal{O}}_{H L V_R} ^{(2)}$&
			$ \color{blue}(N_f^2) \,(H^\dagger_i\, \tau^I\, H^i) \, (\overline{L}_{pj} \, \tau^I\, V_{Rq}^j) $ &
			$\tilde{\mathcal{O}}_{H V_L V_R} ^{(2)}$&
			$ \color{blue}(N_f^2) \,(H^\dagger_i\, \tau^I\, H^i) \, (\overline{V}_{Lpj} \, \tau^I\, V_{Rq}^j) $ \\

			$\tilde{\mathcal{O}}_{H e E_L}$&
			$ \color{blue}(N_f^2)\, (H^\dagger_i \, H^i)\, (\overline{e}_{p} \, E_{Lq}) $&
			$\tilde{\mathcal{O}}_{H N_R}$&
			$ \color{purple}\frac{1}{2}N_f(N_f+1)\, (H^\dagger_i \, H^i) \, (N_{Rp}^T\, C \, N_{Rq}) $ \\

			$\tilde{\mathcal{O}}_{H V_L}$&
			$ \color{purple}\frac{1}{2}N_f(N_f+1)\epsilon_{ij}\, \epsilon_{mn}\,H^i \, H^m \, (V_{Lp}^{j})^TC \, V_{Lq}^{n} $&
			$\tilde{\mathcal{O}}_{H V_R}$&
			$ \color{purple}\frac{1}{2}N_f(N_f+1)\epsilon_{ij}\, \epsilon_{mn}\,H^i \, H^m \, (V_{Rp}^{j})^TC \, V_{Rq}^{n} $ \\

			$\tilde{\mathcal{O}}_{H L V_L}$&
			$ \color{purple}(N_f^2)\epsilon_{ij}\, \epsilon_{mn}\,H^i \, H^m \, (L_{p}^{j})^TC \, V_{Lq}^{n} $&
			$\tilde{\mathcal{O}}_{H N_L}$&
			$ \color{purple}\frac{1}{2}N_f(N_f+1)\, (H^\dagger_i \, H^i) \, (N_{Lp}^T\, C\, N_{Lq}) $ \\

			\hline
			\hline
			\multicolumn{4}{||c||}{$\Psi^2 X $}\\
\hline
	\end{tabular}}
	\caption{SM extended by Vector-like Leptons ($ V_{L,R},\,E_{L,R},\,N_{L,R} $): Additional operators of dimension 5. $i,j,m,n$ are the $ SU(2) $ indices and $ p,q=1,2,\cdots,N_f $ are the flavour indices. $ \tau^I\, (I=1,2,3) $ is the $ SU(2) $ generator. Operators in {\it red} violate lepton and baryon numbers.}
	\label{table:VLL-dim5-ops}
\end{table}

\begin{table}[h]
	\centering
	\renewcommand{\arraystretch}{1.9}
	{\scriptsize\begin{tabular}{||c|c||c|c||}
			\hline
			\hline
			\multicolumn{4}{||c||}{$\Psi^2\Phi^3$}\\
			\hline
			
			$\mathcal{O}_{ H V_L e}$&
			$ \color{blue}(N_f^2)\, (\overline{V}_{Lpi}\, e_{q}\, H^{i})\, (H^\dagger_j\, H^j) $ &
			$\mathcal{O}_{ H V_R E_L }$&
			$ \color{blue}(N_f^2) (\overline{V}_{Rpi}\, E_{Lq}\, H^{i})\, (H^\dagger_j\, H^j) $ \\
			
			$\mathcal{O}_{ H V_L E_R}$&
			$ \color{blue}(N_f^2)\, (\overline{V}_{Lpi}\, E_{Rq}\, H^{i})\, (H^\dagger_j\, H^j) $ &
			$\mathcal{O}_{ H L E_R }$&
			$ \color{blue}(N_f^2)\, (\overline{L}_{pi}\, E_{Rq}\, H^{i})\, (H^\dagger_j\, H^j) $ \\
			
			$\mathcal{O}_{H V_L N_R }$&
			$ \color{blue}(N_f^2)\, \epsilon_{ij}\, (\overline{N}_{Rp}\, V_{Lq}^{i}\, H^{j})\, (H^\dagger_k\, H^k)  $ &
			$\mathcal{O}_{ H L N_R }$&
			$ \color{blue}(N_f^2)\, \epsilon_{ij}\, (\overline{N}_{Rp}\, L_{q}^{i}\, H^{j})\, (H^\dagger_k\, H^k) $ \\

			$\mathcal{O}_{H V_R N_L }$&
			$ \color{blue}(N_f^2)\, \epsilon_{ij}\, (\overline{N}_{Lp}\, V_{Rq}^{i}\, H^{j})\, (H^\dagger_k\, H^k)  $ &
			$\mathcal{O}_{ H L N_L }$&
			$ \color{purple}(N_f^2)\, \epsilon_{ij}\, ((L^{i}_{p})^T\, C\, N_{Lq}\, H^{j})\, (H^\dagger_k\, H^k) $ \\

			$\mathcal{O}_{ H V_L N_L }$&
			$ \color{purple}(N_f^2)\, \epsilon_{ij}\, ((V^{i}_{Lp})^T\, C\, N_{Lq}\, H^{j})\, (H^\dagger_k\, H^k) $ &
			$\mathcal{O}_{ H V_R N_R }$&
			$ \color{purple}(N_f^2)\, \epsilon_{ij}\, ((V^{i}_{Rp})^T\, C\, N_{Rq}\, H^{j})\, (H^\dagger_k\, H^k) $ \\

			\hline
			\hline
			\multicolumn{4}{||c||}{$\Psi^2\Phi X$}\\
			\hline
			
			$\mathcal{O}_{ B H  V_L e }$&
			$ \color{blue}(N_f^2)\, B_{\mu\nu}\, (\overline{V}_{Lpi}\, \sigma^{\mu\nu}\, e_{q})\, H^i   $&
			$\mathcal{O}_{ W H  V_L e }$&
			$ \color{blue} (N_f^2)\, W^I_{\mu\nu}\, (\overline{V}_{Lpi}\, \sigma^{\mu\nu}\, e_{q})\, \tau^I\, H^i  $\\
\hline
	\end{tabular}}
	\caption{SM extended by Vector-like Leptons ($ V_{L,R},\,E_{L,R},\,N_{L,R} $): Additional operators of dimension 6. $i,j,m,n$ are the $ SU(2) $ indices and $ p,q=1,2,\cdots,N_f $ are the flavour indices. $ \tau^I\,  (I=1,2,3) $ is the $ SU(2) $ generator. Operators in {\it red} violate lepton number.}
	\label{table:VLL-dim6-ops1}
\end{table}

\begin{table}[h]
	\centering
	\renewcommand{\arraystretch}{1.9}
	{\scriptsize\begin{tabular}{||c|c||c|c||}
			\hline
			\hline
			\multicolumn{4}{||c||}{$\Psi^4$}\\
			\hline

			$\mathcal{O}_{ d E_L }$&
			$ (N_f^4)\, (\overline{E}_{Lp}\, \gamma^{\mu}\, E_{Lq})\, (\overline{d}_{r\alpha}\, \gamma_{\mu}\, d_s^{\alpha}) $ &
			$\mathcal{O}_{ e E_L }$&
			$ (N_f^4)\, (\overline{E}_{Lp}\, \gamma^{\mu}\, E_{Lq})\, (\overline{e}_{r}\, \gamma_{\mu}\, e_s) $ \\
			
			$\mathcal{O}_{ e E_R }$&
			$ (N_f^4)\, (\overline{e}_{p}\, \gamma^{\mu}\, e_{q})\, (\overline{E}_{Rr}\, \gamma_{\mu}\, E_{Rs}) $ &
			$\mathcal{O}_{ \bar{E}_L \bar{E}_R }$&
			$ (N_f^4)\, ( \overline{E}_{Lp}\, \gamma^{\mu}\, E_{Lq})\, (\overline{E}_{Rr}\, \gamma_{\mu}\, E_{Rs}) $ \\
			
			$\mathcal{O}_{ E_R }$&
			$ \frac{1}{4}N_f^2(N_f +1)^2\, (\overline{E}_{Rp}\, \gamma^{\mu}\, E_{Rq}) (\overline{E}_{Rr}\, \gamma_{\mu}\, E_{Rs}) $ &
			$\mathcal{O}_{ E_L }$&
			$ \frac{1}{4}N_f^2(N_f +1)^2\, (\overline{E}_{Lp}\, \gamma^{\mu}\, E_{Lq}) (\overline{E}_{Lr}\, \gamma_{\mu}\, E_{Ls}) $ \\
			
			$\mathcal{O}_{ L E_L }$&
			$ (N_f^4)\, (\overline{L}_{pi}\, \gamma^{\mu}\, L_{q}^{i}) (\overline{E}_{Lr}\, \gamma_{\mu}\, E_{Ls}) $ &
			$\mathcal{O}_{ L E_R }$&
			$ (N_f^4)\, (\overline{L}_{pi}\, \gamma^{\mu}\, L_{q}^{i}) (\overline{E}_{Rr}\, \gamma_{\mu}\, E_{Rs}) $ \\
			
			$\mathcal{O}_{ L V_L }^{(1)}$&
			$ (N_f^4)\, (\overline{V}_{Lpi}\, \gamma^{\mu}\, V_{Lq}^{i})\, (\overline{L}_{rj}\, \gamma_{\mu} L_{s}^{j})  $ &
			$\mathcal{O}_{ V_L E_L }$&
			$ (N_f^4)\, (\overline{E}_{Lp}\, \gamma^\mu\, E_{Lq} )\, (\overline{V}_{Lri}\, \gamma_{\mu}\,  V_{Ls}^{i}) $ \\
			
			$\mathcal{O}_{ L V_L }^{(2)}$&
			$ (N_f^4)\, (\overline{V}_{Lpi}\, \gamma^{\mu}\, \tau^I\, V_{Lq}^{i})\, (\overline{L}_{rj}\, \gamma_{\mu}\, \tau^I\,  L_{s}^{j})  $ &
			$\mathcal{O}_{ V_L E_R }$&
			$ (N_f^4)\,  (\overline{E}_{Rp}\, \gamma^\mu\, E_{Rq} )\, (\overline{V}_{Lri}\, \gamma_{\mu}\,  V_{Ls}^{i}) $ \\
\hline
			
	\end{tabular}}
	\caption{SM extended by Vector-like Leptons ($ V_{L,R},\,E_{L,R},\,N_{L,R} $): Additional operators of dimension 6. $i,j$ and $ \alpha $ are the $ SU(2) $ and $ SU(3) $ indices respectively. $ p,q,r,s=1,2,\cdots,N_f $ are the flavour indices. $ \tau^I\, (I=1,2,3)$ is the $ SU(2) $ generator. All the operators in this table conserve the lepton and baryon numbers $ (\Delta B=0, \Delta L=0) $.}
	\label{table:VLL-dim6-psi4-ops1}
\end{table}

\begin{table}[h]
	\centering
	\renewcommand{\arraystretch}{1.9}
	{\scriptsize\begin{tabular}{||c|c||c|c||}
			\hline
			\hline
			\multicolumn{4}{||c||}{$\Psi^4$}\\
			\hline
			
			$\mathcal{O}_{ V_L V_R e E_L }^{(1)}$&
			$ \color{blue} (N_f^4)\, (\overline{V}_{Lpi}\,  V_{Rq}^{i})\,(\overline{E}_{Lr}\, e_{s}) $ &
			$\mathcal{O}_{ V_L V_R E_L E_R }^{(1)}$&
			$ \color{blue} (N_f^4)\, (\overline{V}_{Lpi}\, V_{Rq}^{i})\, (\overline{E}_{Lr}\, E_{Rs}) $  \\
			
			$\mathcal{O}_{ V_L V_R e E_L }^{(2)}$&
			$ \color{blue} (N_f^4)\, (\overline{V}_{Lpi}\, \sigma^{\mu\nu}\,  V_{Rq}^{i})\, (\overline{E}_{Lr}\, \sigma_{\mu\nu}\, e_{s}) $ &
			$\mathcal{O}_{ V_L V_R E_L E_R }^{(2)}$&
			$ \color{blue} (N_f^4)\, (\overline{V}_{Lpi}\, \sigma^{\mu\nu}\, V_{Rq}^{i})\,(\overline{E}_{Lr}\, \sigma_{\mu\nu}\, E_{Rs}) $  \\
			
			$\mathcal{O}_{ L V_L e N_R }^{(1)}$ &
			$ \color{blue}(N_f^4)\, \epsilon_{ij}\, (\overline{V}_{Lpi}\,  e_{q} )\, (\overline{L}_{rj}\, \gamma_{\mu}\,  N_{Rs} )  $ &
			$\mathcal{O}_{ L V_L E_R N_R }^{(1)}$ &
			$ \color{blue}(N_f^4)\, \epsilon_{ij}\, (\overline{V}_{Lpi}\, E_{Rq} )\, (\overline{L}_{rj}\, N_{Rs} )  $ \\
			
			$\mathcal{O}_{ L V_L e N_R }^{(2)}$ &
			$ \color{blue}(N_f^4)\, \epsilon_{ij}\, (\overline{V}_{Lpi}\, \sigma^{\mu\nu}\, e_{q} )\, (\overline{L}_{rj}\, \sigma_{\mu\nu}\,  N_{Rs} )  $ &
			$\mathcal{O}_{ L V_L E_R N_R }^{(2)}$ &
			$ \color{blue}(N_f^4)\, \epsilon_{ij}\, (\overline{V}_{Lpi}\, \sigma^{\mu\nu}\, E_{Rq} )\, (\overline{L}_{rj}\, \sigma_{\mu\nu}\,  N_{Rs} )  $ \\
			
			$\mathcal{O}_{ L V_R e E_L }^{(1)}$&
			$ \color{blue} (N_f^4)\, (\overline{L}_{pi}\,  V_{Rq}^{i})\,(\overline{E}_{Lr}\, e_{s}) $ &
			$\mathcal{O}_{ L V_R E_L E_R }^{(1)}$&
			$ \color{blue} (N_f^4)\, (\overline{L}_{pi}\, V_{Rq}^{i})\,(\overline{E}_{Lr}\, E_{Rs}) $  \\
\hline
			
		\end{tabular}}
	\caption{Table \ref{table:VLL-dim6-psi4-ops1} continued.}
\label{table:VLL-dim6-psi4-ops2}
\end{table}

	\begin{table}[h]
		\centering
		\renewcommand{\arraystretch}{1.9}
		{\scriptsize\begin{tabular}{||c|c||c|c||}
		\hline
		\hline
		\multicolumn{4}{||c||}{$\Psi^4$}\\
		\hline

			$\mathcal{O}_{ d e E_R }$&
		$ \color{blue}(N_f^4)\, (\overline{e}_{p}\, \gamma^{\mu}\, E_{Rq})\, (\overline{d}_{r\alpha}\, \gamma_{\mu}\, d_s^{\alpha}) $  &
		$\mathcal{O}_{ d L V_L }$ &
		$ \color{blue} (N_f^4)\, (\overline{d}_{p\alpha}\, \gamma^\mu\, d_{q}^{\alpha}\,)(\overline{L}_{pi}\, \gamma_\mu\, V_{Lq}^i\,)  $  \\
		
		$\mathcal{O}_{ d u e N_R }$&
		$ \color{blue}(N_f^4)\, (\overline{d}_{\alpha p}\, \gamma^{\mu}\, u_{q}^{\alpha})\, (\overline{N}_{Rr}\, \gamma_{\mu}\, e_{s}) $ &
		$\mathcal{O}_{ d u E_R N_R }$&
		$ \color{blue}(N_f^4)\, (\overline{d}_{\alpha p}\, \gamma^{\mu}\, u_{q}^{\alpha})\, (\overline{N}_{Rr}\, \gamma_{\mu}\, E_{Rs}) $ \\

		$\mathcal{O}_{ u L V_L }$ &
		$ \color{blue} (N_f^4)\, (\overline{u}_{p\alpha}\, \gamma^\mu\, u_{q}^{\alpha})\, (\overline{L}_{pi}\, \gamma^\mu\, V_{Lq}^i)  $ &
		$\mathcal{O}_{ u  \overline{e} E_R }$&
		$ \color{blue}(N_f^4)\, (\overline{u}_{p\alpha}\, \gamma^{\mu}\, u_{q}^{\alpha})\,  (\overline{e}_{r}\, \gamma_{\mu}\, E_{Rs}) $ \\

		$\mathcal{O}_{ V_L e E_R }$&
		$ \color{blue}(N_f^4)\, (\overline{e}_{p}\, \gamma^\mu\, E_{Rq} )\, (\overline{V}_{Lri}\, \gamma_{\mu}\,  V_{Ls}^{i})  $  &
		$\mathcal{O}_{ L e E_R}$&
		$ \color{blue}(N_f^4)\, (\overline{L}_{pi}\, \gamma^{\mu}\, L_{q}^{i})\, (\overline{e}_{r}\, \gamma_{\mu}\, E_{Rs})  $ \\
		
		$\mathcal{O}_{ L V_L E_L }$&
		$ \color{blue} (N_f^4)\, (\overline{L}_{pi}\, \gamma^{\mu}\, V_{Lq}^{i}) (\overline{E}_{Lr}\, \gamma_{\mu}\, E_{Ls}) $ &
		$\mathcal{O}_{ \overline{L} V_L \overline{e} E_R }$&
		$ \color{blue} (N_f^4)\, (\overline{L}_{pi}\, \gamma^{\mu}\, V_{Lq}^{i}) (\overline{e}_{r}\, \gamma_{\mu}\, E_{Rs}) $  \\
\hline
			
	\end{tabular}}
	\caption{Table~\ref{table:VLL-dim6-psi4-ops2} continued.}
	\label{table:VLL-dim6-psi4-ops3}
\end{table}

\begin{table}[h]
	\centering
	\renewcommand{\arraystretch}{1.9}
	{\scriptsize\begin{tabular}{||c|c||c|c||}
			\hline
			\hline
			\multicolumn{4}{||c||}{$\Psi^4$}\\
			\hline
			
			$\mathcal{O}_{ d u N_R}$&
			$ \color{purple}(N_f^4)\, \epsilon_{\alpha\beta\gamma}\, [(d^{\alpha}_{p})^T\, C\, d^{\beta}_{q}]\, [(u^{\gamma}_{r})^T\, C\, N_{Rs}]   $ &
			$\mathcal{O}_{ Q d N_R }$&
			$ \color{purple}\frac{1}{2}N_f^3( N_f+1) \epsilon^{\alpha\beta\gamma}\, \epsilon_{ij}\, [(Q^{i\alpha}_{p})^T\, C\, Q^{j\gamma}_{q}]\, [(d^{\beta}_{r})^T\, C\, N_{Rs}] $ \\
			
			$\mathcal{O}_{ Q d u V_L }$&
			$ \color{purple} (N_f^4)\, \epsilon_{\alpha\beta\gamma}\, \epsilon_{ij}\,  [(Q^{i\alpha}_{p})^T\, C\, V^{j}_{Lq}]\, [(d^{\beta}_{r})^T\, C\,  u^{\gamma}_{s}] $ &
			$\mathcal{O}_{ Q V_L }$&
			$ \color{purple}\frac{1}{3}N_f^2(2N_f^2+1) \epsilon_{\alpha\beta\gamma}\,\epsilon_{jn}\,\epsilon_{km}\,[(Q^{j\alpha}_{p})^T\,C\,Q^{k\beta}_{q}]\,[(Q^{m\gamma}_{r})^T\,C\,V^{n}_{Ls}] $\\
			
			$\mathcal{O}_{ u d d N_L}$&
			$ \color{purple}(N_f^4)\, \epsilon_{\alpha\beta\gamma}\, [\overline{N}_{Lp}\, u^{\alpha}_{q}]\, [(d^{\beta}_{r})^T\, C\, d^{\gamma}_{s}]   $ &
			$\mathcal{O}_{ Q u E_R}$&
			$ \color{purple} \frac{1}{2}N_f^3( N_f+1)\, \epsilon^{\alpha\beta\gamma}\, \epsilon_{ij}\, [(Q^{i\alpha}_{p})^T\, C\, Q^{j\beta}_{q}]\, [(u^{\gamma}_{r})^T\, C\, E_{Rs}]  $ \\
			
			$\mathcal{O}_{ u u d E_R}$&
			$ \color{purple}(N_f^4)\, \epsilon_{\alpha\beta\gamma}\, [(u^{\alpha}_{p})^T\, C\, u^{\beta}_{q}]\, [(d^{\gamma}_{r})^T\, C\, E_{Rs}]   $ &
			$\mathcal{O}_{ d E_L }$&
			$ \boxed{ \color{purple}\frac{1}{3}N_f^2(N_f^2-1)\, \epsilon_{\alpha\beta\gamma}\, ((d^{\alpha}_{p})^T\, C\, d^{\beta}_{q})\, (\overline{E}_{Lr}\, d^{\gamma}_{s})} $ \\
\hline
			
	\end{tabular}}
	\caption{SM extended by Vector-like Leptons ($ V_{L,R},\,E_{L,R},\,N_{L,R} $): Additional operators of dimension 6. $i,j,m,n$ and $ \alpha,\beta,\gamma $ are the $ SU(2) $ and $ SU(3) $ indices respectively. $ p,q,r,s=1,2,\cdots,N_f $ are the flavour indices. The operators above the dashed line violate the baryon and lepton number $ (\Delta B=1, \Delta L=\pm 1) $ and below the dashed line violate only the lepton number $ (\Delta B=0, \Delta L= -4) $. }
	\label{table:VLL-dim6-psi4-ops4}
\end{table}

%%%%%%%%%%%%%%%%%%%%%%%%%%%%%%%%%%%%%%%%%%%%%%%%%%%5

\begin{table}[h]
	\centering
	\renewcommand{\arraystretch}{1.9}
	{\scriptsize\begin{tabular}{||c|c||c|c||}
			\hline
			\hline
			\multicolumn{4}{||c||}{$\Psi^4$}\\
			\hline
			
				$\mathcal{O}_{ Q d L N_L }^{(1)}$&
			$ \color{purple}(N_f^4)\, \epsilon_{ij}\, (\overline{d}_{p\alpha}\, Q^{i\alpha}_q)\, ((L^j_r)^T\, C\, N_{Ls}) $&
			$\mathcal{O}_{ Q d V_L N_L }^{(1)}$&
			$ \color{purple}(N_f^4)\, \epsilon_{ij}\, (\overline{d}_{p\alpha}\, Q^{i\alpha}_q)\, ((V^j_{Lr})^T\, C\, N_{Ls}) $ \\
			
			$\mathcal{O}_{ Q d L N_L }^{(2)}$&
			$ \color{purple}(N_f^4)\, \epsilon_{ij}\, (\overline{d}_{p\alpha}\, \sigma_{\mu\nu}\, Q^{i\alpha}_q)\, ((L^j_r)^T\, C\, \sigma^{\mu\nu}\, N_{Ls}) $&
			$\mathcal{O}_{ Q d V_L N_L }^{(2)}$&
			$ \color{purple}(N_f^4)\, \epsilon_{ij}\, (\overline{d}_{p\alpha}\, \sigma_{\mu\nu}\, Q^{i\alpha}_q)\, ((V^j_{Lr})^T\, C\, \sigma^{\mu\nu}\, N_{Ls}) $\\
			
			$\mathcal{O}_{ Q N_L N_R } $&
			$ \color{purple}(N_f^4)\, (\overline{Q}_{p\alpha i}\, N_{Rq})\, ((Q^{i\alpha}_{r})^T\, C\, N_{Ls}) $&
			$\mathcal{O}_{ Q u L N_L } $&
			$ \color{purple}(N_f^4)\, (\overline{Q}_{p\alpha i}\, u^{\alpha}_q)\, ((L^{i}_{r})^T\, C\, N_{Ls}) $ \\
			
			$\mathcal{O}_{ Q u V_L N_L } $ &
			$ \color{purple}(N_f^4)\, (\overline{Q}_{p\alpha i}\, u^{\alpha}_q)\, ((V^{i}_{Lr})^T\, C\, N_{Ls}) $ &
			$\mathcal{O}_{ Q u V_R N_R }^{(1)} $ &
			$ \color{purple}(N_f^4)\, (\overline{Q}_{p\alpha i}\, u^{\alpha}_q)\, ((V^{i}_{Rr})^T\, C\, N_{Rs}) $ \\
			
			$\mathcal{O}_{ Q d V_R N_R } $ &
			$ \color{purple}(N_f^4)\, \epsilon_{ij}\, (\overline{d}_{p\alpha}\, Q^{i\alpha}_q)\, ((V^j_{Rr})^T\, C\, N_{Rs}) $  &
			$\mathcal{O}_{ Q u V_R N_R }^{(2)} $ &
			$ \color{purple}(N_f^4)\, (\overline{Q}_{p\alpha i}\, \sigma_{\mu\nu}\,  u^{\alpha}_q)\, ((V^{i}_{Rr})^T\, C\, \sigma^{\mu\nu}\,  N_{Rs}) $ \\
			
			$\mathcal{O}_{ u d V_L V_R } $&
			$ \color{purple}(N_f^4)\, \epsilon_{ij}\, (\overline{d}_{p\alpha}\,  V_{Lq}^{i})\, ((u^{j\alpha}_{r})^T\, C\, V_{Rs}^j) $ &
			$\mathcal{O}_{ u d L V_R } $&
			$ \color{purple}(N_f^4)\, \epsilon_{ij}\, (\overline{d}_{p\alpha}\,  L^i_q)\, ((u^\alpha_r)^T\, C\, V_{Rs}^j) $ \\
			
			$\mathcal{O}_{ u d e N_L } $&
			$ \color{purple}(N_f^4)\, (\overline{d}_{p\alpha}\, N_{Lq})\, ((u_{r}^{\alpha})^T\, C\, e_{s}) $ &
			$\mathcal{O}_{ u d E_R N_L } $&
			$ \color{purple}(N_f^4)\, (\overline{d}_{p\alpha}\, N_{Lq})\, ((u_{r}^{\alpha})^T\, C\, E_{Rs}) $ \\
			
			$\mathcal{O}_{ u d E_L N_R } $&
			$ \color{purple}(N_f^4)\, (\overline{d}_{p\alpha}\, E_{Lq})\, ((u_{r}^{\alpha})^T\, C\, N_{Rs} ) $ &
			$\mathcal{O}_{ d N_L N_R }$&
			$ \color{purple}(N_f^4)\, (\overline{d}_{p\alpha}\,  N_{Lq} )\, ((d_{r}^{\alpha})^T\, C\,   N_{Rs})  $ \\

			\hline
			
		\end{tabular}}
	\caption{SM extended by Vector-like Leptons ($ V_{L,R},\,E_{L,R},\,N_{L,R} $): Additional operators of dimension 6. $i,j$ and $ \alpha $ are the $ SU(2) $ and $ SU(3) $ indices respectively. $ p,q,r,s=1,2,\cdots,N_f $ are the flavour indices. All the operators violate only the lepton number $ (\Delta B=0, \Delta L= -2) $. }
\label{table:VLL-dim6-psi4-ops5}
\end{table}

\begin{table}[h]
	\centering
	\renewcommand{\arraystretch}{1.9}
		{\scriptsize\begin{tabular}{||c|c||c|c||}
		\hline
		\hline
		\multicolumn{4}{||c||}{$\Psi^4$}\\
		\hline
			
			$\mathcal{O}_{L e N_L }$&
			$ \color{purple}(N_f^4)\, \epsilon_{ij}\, [\overline{e}_{p}\,  N_{Lq}]\, [(L^{i}_{r})^T\, C\, L^{j}_{s}] $ &
			$\mathcal{O}_{ V_R e  N_L }$&
			$\boxed{ \color{purple}\frac{1}{2}N_f^3(N_f-1) \epsilon_{ij}\, [\overline{e}_{p}\,  N_{Lq}]\, [(V^{i}_{Rr})^T\, C\, V^{j}_{Rs}]} $\\
			
			$\mathcal{O}_{ L V_L e N_L }^{(1)}$&
			$ \color{purple}(N_f^4)\, \epsilon_{ij}\, [\overline{e}_{p}\,   L^{i}_{q}]\, [(V^{j}_{Lr})^T\, C\, N_{Ls}] $ &
			$\mathcal{O}_{ V_L e N_L }$&
			$ \color{purple}(N_f^4)\, \epsilon_{ij}\, [\overline{e}_{p}\, V^{i}_{Lq}]\, [(V^{j}_{Lr})^T\, C\, N_{Ls}] $ \\
			
			$\mathcal{O}_{ L V_L e N_L }^{(2)}$&
			$ \color{purple}(N_f^4)\, \epsilon_{ij}\, [\overline{e}_{p}\, \sigma_{\mu\nu}\, L^{i}_{q}]\, [(V^{j}_{Lr})^T\, C\, \sigma^{\mu\nu} N_{Ls}] $ &
			$\mathcal{O}_{ \overline{e} E_L N_L }$&
			$ \color{purple}(N_f^4)\, (\overline{e}_{p}\, E_{Lq})\, (N_{Rr}^T\, C\, N_{Rs})  $ \\
			
			$\mathcal{O}_{ L V_R e N_R } $&
			$ \color{purple}(N_f^4)\, \epsilon_{ij}\, [\overline{e}_{p}\,  L^{i}_{q}]\, [(V^{j}_{Rr})^T\, C\, N_{Rs}] $ &
			$\mathcal{O}_{ V_L V_R e N_R } $&
			$ \color{purple}(N_f^4)\, \epsilon_{ij}\, [\overline{e}_{p}\, V^{i}_{Lq}]\, [(V^{j}_{Rr})^T\, C\, N_{Rs}] $ \\
			
			$\mathcal{O}_{ e N_L N_R }$&
			$ \color{purple}(N_f^4) (\overline{e}_{p}\, N_{Lq})\, (e_{r}^T\, C\, N_{Rs}) $&
			$\mathcal{O}_{ \overline{e} E_R N_L N_R }$&
			$ \color{purple}(N_f^4) (\overline{e}_{p}\,N_{Lq})\, (E_{Rr}^T\, C\, N_{Rs}) $ \\
			
			$\mathcal{O}_{ L V_R E_L N_L }$&
			$ \color{purple}(N_f^4)\, \epsilon_{ij}\, [\overline{E}_{Lp}\, V^{i}_{Rq}]\, [(L^{j}_{r})^T\, C\, N_{Ls}]  $&
			$\mathcal{O}_{ \overline{e} E_L N_R}$&
			$ \color{purple}\frac{1}{2}N_f^3(N_f+1) (\overline{e}_{p}\, E_{Lq})\, (N_{Rr}^T\, C\, N_{Rs}) $ \\
			
			$\mathcal{O}_{ V_L V_R E_L N_L }$&
			$ \color{purple}(N_f^4)\, \epsilon_{ij}\, [\overline{E}_{Lp}\, V^{i}_{Rq}]\, [(V^{j}_{Lr})^T\, C\, N_{Ls}]  $&
			$\mathcal{O}_{ e \overline{E}_L N_L }$&
			$ \color{purple}\frac{1}{2}N_f^3(N_f+1) (\overline{E}_{Lp}\, e_{q})\, (N_{Lr}^T\, C\, N_{Ls}) $\\
			
			$\mathcal{O}_{ \overline{E}_L E_R N_L }$&
			$ \color{purple}\frac{1}{2}N_f^3(N_f+1) (\overline{E}_{Lp}\, E_{Rq})\, (N_{Lr}^T\, C\, N_{Ls}) $ &
			$\mathcal{O}_{ L E_L N_R }$&
			$ \boxed{ \color{purple} \frac{1}{2}N_f^3(N_f-1)\, \epsilon_{ij}\, [\overline{E}_{Lp}\,  N_{Rq}]\, [(L^{i}_{r})^T\, C\, L^{j}_{s}]} $ \\
			
			$\mathcal{O}_{ L V_L E_L N_R }$&
			$ \color{purple}(N_f^4)\, \epsilon_{ij}\, [\overline{E}_{Lp}\, N_{Rq}]\, [(L^{i}_{r})^T\, C\,  V^{j}_{Ls}]  $ &
			$\mathcal{O}_{ V_L E_L N_R }$&
			$ \boxed{ \color{purple} \frac{1}{2}N_f^3(N_f-1)\, \epsilon_{ij}\, [\overline{E}_{Lp}\,  N_{Rq}]\, [(V^{i}_{Lr})^T\, C\, V^{j}_{Ls}] }$ \\
			
			$\mathcal{O}_{ V_R E_L N_R }$&
			$ \color{purple}(N_f^4)\, \epsilon_{ij}\, [\overline{E}_{Lp}\,  V^{i}_{Rq}]\, [(V^{j}_{Rr})^T\, C\, N_{Rs}]  $ &
			$\mathcal{O}_{ E_L N_L N_R}$&
			$ \color{purple}(N_f^4)\, (\overline{E}_{Lp}\, N_{Rq})\, (E_{Lr}^T\, C\, N_{Ls}) $ \\
			
			$\mathcal{O}_{ e \overline{E}_L N_R}$&
			$ \color{purple}(N_f^4)\,(\overline{E}_{Lp}\, N_{Rq})\, (e_{r}^T\, C\, N_{Rs}) $ &
			$\mathcal{O}_{ \overline{E}_L E_R N_R}$&
			$ \color{purple}(N_f^4)\,(\overline{E}_{Lp}\, N_{Rq})\,(E_{Rr}^T\, C\, N_{Rs}) $ \\
			
			$\mathcal{O}_{ L E_R N_L }$&
			$ \color{purple}(N_f^4)\, \epsilon_{ij}\, [\overline{E}_{Rp}\, L^{i}_{q}]\, [(L^{j}_{r})^T\, C\, N_{Ls}] $ &
			$\mathcal{O}_{ V_R E_R N_L }$&
			$ \boxed{\color{purple}\frac{1}{2}N_f^3(N_f-1)\, \epsilon_{ij}\, [\overline{E}_{Rp}\,  N_{Lq}]\, [(V^{i}_{Rr})^T\, C\, V_{Rs}^{j}]} $ \\

			%%%%%%%%%%%%%%%%%%%%%%%%%%%%%%%%%%%%%%%%%%%%%%%%%%%%%%%%%%%%

		$\mathcal{O}_{ L V_L E_R N_L }^{(1)}$&
		$ \color{purple}(N_f^4)\, \epsilon_{ij}\, [\overline{E}_{Rp}\, L^{i}_{q}]\, [(V^{j}_{Lr})^T\, C\,  N_{Ls}] $ &
		$\mathcal{O}_{ V_L E_R N_L }$&
		$ \color{purple}(N_f^4)\, \epsilon_{ij}\, [\overline{E}_{Rp}\, V^{i}_{Lq}]\, [(V^{j}_{Lr})^T\, C\, N_{Ls}] $ \\
		
		$\mathcal{O}_{ L V_L E_R N_L }^{(2)}$&
		$ \color{purple}(N_f^4)\,\epsilon_{ij}\, [\overline{E}_{Rp}\, \sigma_{\mu\nu}\, L^{i}_{q}]\, [(V^{j}_{Lr})^T\, C\, \sigma^{\mu\nu}\,  N_{Ls}] $ &
		$\mathcal{O}_{ E_L \overline{E}_R N_L }$ &
		$ \color{purple}(N_f^4)\,[\overline{E}_{Rp}\, N_{Lq}]\, [(E_{Lr})^T\,C\,N_{Ls}] $ \\
		
		$\mathcal{O}_{ L V_R E_R N_R }$&
		$ \color{purple}(N_f^4)\, \epsilon_{ij}\, [\overline{E}_{Rp}\, L^{i}_{q}]\, [(V^{j}_{Rr})^T\, C\, N_{Rs}] $ &
		$\mathcal{O}_{ V_L V_R E_R N_R }$&
		$ \color{purple}(N_f^4)\, \epsilon_{ij}\, [\overline{E}_{Rp}\,  V^{i}_{Lq}]\, [(V^{j}_{Rr})^T\, C\, N_{Rs}] $ \\
		
		$\mathcal{O}_{ e \overline{E}_R N_L N_R }$ &
		$ \color{purple}(N_f^4)\,(\overline{E}_{Rp}\,N_{Lq}\,)\,(e_{r}^T\, C\, N_{Rs}) $ &
		$\mathcal{O}_{ E_R N_L N_R }$ &
		$ \color{purple}(N_f^4)\, (\overline{E}_{Rp}\, N_{Lq}\,)\, (E_{Rr}^T\, C\, N_{Rs}) $ \\
		
		$\mathcal{O}_{ \overline{L} V_R N_L }$ &
		$ \color{purple}\frac{1}{2}N_f^3(N_f+1)\, (\overline{L}_{pi}\, V_{Rq}^i\,) (N_{Lr}^T\, C\, N_{Ls}) $ &
		$\mathcal{O}_{ E_L \overline{E}_R N_R }$ &
		$ \color{purple}\frac{1}{2}N_f^3(N_f+1)\, (\overline{E}_{Rp}\, E_{Lq}\,) (N_{Rr}^T\, C\, N_{Rs}) $ \\
		
		$\mathcal{O}_{ L N_L N_R }$ &
		$ \color{purple} (N_f^4)\, (\overline{L}_{pi}\, N_{Rq})\,  ((L_{r}^i)^T\, C\,  N_{Ls}) $ &
		$\mathcal{O}_{ \overline{L} V_L N_R N_L }$ &
		$ \color{purple} (N_f^4)\, (\overline{L}_{pi}\, N_{Rq}) \, ((V_{Lr}^i)^T\, C\,  N_{Ls}) $ \\
		
		$\mathcal{O}_{ \overline{L} V_R N_R }$ &
		$ \color{purple} (N_f^4)\, (\overline{L}_{pi}\, N_{Rq})\,  ((V_{Rr}^i)^T\, C\,  N_{Rs}) $ &
		$\mathcal{O}_{ \overline{V}_L V_R N_L }$&
		$ \color{purple}\frac{1}{2}N_f^3(N_f+1)\, (\overline{V}_{Lpi}\, V_{Rq}^i\,) (N_{Lr}^T\, C\, N_{Ls})  $ \\
		
		$\mathcal{O}_{ L \overline{V}_l N_L N_R }$ &
		$ \color{purple} (N_f^4)\, (\overline{V}_{Lpi}\, N_{Rq}) \, ((L_{r}^i)^T\, C\,  N_{Ls}) $  &
		$\mathcal{O}_{ V_L  N_L N_R }$&
		$ \color{purple} (N_f^4)\, (\overline{V}_{Lpi}\, N_{Rq})\, ((V_{Lr}^i)^T\, C\, N_{Ls})  $ \\
		
		$\mathcal{O}_{ \overline{V}_L V_R N_R }$ &
		$ \color{purple} (N_f^4)\, (\overline{V}_{Lpi}\, V_{Rq}^{i}\,) \, (N_{Rr}^T\, C\,  N_{Rs}) $ &
		$\mathcal{O}_{ \overline{V}_R V_L N_L}$&
		$ \color{purple}(N_f^4)\, (\overline{V}_{Rpi}\,  V_{Lq}^i\,)\, (N_{Lr}^T\, C\, N_{Ls})  $ \\
		
		$\mathcal{O}_{ V_R N_L N_R } $&
		$ \color{purple}(N_f^4)\, (\overline{V}_{Rpi}\, N_{Lq})\, ((V_{Rr}^i)^T\, C\, N_{Rs})  $ &
		$\mathcal{O}_{ L \overline{V}_R N_L}$&
		$ \color{purple}(N_f^4)\, (\overline{V}_{Rpi}\, L_{q}^i\,) (N_{Lr}^T\, C\, N_{Ls})  $ \\
		
		$\mathcal{O}_{ L \overline{V}_R N_R }$&
		$ \color{purple}\frac{1}{2}N_f^3(N_f+1)\, (\overline{V}_{Rpi}\, L_{q}^i) \, (N_{Rr}^T\, C\, N_{Rs}) $ &
		$\mathcal{O}_{ V_L \overline{V}_R  N_R }$&
		$ \color{purple}\frac{1}{2}N_f^3(N_f+1)\, (\overline{V}_{Rpi}\, V_{Lq}^i) \, (N_{Rr}^T\, C\, N_{Rs}) $ \\

		$\mathcal{O}_{ \overline{N}_L N_L N_R}$&
		$ \color{purple}\frac{1}{2}N_f^3(N_f+1)\, (\overline{N}_{Lpi}\, N_{Rq}^i) \, (N_{Lr}^T\, C\, N_{Ls}) $ &
		$\mathcal{O}_{ \overline{N}_L N_R N_R}$&
		$ \boxed{ \color{purple}\frac{1}{3}N_f^2(N_f^2-1)\, (\overline{N}_{Lpi}\, N_{Rq}^i) \, (N_{Rr}^T\, C\, N_{Rs}) } $ \\
		
		$\mathcal{O}_{ \overline{N}_R N_L N_L}$&
		$ \boxed{ \color{purple}\frac{1}{3}N_f^2(N_f^2-1)\, (\overline{N}_{Rp}\, N_{Lq}) \, (N_{Lr}^T\, C\, N_{Ls}) }$ &
		$\mathcal{O}_{ N_L \overline{N}_R N_R }$&
		$ \color{purple}\frac{1}{2}N_f^3(N_f+1)\, (\overline{N}_{Rp}\, N_{Lq}) \, (N_{Rr}^T\, C\, N_{Rs}) $ \\
		
		$\mathcal{O}_{ u N_L N_R }$&
		$ \color{purple}(N_f^4)\, (\overline{u}_{p\alpha}\, N_{Lq})\,  ((u_{r}^\alpha)^T\, C\, N_{Rs}) $ &
		&
		\\
			
			\hline
			
	\end{tabular}}
	\caption{Table~\ref{table:VLL-dim6-psi4-ops5} continued.}
	\label{table:VLL-dim6-psi4-ops5-contd}
\end{table}

\clearpage

\clearpage
\subsection{Standard Model extended by colored particles}
So far we have discussed the possible choices of lighter DOFs which are color singlets.
Next, we have considered a few cases where the light BSM particles are Lepto-Quark scalars that transform non-trivially under $SU(3)_C$, see Table~\ref{table:Lepto-Quark-quantum-no}. These particles possess attractive phenomenological features due to their participation in color interactions 
 \cite{Dey:2017ede,Bandyopadhyay:2018syt,Das:2017kkm,Arnold_2013,Dorsner:2014axa,Kohda:2012sr,Dorsner:2016wpm} and to be precise for their role in explaining the B-physics anomalies \cite{Monteux:2018ufc,Bauer:2015knc}. The Lepto-Quarks may belong to multiplets of a rather complete theory, e.g., Pati-Salam model \cite{Pati:1974yy,PhysRevD.8.1240}, unified scenarios \cite{FileviezPerez:2008dw,Dorsner:2005fq,Buchmuller:1986iq}\footnote{Supersymmetric theories also naturally contain colored scalars very similar to this IR DOF.} etc. In most of the UV complete  theories, the colored scalars are accompanied by other particles. To capture their impact in low energy predictions, it is suggestive to consider the effective operators involving these colored scalars. To encapsulate that, we have constructed the effective operator basis beyond SMEFT including these Lepto-Quarks.

\begin{table}[h]
	\centering
	\renewcommand{\arraystretch}{1.6}
	{\scriptsize\begin{tabular}{|c|c|c|c|c|c|c|c|}
			\hline
			\multirow{2}{*}{\textbf{Model No.}}&
			\textbf{Non-SM IR DOFs}&
			\multirow{2}{*}{$SU(3)_C$}&
			\multirow{2}{*}{$SU(2)_L$}&
			\multirow{2}{*}{$U(1)_Y$}&
			\multirow{2}{*}{\textbf{Spin}}&
			\multirow{2}{*}{\textbf{Baryon No.}}&
			\multirow{2}{*}{\textbf{Lepton No.}}\\
			
			&
			\textbf{(Lepto-Quarks)}&
			&
			&
			&
			&
			&
			\\			
			\hline

			1&
			$\chi_1$&
			3&
			2&
			1/6&
			0&
			1/3&
			-1\\
			\hline
						
			2&
			$\varphi_1$&
			3&
			1&
			2/3&
			0&
			1/3&
			-1\\
			\hline
	\end{tabular}}
	\caption{\small Additional IR DOFs (Lepto-Quarks) as representations of the SM gauge groups along with their spin and baryon and lepton numbers.}
	\label{table:Lepto-Quark-quantum-no}
\end{table} 
\noindent
We would like to mention that due to their non-trivial transformation properties under $SU(3)_C$,  while computing the effective operators in covariant forms we may require following tensors $f^{ABC}$ and $d^{ABC}$, defined as:

\vspace{-0.6cm}
{\small\begin{eqnarray}
	[T^{A},\,T^{B}] = \sum_{C=1}^{8}\,f^{ABC}\,T^{C}, \hspace{0.7cm}
	\{T^{A},\,T^{B}\} = \frac{1}{3}\delta^{AB}\,\mathbb{I}_3 + \sum_{C=1}^{8}\,d^{ABC}\,T^{C}.
	\end{eqnarray}}
Here, $\delta^{AB}$ is the Kronecker delta and $\mathbb{I}_3$ is the $3\times 3$ unit matrix.  We have also used specific forms of the derivatives as:

\vspace{-0.6cm}
{\small\begin{eqnarray}
	i\overleftrightarrow{\mathcal{D}}_{\mu}^{I} = \tau^{I}\,i\mathcal{D}_{\mu} - i\overleftarrow{\mathcal{D}}_{\mu}\,\tau^{I}\hspace{0.5cm} \text{and} \hspace{0.5cm} 
	i\overleftrightarrow{\mathcal{D}}_{\mu}^{A} = T^{A}\,i\mathcal{D}_{\mu} - i\overleftarrow{\mathcal{D}}_{\mu}\,T^{A}.
	\end{eqnarray}}

\vspace{-0.8cm}
\subsubsection*{\underline{SM + Lepto-Quark ($\chi_1$)}}

To start with,  we have considered a color triplet, iso-spin doublet scalar with hypercharge 1/6, and specific baryon and lepton numbers, see Table~\ref{table:Lepto-Quark-quantum-no}. This particle possesses similar gauge charges  as the SM quark doublet and allows mixing between  quarks and leptons. The effective operators of dimensions 5 and 6 containing $\chi_{1}$ have been catalogued in Tables~\ref{table:Lepto-QuarkModel1-dim5-ops} and~\ref{table:Lepto-QuarkModel1-dim6-ops-1}-\ref{table:Lepto-QuarkModel1-dim6-ops-1-contd} respectively. The operators with distinct hermitian conjugates have been highlighted in blue colour.

\newpage
\noindent\textbf{Features of the additional operators:}
\begin{itemize}
	\item We have noted the presence of lepton and baryon number conserving as well as violating operators in the $\Psi^2\Phi^2$ class only. The mixing between quark and lepton sectors is  induced through operators like $\tilde{\mathcal{O}}_{QLH\chi_1}^{(1), (2)}$, $\tilde{\mathcal{O}}_{ueH\chi_{1}}$, $\tilde{\mathcal{O}}_{ue\chi_{1}}$ and  $\tilde{\mathcal{O}}_{QL\chi_{1}}$.

	\item As $\chi_1$ transforms non-trivially under all three gauge groups, there exist multiple operators with similar structures belonging to the $\Phi^6$, $\Phi^4\mathcal{D}^2$, $\Psi^2\Phi^2\mathcal{D}$, and $\Psi^2\Phi^3$ classes.  In addition, $\chi_1$  offers multiple ways to contract the gauge indices to form the invariant operators. Thus a naive construction may lead to erroneous results and one may end up with an overcomplete set of operators. To avoid this, we have suitably taken care of the constraints and identities discussed in section \ref{sec:op-constr-procedure} to get rid of the redundant operators. For example, the set $\mathcal{O}^{(1)}_{Q\chi_{1}\mathcal{D}}$ -  $\mathcal{O}^{(4)}_{Q\chi_{1}\mathcal{D}}$ exhausts the list of independent operators.  All other structures are related to these operators as shown in Eqns.~\eqref{eq:fierz-psi-phi-D-cls-1}-\eqref{eq:fierz-psi-phi-D-cls-3}.

	\item There are $new$ lepton and baryon number violating operators (in red colour) in the $\Phi^4\mathcal{D}^2$, $\Psi^2\Phi^2\mathcal{D}$, $\Psi^2\Phi^3$ and $\Psi^2\Phi X$ classes.  There is notable mixing between the quark and lepton sectors within the $\Psi^2\Phi^2\mathcal{D}$, $\Psi^2\Phi^3$, and $\Psi^2\Phi X$ classes.
	
	\item Within the $\Phi^2X^2$ class, we have observed the mixing of color field strength tensor ($G^{A}_{\mu\nu}$) with the electroweak ones ($W^{I}_{\mu\nu}$, $B_{\mu\nu}$). This feature is specific to this particular model.
\end{itemize}

\begin{table}[h]
	\centering
	\renewcommand{\arraystretch}{1.9}
	{\scriptsize\begin{tabular}{||c|c||c|c||}
			\hline
			\hline
			\multicolumn{4}{||c||}{$\Psi^2\Phi^2$}\\
			\hline
			$\tilde{\mathcal{O}}_{Q L H \chi_1}^{(1)}$&
			$\color{blue}{(N_f^2) \,\epsilon_{ij} \,((Q^{\alpha i}_{p})^T\, C \,L^j_{q}) \,H^k \,\chi^{\dagger}_{1,\alpha k}}$
			&$\tilde{\mathcal{O}}_{Q L H \chi_1}^{(2)}$&
			$\color{blue}{(N_f^2) \,\epsilon_{ij} \,((Q^{\alpha i}_{p})^T\,C \,\tau^I \,L^j_{q}) \,(H^k \,\tau^I\,\chi^{\dagger}_{1,\alpha k})}$
			\\
			
			$\tilde{\mathcal{O}}_{u e H \chi_1}$&
			$\color{blue}{(N_f^2) \,((u^{\alpha}_{p})^T\,C \,e_{q}) \,(H^i \,\chi^{\dagger}_{1,\alpha i})}$&
			$\tilde{\mathcal{O}}_{Q \chi_1}^{(1)}$&
			$\color{purple}{\frac{1}{2} (N_f^2+N_f) \,((Q^{\alpha i}_{p})^T \,C \,Q^{\beta j}_{q}) \,(\chi^{\dagger}_{1,\alpha i} \,\chi^{\dagger}_{1,\beta j})}$
			\\

			$\tilde{\mathcal{O}}_{u d \chi_1}$&
			$\color{purple}{(N_f^2) \,\epsilon_{ij} \,(I(u^{\alpha}_{p})^T\,C \,d^{\beta}_{q}) \,(\chi^{\dagger}_{1,\alpha i} \,\chi^{\dagger}_{1,\beta j})}$&
			$\tilde{\mathcal{O}}_{Q \chi_1}^{(2)}$&
			$\color{purple}{\frac{1}{2} (N_f^2+N_f) \,((Q^{\alpha i}_{p})^T \,C \,T^A\,Q^{\beta j}_{q}) \,(\chi^{\dagger}_{1,\alpha i} \,T^A\,\chi^{\dagger}_{1,\beta j})}$\\
			
			$\tilde{\mathcal{O}}_{u d H \chi_1}$&
			$\color{purple}{(N_f^2) \,\epsilon_{\alpha\beta\gamma} \,((u^{\alpha}_{p})^T\, C \,d^{\beta}_{q}) \,(\tilde{H}_i \,\chi_1^{\gamma i})}$
			&$\tilde{\mathcal{O}}_{Q H \chi_1}$&
			$\color{purple}{(N_f^2) \,\epsilon_{\alpha\beta\gamma} \,\epsilon_{ij} \,((Q^{\alpha i}_{p})^T \,C \,Q^{\beta j}_{q}) \,(\tilde{H}_k \,\chi_1^{\gamma k})}$
			\\

			$\tilde{\mathcal{O}}_{Q L \chi_1}$&
			$\color{purple}{(N_f^2) \,\epsilon_{\alpha\beta\gamma} \,\epsilon_{ij} \,\epsilon_{kl} \,((Q^{\alpha i}_{p})^T\, C \,L^j_{q}) \,(\chi^{\beta k}_1 \,\chi^{\gamma l}_1)}$
			&$\tilde{\mathcal{O}}_{u e \chi_1}$&
			$\color{purple}{(N_f^2) \,\epsilon_{\alpha\beta\gamma} \,\epsilon_{ij} \,((u^{\alpha}_{p})^T\, C \,e_{q}) \,(\chi^{\beta i}_1 \,\chi^{\gamma j}_1)}$
			\\
			
			$\tilde{\mathcal{O}}_{d H \chi_1}$&
			$\boxed{\color{purple}{\frac{1}{2} (N_f^2 -N_f) \,\epsilon_{\alpha\beta\gamma} \,\epsilon_{ij} \,((d^{\alpha}_{p})^T \,C \,d^{\beta}_{q}) \,(H^i \,\chi_1^{\gamma j})}}$&
			&
			\\
			\hline
	\end{tabular}}
	\caption{SM extended by Lepto-Quark ($\chi_1$): Additional operators of  dimension 5. Here $i, j, k, l$ and $\alpha, \beta, \gamma$ are the $SU(2)$ and $SU(3)$ indices respectively. $T^A$ and $\tau^I$ are $SU(3)$ and $SU(2)$ generators respectively. $A=1,2,\cdots,8$ and $I=1,2,3$. $p, q=1,2,\cdots,N_f$ are the flavour indices. Operators in {\it red} violate lepton and baryon numbers. }
	\label{table:Lepto-QuarkModel1-dim5-ops}
\end{table}

\begin{table}[h]
	\centering
	\renewcommand{\arraystretch}{1.9}
	{\scriptsize\begin{tabular}{||c|c||c|c||}
			\hline
			\hline
			\multicolumn{2}{||c||}{$\Phi^4\mathcal{D}^2$}&
			\multicolumn{2}{c||}{$\Phi^6$}\\
			\hline
			$\mathcal{O}_{\chi_1 \square}^{(1)}$&
			$(\chi^{\dagger}_1 \,\chi_1) \,\square \,(\chi^{\dagger}_1 \,\chi_1)$&
			$\mathcal{O}_{\chi_1}^{(1)}$&
			$(\chi_1^{\dagger} \,\chi_1)^3$\\
			
			$\mathcal{O}_{\chi_1 \square}^{(2)}$&
			$(\chi^{\dagger}_1 \,T^A \,\chi_1) \,\square \,(\chi^{\dagger}_1 \,T^A \,\chi_1)$&
			$\mathcal{O}_{\chi_1}^{(2)}$&
			$(\chi_1^{\dagger} \,T^A \,\chi_1) \,(\chi_1^{\dagger} \,T^A \,\chi_1) \,(\chi_1^{\dagger} \,\chi_1)$
			\\
			
			$\mathcal{O}_{\chi_1 \mathcal{D}}^{(1)}$&
			$(\chi^{\dagger}_1 \,\overleftrightarrow{\mathcal{D}}_{\mu} \,\chi_1)\,(\chi^{\dagger}_1 \,\overleftrightarrow{\mathcal{D}}^{\mu} \,\chi_1)$&
			$\mathcal{O}_{H^2 \chi_1^4}^{(1)}$&
			$(\chi^{\dagger}_1 \,\chi_1)^2 \,(H^{\dagger} \,H)$
			\\
			
			$\mathcal{O}_{\chi_1 \mathcal{D}}^{(2)}$&
			$(\chi^{\dagger}_1 \,\overleftrightarrow{\mathcal{D}}_{\mu}^{A} \,\chi_1)\,(\chi^{\dagger}_1 \,\overleftrightarrow{\mathcal{D}}^{\mu A} \,\chi_1)$&
			$\mathcal{O}_{H^2 \chi_1^4}^{(2)}$&
			$(\chi^{\dagger}_1 \,T^A \,\chi_1) \,(\chi^{\dagger}_1 \,T^A \,\chi_1) \,(H^{\dagger} \,H)$
			\\
			
			$\mathcal{O}_{H \chi_1 \square}$&
			$(\chi^{\dagger}_1 \,\chi_1) \,\square \,(H^{\dagger} \,H)$&
			$\mathcal{O}_{H^2 \chi_1^4}^{(3)}$&
			$(\chi^{\dagger}_1 \,\chi_1) \,(\chi^{\dagger}_1 \,\tau^I \,\chi_1) \,(H^{\dagger} \,\tau^I\,H)$\\
			
			$\mathcal{O}_{H \chi_1 \mathcal{D}}^{(1)}$&
			$(\chi^{\dagger}_1 \,\overleftrightarrow{\mathcal{D}}_{\mu}^I \,\chi_1) \,(H^{\dagger} \,\overleftrightarrow{\mathcal{D}}^{\mu I}\,H)$&
			$\mathcal{O}_{H^4 \chi_1^2}^{(1)}$&
			$(\chi^{\dagger}_1 \,\chi_1) \,(H^{\dagger} \,H)^2$\\
			
			$\mathcal{O}_{H \chi_1 \mathcal{D}}^{(2)}$&
			$(H^{\dagger} \,H) \,[(\mathcal{D}^{\mu} \,\chi_1)^{\dagger} \,(\mathcal{D}_{\mu} \,\chi_1)]$&
			$\mathcal{O}_{H^4 \chi_1^2}^{(2)}$&
			$(\chi^{\dagger}_1 \,\tau^I \,\chi_1) \,(H^{\dagger} \,\tau^I \,H) \,(H^{\dagger} \,H)$\\
			
			$\mathcal{O}_{H \chi_1 \mathcal{D}}^{(3)}$&
			$(\chi^{\dagger}_1 \,\chi_1) \,[(\mathcal{D}^{\mu} \,H)^{\dagger} \,(\mathcal{D}_{\mu} \,H)]$&			
			&
			\\
			
			$\mathcal{O}_{H \chi_1^{3} \mathcal{D}}$&
			$\textcolor{purple}{\epsilon_{\alpha\beta\gamma}\,\epsilon_{ij} \,(\chi_1^{\alpha i} \,\chi_1^{\beta j}) \,[(\mathcal{D}^{\mu} \,H^k)^{\dagger}\,(\mathcal{D}_{\mu} \,\chi_1^{\gamma k})]}$&
			&
			\\
			\hline
			\hline
			\multicolumn{4}{||c||}{$\Psi^2\Phi^2\mathcal{D}$}\\
			\hline
			
			$\mathcal{O}_{Q \chi_1 \mathcal{D}}^{(1)}$&
			$(N_f^2) \,(\overline{Q}_{p\alpha i} \,\gamma^{\mu} \,Q^{\alpha i}_{q}) \,(\chi^{\dagger}_1 \,i\overleftrightarrow{\mathcal{D}}_{\mu} \,\chi_1)$
			&$\mathcal{O}_{Q \chi_1 \mathcal{D}}^{(2)}$&
			$(N_f^2) \,(\overline{Q}_{p\alpha i} \,T^A \,\gamma^{\mu} \,Q^{\alpha i}_{q}) \,(\chi^{\dagger}_1 \,i\overleftrightarrow{\mathcal{D}}_{\mu}^A \,\chi_1)$
			\\
			
			$\mathcal{O}_{Q \chi_1 \mathcal{D}}^{(3)}$&
			$(N_f^2) \,(\overline{Q}_{p\alpha i} \,\tau^I \,\gamma^{\mu} \,Q^{\alpha i}_{q}) \,(\chi^{\dagger}_1 \,i\overleftrightarrow{\mathcal{D}}_{\mu}^I \,\chi_1)$
			&$\mathcal{O}_{Q \chi_1 \mathcal{D}}^{(4)}$&
			$(N_f^2) \,(\overline{Q}_{p\alpha i} \,T^A \,\tau^I \,\gamma^{\mu} \,Q^{\alpha i}_{i}) \,(\chi^{\dagger}_1 \,T^A \,i\overleftrightarrow{\mathcal{D}}_{\mu}^I \,\chi_1)$
			\\
			
			$\mathcal{O}_{L \chi_1 \mathcal{D}}^{(1)}$&
			$(N_f^2) \,(\overline{L}_{pi} \,\gamma^{\mu} \,L^{i}_{q}) \,(\chi^{\dagger}_1 \,i\overleftrightarrow{\mathcal{D}}_{\mu} \,\chi_1)$
			&$\mathcal{O}_{L \chi_1 \mathcal{D}}^{(2)}$&
			$(N_f^2) \,(\overline{L}_{pi} \,\tau^I \,\gamma^{\mu} \,L^{i}_{q}) \,(\chi^{\dagger}_1 \,i\overleftrightarrow{\mathcal{D}}_{\mu}^I \,\chi_1)$
			\\
			
			$\mathcal{O}_{u \chi_1 \mathcal{D}}^{(1)}$&
			$(N_f^2) \,(\overline{u}_{p\alpha} \,\gamma^{\mu} \,u^{\alpha}_{q}) \,(\chi^{\dagger}_1 \,i\overleftrightarrow{\mathcal{D}}_{\mu} \,\chi_1)$
			&$\mathcal{O}_{u \chi_1 \mathcal{D}}^{(2)}$&
			$(N_f^2) \,(\overline{u}_{p\alpha} \,T^A \,\gamma^{\mu} \,u^{\alpha}_{q}) \,(\chi^{\dagger}_1 \,i\overleftrightarrow{\mathcal{D}}_{\mu}^A \,\chi_1)$
			\\
			
			$\mathcal{O}_{d \chi_1 \mathcal{D}}^{(1)}$&
			$(N_f^2) \,(\overline{d}_{p\alpha} \,\gamma^{\mu} \,d^{\alpha}_{q}) \,(\chi^{\dagger}_1 \,i\overleftrightarrow{\mathcal{D}}_{\mu} \,\chi_1)$
			&$\mathcal{O}_{d \chi_1 \mathcal{D}}^{(2)}$&
			$(N_f^2) \,(\overline{d}_{p\alpha} \,T^A \,\gamma^{\mu} \,d^{\alpha}_{q}) \,(\chi^{\dagger}_1 \,i\overleftrightarrow{\mathcal{D}}_{\mu}^A \,\chi_1)$
			\\

			$\mathcal{O}_{e \chi_1 \mathcal{D}}$&
			$(N_f^2) \,(\overline{e}_{p} \,\gamma^{\mu} \,e_{q}) \,(\chi^{\dagger}_1 \,i\overleftrightarrow{\mathcal{D}}_{\mu} \,\chi_1)$&
			$\mathcal{O}_{d e H \chi_1 \mathcal{D}}$&
			$\color{purple}{(N_f^2) \,(\overline{d}_{p\alpha} \,\gamma^{\mu} \,e_{q}) \,(\tilde{H}^{\dagger}_i\,i\mathcal{D}_{\mu} \chi^i_{1,\alpha})}$
			\\
			
			$\mathcal{O}_{Q L H \chi_1 \mathcal{D}}^{(1)}$&
			$\textcolor{purple}{(N_f^2) \,(\overline{Q}_{p\alpha i} \,\gamma^{\mu} \,L^{i}_{q}) \,(\tilde{H}^{\dagger}_j \,i\mathcal{D}_{\mu}\chi^{\alpha j}_{1})}$
			&$\mathcal{O}_{Q L H \chi_1 \mathcal{D}}^{(2)}$&
			$\textcolor{purple}{(N_f^2) \,(\overline{Q}_{p\alpha i} \,\tau^I \,\gamma^{\mu} \,L^{i}_{q}) \,(\tilde{H}^{\dagger}_j \,i\mathcal{D}_{\mu}^I \chi^{\alpha j}_{1})}$
			\\
			\hline
			\hline
			\multicolumn{4}{||c||}{$\Phi^2X^2$}\\
			\hline
			
			$\mathcal{O}_{B \chi_1}$&
			$B_{\mu\nu} \,B^{\mu\nu} \,(\chi^{\dagger}_1 \,\chi_1)$&$\mathcal{O}_{\tilde{B} \chi_1}$&
			$\tilde{B}_{\mu\nu} \,B^{\mu\nu} \,(\chi^{\dagger}_1 \,\chi_1)$\\
			
			$\mathcal{O}_{G \chi_1}^{(1)}$&
			$G^{A}_{\mu\nu}\,G^{A\mu\nu}\,(\chi^{\dagger}_1 \,\chi_1)$&
			$\mathcal{O}_{G \chi_1}^{(2)}$&
			$d_{ABC} \,G^{A}_{\mu\nu} \,G^{B\mu\nu} \,(\chi^{\dagger}_1 \,T^C \,\chi_1)$\\
			
			$\mathcal{O}_{\tilde{G} \chi_1}^{(1)}$&
			$\tilde{G}^{A}_{\mu\nu}\,G^{A\mu\nu}\,(\chi^{\dagger}_1 \,\chi_1)$&$\mathcal{O}_{\tilde{G} \chi_1}^{(2)}$&
			$d_{ABC} \,\tilde{G}^{A}_{\mu\nu} \,G^{B\mu\nu} \,(\chi^{\dagger}_1 \,T^C \,\chi_1)$\\
			
			$\mathcal{O}_{W \chi_1}$&
			$W^{I}_{\mu\nu} \,W^{I\mu\nu} \,(\chi^{\dagger}_1 \,\chi_1)$&$\mathcal{O}_{\tilde{W} \chi_1}$&
			$\tilde{W}^{I}_{\mu\nu} \,W^{I\mu\nu} \,(\chi^{\dagger}_1 \,\chi_1)$
			\\
			
			$\mathcal{O}_{B G \chi_1}$&
			$B_{\mu\nu} \,G^{A\mu\nu} \,(\chi^{\dagger}_1 \,T^A \,\chi_1)$&
			$\mathcal{O}_{B \tilde{G} \chi_1}$&
			$B_{\mu\nu} \,\tilde{G}^{A\mu\nu} \,(\chi^{\dagger}_1 \,T^A \,\chi_1)$\\
			
			$\mathcal{O}_{B W \chi_1}$&
			$B_{\mu\nu} \,W^{I\mu\nu} \,(\chi^{\dagger}_1 \,\tau^I \,\chi_1)$&
			$\mathcal{O}_{B \tilde{W} \chi_1}$&
			$B_{\mu\nu} \,\tilde{W}^{I\mu\nu} \,(\chi^{\dagger}_1 \,\tau^I \,\chi_1)$
			\\
			
			$\mathcal{O}_{W G \chi_1}$&
			$W_{\mu\nu}^I \,G^{A\mu\nu} \,(\chi^{\dagger}_1 \,T^A \,\tau^I \,\chi_1)$&
			$\mathcal{O}_{W \tilde{G} \chi_1}$&
			$W_{\mu\nu}^I \,\tilde{G}^{A\mu\nu} \,(\chi^{\dagger}_1 \,T^A \,\tau^I \,\chi_1)$\\
			\hline
			\hline
			\multicolumn{4}{||c||}{$\Psi^2\Phi X$}\\
			\hline
			
			$\mathcal{O}_{B L d \chi_1}$&
			$\textcolor{purple}{(N_f^2) \,\epsilon_{ij} \,B_{\mu\nu} \,( \overline{d}_{p\alpha}\,\sigma^{\mu\nu} \,L^i_{q}) \,\chi_1^{\alpha j}}$&
			$\mathcal{O}_{G L d \chi_1}$&
			$\textcolor{purple}{(N_f^2) \,\epsilon_{ij} \,G^A_{\mu\nu} \,(\overline{d}_{p\alpha}\,\sigma^{\mu\nu} \,L^i_{q}) \,T^A \,\chi_1^{\alpha j}}$\\
			
			$\mathcal{O}_{W L d \chi_1}$&
			$\textcolor{purple}{(N_f^2) \,\epsilon_{ij} \,W^I_{\mu\nu} \,(\overline{d}_{p\alpha}\,\sigma^{\mu\nu} \,L^i_{q}) \,\tau^I \,\chi_1^{\alpha j}}$&
			&
			\\
			\hline
	\end{tabular}}
	\caption{SM extended by Lepto-Quark ($\chi_1$): Additional operators of  dimension 6. Here $i, j$ and $\alpha, \beta, \gamma$ are the $SU(2)$ and $SU(3)$ indices respectively. $T^A$ and $\tau^I$ are $SU(3)$ and $SU(2)$ generators respectively. $A,B,C=1,2,\cdots,8$ and $I=1,2,3$. $p, q=1,2,\cdots,N_f$ are the flavour indices. Operators in  {\it red} violate lepton and baryon numbers.}
	\label{table:Lepto-QuarkModel1-dim6-ops-1}
\end{table}
\clearpage
\begin{table}[h]
	\centering
	\renewcommand{\arraystretch}{1.9}
	{\scriptsize\begin{tabular}{||c|c||c|c||}
			
			\hline
			\hline
			\multicolumn{4}{||c||}{$\Psi^2\Phi^3$}\\
			\hline
			
			$\mathcal{O}_{Q d H \chi_1}^{(1)}$&
			$\textcolor{blue}{(N_f^2) \,(\overline{Q}_{p\alpha i} \,d^{\alpha}_{q}) \,H^i \,(\chi^{\dagger}_1 \,\chi_1)}$&
			$\mathcal{O}_{Q d H \chi_1}^{(2)}$&
			$\color{blue}{(N_f^2) \,(\overline{Q}_{p\alpha i} \,T^A \,d^{\alpha}_{q}) \,H^i \,(\chi^{\dagger}_1 \,T^A \,\chi_1)}$
			\\
			
			$\mathcal{O}_{Q d H \chi_1}^{(3)}$&
			$\color{blue}{(N_f^2) \,(\overline{Q}_{p\alpha i} \,d^{\alpha}_{q}) \,\tau^I \,H^i \,(\chi^{\dagger}_1 \,\tau^I \,\chi_1)}$
			&$\mathcal{O}_{Q d H \chi_1}^{(4)}$&
			$\color{blue}{(N_f^2) \,(\overline{Q}_{p\alpha i} \,T^A \,d^{\alpha}_{q}) \,\tau^I \,H^i \,(\chi^{\dagger}_1 \,T^A \,\tau^I \,\chi_1)}$
			\\
			
			$\mathcal{O}_{Q u H \chi_1}^{(1)}$&
			$\textcolor{blue}{(N_f^2) \,\epsilon_{ij} \,(\overline{Q}_{p\alpha i} \,u^{\alpha}_{q}) \,\tilde{H}_j \,(\chi^{\dagger}_1 \,\chi_1)}$&
			$\mathcal{O}_{Q u H \chi_1}^{(2)}$&
			$\textcolor{blue}{(N_f^2) \,\epsilon_{ij}  \,(\overline{Q}_{p\alpha i} \,T^A \,u^{\alpha}_{q}) \,\tilde{H}_j \,(\chi^{\dagger}_1 \,T^A \,\chi_1)}$
			\\
			
			$\mathcal{O}_{Q u H \chi_1}^{(3)}$&
			$\textcolor{blue}{(N_f^2) \,\epsilon_{ij}  \,(\overline{Q}_{p\alpha i} \,u^{\alpha}_{q}) \,\tau^I \,\tilde{H}_j \,(\chi^{\dagger}_1 \,\tau^I \,\chi_1)}$&
			$\mathcal{O}_{Q u H \chi_1}^{(4)}$&
			$\textcolor{blue}{(N_f^2) \,\epsilon_{ij}  \,(\overline{Q}_{p\alpha i} \,T^A \,u^{\alpha}_{q}) \,\tau^I \,\tilde{H}_j \,(\chi^{\dagger}_1 \,T^A \,\tau^I \,\chi_1)}$
			\\
			
			$\mathcal{O}_{L e H \chi_1}^{(1)}$&
			$\textcolor{blue}{(N_f^2) \,(\overline{L}_{pi} \,e_{q}) \,H_i \,(\chi_1^{\dagger} \,\chi_1)}$&
			$\mathcal{O}_{L e H \chi_1}^{(2)}$&
			$\textcolor{blue}{(N_f^2) \,(\overline{L}_{pi}\,e_{q}) \,\tau^I \,H_i \,(\chi_1^{\dagger} \,\tau^I \,\chi_1)}$
			\\
						
			$\mathcal{O}_{Q d \chi_1}$
			&$\color{purple}{(N_f^2) \,\epsilon_{\alpha\beta\gamma} \,\epsilon_{kl} \,(\overline{Q}_{p\delta i} \,d^{\alpha}_{q}) \,\chi_{1}^{\delta i} \,(\chi_1^{\beta k} \,\chi_1^{\gamma l})}$&
			$\mathcal{O}_{Q u \chi_1}$&
			$\textcolor{purple}{(N_f^2) \,\epsilon_{\alpha\beta\gamma} \,\epsilon_{ij} \,\epsilon_{kl} \,(\overline{u}_{p\delta}\, Q^{\alpha i}_{q}) \,\chi^{j\delta}_{1} \,(\chi_1^{\beta k} \,\chi_1^{\gamma l})}$
			\\
			
			$\mathcal{O}_{Q e H \chi_1}$&
			$\color{purple}{(N_f^2) \,\epsilon_{kl} \,(\overline{Q}_{p\alpha i} \,e_{q}) \,\chi^{k\alpha}_{1} \,(H_i \,H^l)}$&
			$\mathcal{O}_{L d H \chi_1^2}$&
			$\color{purple}{(N_f^2) \,\epsilon_{\alpha\beta\gamma} \,(\overline{L}_{pi} \,d^{\alpha}_{q}) \,\tilde{H}_j \,(\chi_1^{\beta i} \,\chi_1^{\gamma j})}$
			\\
			
			$\mathcal{O}_{L d H^2 \chi_1}^{(1)}$&
			$\color{purple}{(N_f^2) \,\epsilon_{ij} \,(\overline{d}_{p\alpha} \,L_{q}^i ) \,\chi_1^{\alpha j} \,(H^{\dagger} \,H)}$&$\mathcal{O}_{L d H^2 \chi_1}^{(2)}$&
			$\color{purple}{(N_f^2) \,\epsilon_{ij} \,(\overline{d}_{p\alpha} \,L_{q}^i ) \,\tau^I \,\chi_1^{\alpha j} \,(H^{\dagger} \,\tau^I \,H)}$
			\\
			
			$\mathcal{O}_{L d \chi_1}^{(1)}$&
			$\color{purple}{(N_f^2) \,\epsilon_{ij} \,(\overline{d}_{p\alpha} \,L_{q}^i ) \,\chi_1^{\alpha j} \,(\chi^{\dagger}_1 \,\chi_1)}$&
			$\mathcal{O}_{L d \chi_1}^{(2)}$&
			$\color{purple}{(N_f^2) \,\epsilon_{ij} \,(\overline{d}_{p\alpha} \,L_{q}^i ) \,\tau^I \,\chi_1^{\alpha j} \,(\chi^{\dagger}_1 \,\tau^I \,\chi_1)}$
			\\
			
			$\mathcal{O}_{L u H \chi_1}$&
			$\color{purple}{(N_f^2) \,\epsilon_{ij} \,\epsilon_{kl} \,(\overline{u}_{p\alpha}\,L^i_{q}) \,\chi_1^{\alpha k} \,(H^j \,H^l)}$&
			&
			\\
			\hline
	\end{tabular}}
	\caption{Table \ref{table:Lepto-QuarkModel1-dim6-ops-1} continued.}
	\label{table:Lepto-QuarkModel1-dim6-ops-1-contd}
\end{table}

\subsubsection*{\underline{SM + Lepto-Quark ($\varphi_1$)}}

We have considered another example of a Lepto-Quark that has similar gauge quantum numbers as the up-type $SU(2)$ singlet quark within SM, see Table~\ref{table:Lepto-Quark-quantum-no}. Here, we have computed the effective operators to grab the features of  full theories containing $\varphi_1$ for reasons similar to those discussed in the previous section. 
 The  operators of dimensions 5 and 6 containing $\varphi_{1}$ have been collected in Tables~\ref{table:Lepto-QuarkModel2-dim5-ops-1} and~\ref{table:Lepto-QuarkModel2-dim6-ops-1} respectively. The operators with distinct hermitian conjugates have been coloured blue.

\begin{table}[h]
	\centering
	\renewcommand{\arraystretch}{1.9}
	{\scriptsize\begin{tabular}{||c|c||c|c||}
			\hline
			\hline
			\multicolumn{4}{||c||}{$\Psi^2\Phi^2$}\\
			\hline
			$\tilde{\mathcal{O}}_{u \varphi_1}$&
			$\color{purple}{\frac{1}{2}(N_f^2+N_f) \,((u^{\alpha}_{p})^T \,C \,u^{\beta}_{q}) \,(\varphi^{\dagger}_{1,\alpha} \,\varphi^{\dagger}_{1,\beta})}$
			&$\tilde{\mathcal{O}}_{L d H \varphi_1}$&
			$\color{purple}{(N_f^2) \,(\overline{d}_{p\alpha} \,L_{qi}) \,(\tilde{H}_i \,\varphi_1^{\alpha})}$
			\\
			
			$\tilde{\mathcal{O}}_{Q e H \varphi_1}$&
			$\color{purple}{(N_f^2) \,(\overline{Q}_{p\alpha i} \,e_{q}) \,(H^i \,\varphi^{\alpha}_1)}$&
			$\tilde{\mathcal{O}}_{L u H \varphi_1}$&
			$\color{purple}{(N_f^2) \,\epsilon_{ij} \,(\overline{u}_{p\alpha} \,L^i_{q}) \,(H^j \,\varphi^{\alpha}_1)}$
			\\
			\hline
	\end{tabular}}
	\caption{SM extended by Lepto-Quark ($\varphi_{1}$): Additional operators of  dimension 5. Here $i, j$ and $\alpha, \beta$ are the $SU(2)$ and $SU(3)$ indices respectively. $p, q=1,2,\cdots,N_f$ are the flavour indices. All the  operators of this class violate lepton number.}
	\label{table:Lepto-QuarkModel2-dim5-ops-1}
\end{table}

\noindent\textbf{Features of the additional operators:}
\begin{itemize}
	\item At dimension 5, $\varphi_1$ offers only baryon and lepton number violating operators belonging to 
	$\Psi^2\Phi^2$ class, unlike the previous ($\chi_{1}$) case. 
	
	\item Although there is mixing between $B_{\mu\nu}$ and $G^{A}_{\mu\nu}$ within the $\Phi^2X^2$ class, there is no mixing between $W^{I}_{\mu\nu}$ and $G^{A}_{\mu\nu}$ on account of $\varphi_1$ being an $SU(2)_L$ singlet.

	\item For the same reason, $\varphi_1$ allows fewer possible ways to construct invariant operators than the $SU(2)_L$ doublet $\chi_1$. This is quite evident from the number of operators listed in Table~\ref{table:Lepto-QuarkModel2-dim6-ops-1}.
\end{itemize}

\begin{table}[h]
	\centering
	\renewcommand{\arraystretch}{1.9}
	{\scriptsize\begin{tabular}{||c|c||c|c||}
			\hline
			\hline
			\multicolumn{2}{||c||}{$\Phi^4\mathcal{D}^2$}&
			\multicolumn{2}{c||}{$\Phi^6$}\\
			\hline
			
			$\mathcal{O}_{\varphi_1 \mathcal{D}}^{(1)}$&
			$(\varphi^{\dagger}_1 \,\varphi_1) \,\square \,(\varphi^{\dagger}_1 \,\varphi_1)$&$\mathcal{O}_{\varphi_1}$&
			$(\varphi_1^{\dagger} \,\varphi_1)^3$\\
			
			$\mathcal{O}_{\varphi_1 \mathcal{D}}^{(2)}$&
			$(\varphi^{\dagger}_1 \,\overleftrightarrow{\mathcal{D}}_{\mu} \,\varphi_1)\,(\varphi^{\dagger}_1 \,\overleftrightarrow{\mathcal{D}}^{\mu} \,\varphi_1)$&
			$\mathcal{O}_{H^2 \varphi_1^4}$&
			$(\varphi^{\dagger}_1 \,\varphi_1)^2 \,(H^{\dagger} \,H)$
			\\
			
			$\mathcal{O}_{ H \varphi_1 \mathcal{D}}^{(1)}$&
			$(\varphi^{\dagger}_1 \,\varphi_1) \,\left[(\mathcal{D}^{\mu}\,H)^{\dagger}(\mathcal{D}_{\mu}\,H)\right]$&
			$\mathcal{O}_{H^4 \varphi_1^2}$&
			$ (\varphi^{\dagger}_1 \,\varphi_1) \,(H^{\dagger} \,H)^2$
			\\
			
			$\mathcal{O}_{H \varphi_1 \mathcal{D}}^{(2)}$&
			$(H^{\dagger}\,H)\,\left[(\mathcal{D}^{\mu}\,\varphi_1)^{\dagger}(\mathcal{D}_{\mu}\,\varphi_1)\right]$&
			&
			\\
			\hline
			\hline
			\multicolumn{4}{||c||}{$\Psi^2\Phi^2\mathcal{D}$}\\
			\hline
			
			$\mathcal{O}_{Q \varphi_1 \mathcal{D}}^{(1)}$
			&$(N_f^2) \,(\overline{Q}_{p\alpha i} \,\gamma^{\mu} \,Q^{\alpha i}_{q}) \,(\varphi^{\dagger}_1 \,i\overleftrightarrow{\mathcal{D}}_{\mu} \,\varphi_1)$&
			$\mathcal{O}_{Q \varphi_1 \mathcal{D}}^{(2)}$&
			$(N_f^2) \,(\overline{Q}_{p\alpha i} \,T^A \,\gamma^{\mu} \,Q^{\alpha i}_{q}) \,(\varphi^{\dagger}_1 \,i\overleftrightarrow{\mathcal{D}}_{\mu}^A \,\varphi_1)$
			\\
			
			$\mathcal{O}_{L \varphi_1 \mathcal{D}}$&
			$(N_f^2) \,(\overline{L}_{pi} \,\gamma^{\mu} \,L^{i}_{q}) \,(\varphi^{\dagger}_1 \,i\overleftrightarrow{\mathcal{D}}_{\mu} \,\varphi_1)$
			&$\mathcal{O}_{u \varphi_1 \mathcal{D}}^{(1)}$&
			$(N_f^2) \,(\overline{u}_{p\alpha} \,\gamma^{\mu} \,u^{\alpha}_{q}) \,(\varphi^{\dagger}_1 \,i\overleftrightarrow{\mathcal{D}}_{\mu} \,\varphi_1)$\\
			
			$\mathcal{O}_{u \varphi_1 \mathcal{D}}^{(2)}$&
			$(N_f^2) \,(\overline{u}_{p\alpha} \,T^A \,\gamma^{\mu} \,u^{\alpha}_{q}) \,(\varphi^{\dagger}_1 \,i\overleftrightarrow{\mathcal{D}}_{\mu}^A \,\varphi_1)$&
			$\mathcal{O}_{d \varphi_1 \mathcal{D}}^{(1)}$&
			$(N_f^2) \,(\overline{d}_{p\alpha} \,\gamma^{\mu} \,d^{\alpha}_{q}) \,(\varphi^{\dagger}_1 \,i\overleftrightarrow{\mathcal{D}}_{\mu} \,\varphi_1)$
			\\
			
			$\mathcal{O}_{d \varphi_1 \mathcal{D}}^{(2)}$&
			$(N_f^2) \,(\overline{d}_{p\alpha} \,T^A \,\gamma^{\mu} \,d^{\alpha}_{q}) \,(\varphi^{\dagger}_1 \,i\overleftrightarrow{\mathcal{D}}_{\mu}^A \,\varphi_1)$
			&$\mathcal{O}_{e \varphi_1 \mathcal{D}}$&
			$(N_f^2) \,(\overline{e}_{p} \,\gamma^{\mu} \,e_{q}) \,(\varphi^{\dagger}_1 \,i\overleftrightarrow{\mathcal{D}}_{\mu} \,\varphi_1)$
			\\
			
			$\mathcal{O}_{Q d H \varphi_1 \mathcal{D}}$&
			$\textcolor{purple}{(N_f^2) \,\epsilon_{\alpha\beta\gamma} \,((Q^{\alpha i}_{p})^T\, C \,\gamma^{\mu} \,d^{\beta}_{q}) \,(H^{\dagger}_i \,i\mathcal{D}_{\mu} \varphi_1^{\gamma})}$&
			$\mathcal{O}_{L u H \varphi_1 \mathcal{D}}$
			&$\color{purple}{(N_f^2) \,\epsilon_{ij} \,((L^i_{p})^T\, C \,\gamma^{\mu} \,u^{\alpha}_{q}) \,(\varphi_{1,\alpha}^{\dagger}\,i\mathcal{D}_{\mu}H^j)}$
			\\
			\hline
			\hline
			\multicolumn{4}{||c||}{$\Phi^2X^2$}\\
			\hline
			
			$\mathcal{O}_{B \varphi_1}$&
			$B_{\mu\nu} \,B^{\mu\nu} \,(\varphi^{\dagger}_1 \,\varphi_1)$&
			$\mathcal{O}_{\tilde{B} \varphi_1}$&
			$\tilde{B}_{\mu\nu} \,B^{\mu\nu} \,(\varphi^{\dagger}_1 \,\varphi_1)$\\
			
			$\mathcal{O}_{G \varphi_1}^{(1)}$&
			$G^{A}_{\mu\nu}\,G^{A\mu\nu}\,(\varphi^{\dagger}_1 \,\varphi_1)$&
			$\mathcal{O}_{G \varphi_1}^{(2)}$&
			$d_{ABC} \,G^{A}_{\mu\nu} \,G^{B\mu\nu} \,(\varphi^{\dagger}_1 \,T^C \,\varphi_1) $\\
			
			$\mathcal{O}_{\tilde{G} \varphi_1}^{(1)}$&
			$\tilde{G}^{A}_{\mu\nu}\,G^{A\mu\nu}\,(\varphi^{\dagger}_1 \,\varphi_1)$&
			$\mathcal{O}_{\tilde{G} \varphi_1}^{(2)}$&
			$d_{ABC} \,\tilde{G}^{A}_{\mu\nu} \,G^{B\mu\nu} \,(\varphi^{\dagger}_1 \,T^C \,\varphi_1) $\\

			$\mathcal{O}_{W \varphi_1}$&
			
			$W^{I}_{\mu\nu} \,W^{I\mu\nu} \,(\varphi^{\dagger}_1 \,\varphi_1)$&
			
			$\mathcal{O}_{\tilde{W} \varphi_1}$&
			$\tilde{W}^{I}_{\mu\nu} \,W^{I\mu\nu} \,(\varphi^{\dagger}_1 \,\varphi_1)$\\

			$\mathcal{O}_{B G \varphi_1}$&
			$B_{\mu\nu} \,G^{A\mu\nu} \,(\varphi^{\dagger}_1 \,T^A \,\varphi_1)$&
			$\mathcal{O}_{B \tilde{G} \varphi_1}$&
			$B_{\mu\nu} \,\tilde{G}^{A\mu\nu} \,(\varphi^{\dagger}_1 \,T^A \,\varphi_1)$\\			
			
			\hline
			\hline
			\multicolumn{4}{||c||}{$\Psi^2\Phi X$}\\
			\hline

			$\mathcal{O}_{B d \varphi_1}$&
			$\textcolor{purple}{\frac{1}{2}(N_f^2+N_f) \,\epsilon_{\alpha\beta\gamma} \,B_{\mu\nu} \,((d^{\alpha}_{p})^T \,C \,\sigma^{\mu\nu} \,d^{\beta}_{q}) \,\varphi_1^{\gamma}}$&
			$\mathcal{O}_{G d \varphi_1}$&
			$\color{purple}{(N_f^2) \,\epsilon_{\alpha\beta\gamma} \,G^A_{\mu\nu} \,((d^{\alpha}_{p})^T \,C \,\sigma^{\mu\nu} \,d^{\beta}_{q}) \,T^A \,\varphi_1^{\gamma}}$\\	
			\hline
			\hline
			\multicolumn{4}{||c||}{$\Psi^2\Phi^3$}\\
			\hline
			
			$\mathcal{O}_{Q d H \varphi_1}^{(1)}$&
			$\textcolor{blue}{(N_f^2) \,(\overline{Q}_{p\alpha i} \,d^{\alpha}_{q}) \,H^i \,(\varphi^{\dagger}_1 \,\varphi_1)}$&
			$\mathcal{O}_{Q d H \varphi_1}^{(2)}$&
			$\color{blue}{(N_f^2) \,(\overline{Q}_{p\alpha i} \,T^A \,d^{\alpha}_{q}) \,H^i \,(\varphi^{\dagger}_1 \,T^A \,\varphi_1)}$
			\\
			
			$\mathcal{O}_{Q u H \varphi_1}^{(1)}$&
			$\textcolor{blue}{(N_f^2) \,\epsilon_{ij} \,(\overline{Q}_{p\alpha i} \,u^{\alpha}_{q}) \,\tilde{H}^j \,(\varphi^{\dagger}_1 \varphi_1)}$&
			$\mathcal{O}_{Q u H \varphi_1}^{(2)}$&
			$\textcolor{blue}{(N_f^2) \,\epsilon_{ij} \,(\overline{Q}_{p\alpha i} \,T^A \,u^{\alpha}_{q}) \,\tilde{H}^j \,(\varphi^{\dagger}_1 \,T^A \,\varphi_1)}$
			\\
			
			$\mathcal{O}_{L e H \varphi_1}$
			&$\color{blue}{(N_f^2) \,(\overline{L}_{pi} \,e_{q}) \,H^i \,(\varphi^{\dagger}_1 \,\varphi_1)}$&
			$\mathcal{O}_{Q L H \varphi_1}$&
			$\textcolor{blue}{(N_f^2) \,\epsilon_{ij} \,\epsilon_{kl} \,((Q^{p\alpha i})^T\, C \,L^k_{q}) \,(H^j \,H^l) \,\varphi^{\dagger}_{1,\alpha}}$\\
			
			$\mathcal{O}_{d \varphi_1}$&
			$\boxed{\color{purple}{\frac{1}{2}(N_f^2-N_f) \,\epsilon_{\alpha\beta\gamma} \,((d^{\alpha}_{p})^T \,C \,d^{\beta}_{q}) \,\varphi^{\gamma}_1 \,(\varphi^{\dagger}_1 \,\varphi_1)}}$&
			$\mathcal{O}_{d H \varphi_1}$&
			$\boxed{\textcolor{purple}{\frac{1}{2}(N_f^2-N_f) \,\epsilon_{\alpha\beta\gamma} \,((d^{\alpha}_{p})^T \,C \,d^{\beta}_{q}) \,\varphi^{\gamma}_1 \,(H^{\dagger} \,H)}}$\\
			
			$\mathcal{O}_{Q H \varphi_1}$&
			$\boxed{\color{purple}{\frac{1}{2}(N_f^2-N_f) \,\epsilon_{\alpha\beta\gamma} \,((Q^{\alpha i}_{p})^T \,C \,Q^{\beta j}_{q}) \,(\tilde{H}_i \,\tilde{H}_j)  \,\varphi_1^{\gamma}}}$&
			&
			\\
			
			\hline
	\end{tabular}}
	\caption{SM extended by Lepto-Quark ($\varphi_1$): Additional operators of  dimension 6. Here $i, j$ and $\alpha,\beta,\gamma$ are the $SU(2)$ and $SU(3)$ indices respectively. $T^A$ are the $SU(3)$ generators. $A,B,C=1,2,\cdots,8$ and $I=1,2,3$. $p, q=1,2,\cdots,N_f$ are the flavour indices. Operators in {\it red} violate lepton and baryon numbers.}
	\label{table:Lepto-QuarkModel2-dim6-ops-1}
\end{table}
\clearpage
\subsection{Standard Model extended by abelian gauge symmetries}

It is believed that at a very high scale there is a unified gauge theory (GUT) and from there the SM is originated through a cascade of symmetry breaking. As the rank of the viable unified gauge groups are larger than that of the SM, in the process of symmetry breaking the desert region between the electroweak and unified scales may be filled up with multiple intermediate symmetries. Most of the consistent GUT breaking chains lead to the presence of multiple abelian ($U(1)$) gauge symmetries around the electroweak scale, i.e., the SM \cite{Chakrabortty:2010az, Chakrabortty:2009xm, Chakrabortty:2017mgi, Chakrabortty:2019fov}.  In addition, there are many phenomenological attempts to extend the SM using multiple additional abelian gauge symmetries, e.g., $ U(1)_B \otimes U(1)_L $ \cite{FileviezPerez:2010gw}, $ U(1)_{B-L} \otimes U(1)_{L_\mu -L_\tau} $ ($ L_\alpha $ denotes lepton flavour number) \cite{Heeck_2011,Heeck:2018nzc,Kahn:2018cqs}. All such scenarios are expected to be effective ones. Thus we need to compute the complete set of effective operators to encapsulate the footprints of the heavier modes which  are already integrated out. 
 Instead of considering all such possible scenarios, we have worked out a specific example model, see Table~\ref{table:abelian-extension-quantum-no}. The other possible cases can be addressed in a similar spirit and using the same methodology. We have listed all the operators of mass dimension 6 in Table \ref{table:abelian-gauge-extension}. The operators with distinct hermitian conjugates have been coloured blue.\\

\begin{table}[h!]
	\centering
	\renewcommand{\arraystretch}{1.7}
	{\scriptsize\begin{tabular}{|c|c|c|c|c|c|c|c|c|}
			\hline
			\textbf{Field} & \textbf{$SU(3)_C$} & \textbf{$SU(2)_{L}$}&
			\textbf{$U(1)_{Y}$}&
			\textbf{$U(1)^{\prime}$}&
			\textbf{$U(1)^{\prime\prime}$}&
			\textbf{Baryon No.}&
			\textbf{Lepton No.}&
			\textbf{Spin}\\
			\hline
			$H$    		& 1 & 2 & 1/2 & 0 & 0  &  0  & 0 & 0 \\
			$Q^p_L$     & 3 & 2 & 1/6 & 0 & 0  & 1/3 & 0 &1/2\\
			$u^p_R$     & 3 & 1 & 2/3 & 0 & 0  & 1/3 & 0 &1/2\\
			$d^p_R$     & 3 & 1 &-1/3 & 0 & 0  & 1/3 & 0 &1/2\\
			$L^p_L$     & 1 & 2 &-1/2 & 0 & 0  &  0  & -1&1/2\\
			$e^p_R$     & 1 & 1 & -1  & 0 & 0  &  0  & -1 & 1/2\\
			\hline
			$G^A_{\mu}$ & 8 & 1 & 0 & 0 & 0 & 0 & 0 & 1\\
			$W^I_{\mu}$ & 1 & 3 & 0 & 0 & 0 & 0 & 0 & 1\\
			$B_{\mu}$   & 1 & 1 & 0 & 0 & 0 & 0 & 0 &1\\
			$X_{\mu}$   & 1 & 1 & 0 & 0 & 0 & 0 & 0 &1\\
			$Y_{\mu}$   & 1 & 1 & 0 & 0 & 0 & 0 & 0 &1\\
			\hline
	\end{tabular}}
	\caption{\small SM extended by two abelian gauge symmetries: Quantum numbers of the fields.} 
	\label{table:abelian-extension-quantum-no}
\end{table}

\noindent\textbf{Features of the additional operators:}
\begin{itemize}
	\item There is  no dimension 5 operator unlike the previous cases.
	
	\item Abelian mixing among $ B_{\mu\nu},\, X_{\mu\nu} $ and $ Y_{\mu\nu} $ has been noted in $\Phi^2X^2 $ and $ X^3 $ classes. These operators can generate kinetic mixing even if it is switched off at tree-level.
	
	\item There are no additional baryon and (or) lepton number violating operators as expected.
\end{itemize}

\clearpage

\begin{table}[h]
	\centering
	\renewcommand{\arraystretch}{1.9}
	{\scriptsize\begin{tabular}{||c|c||c|c||}
			\hline
			\hline
			\multicolumn{4}{||c||}{$\Phi^2X^2$}\\
			\hline

			$\mathcal{O}_{W X H}$&
			$ W_{\mu\nu}^I \,X^{\mu\nu} \,(H^\dagger\,\tau^I \,H) $ &
			$\mathcal{O}_{\tilde{W} X H}$&
			$\tilde{W}_{\mu\nu}^I \,X^{\mu\nu} \,(H^\dagger \tau^I H)$ \\

			$\mathcal{O}_{B X H}$&
			$ B_{\mu\nu}\,X^{\mu\nu} \,(H^\dagger H)$ &
			$\mathcal{O}_{\tilde{B} X H}$&
			$ \tilde{B}_{\mu\nu}\,X^{\mu\nu} \,(H^\dagger H)$\\

			$\mathcal{O}_{X H}$&
			$  X_{\mu\nu} \,X^{\mu\nu} \,(H^\dagger H) $&
			$\mathcal{O}_{\tilde{X} X H}$&
			$ \tilde{X}_{\mu\nu}\,X^{\mu\nu} \,(H^\dagger H)$\\
			
			$\mathcal{O}_{W Y H}$&
			$W_{\mu\nu}^I \,Y^{\mu\nu} \,(H^\dagger \tau^I H)$ &
			$\mathcal{O}_{\tilde{W} Y H}$&
			$\tilde{W}_{\mu\nu}^I \,Y^{\mu\nu} \,(H^\dagger \tau^I H)$ \\

			$\mathcal{O}_{B Y H}$&
			$ B_{\mu\nu} \,Y^{\mu\nu} \,(H^\dagger H) $&
			$\mathcal{O}_{\tilde{B} Y H}$&
			$ \tilde{B}_{\mu\nu}\,Y^{\mu\nu} \,(H^\dagger H)$\\

			$\mathcal{O}_{Y H}$&
			$ Y_{\mu\nu}\,Y^{\mu\nu} \,(H^\dagger H)$ &
			$\mathcal{O}_{\tilde{Y} Y H}$&
			$ \tilde{Y}_{\mu\nu}\,Y^{\mu\nu} \,(H^\dagger H)$\\

			$\mathcal{O}_{X Y H}$&
			$ X_{\mu\nu}\,Y^{\mu\nu} \,(H^\dagger H)$&
			$\mathcal{O}_{\tilde{X} Y H}$&
			$ \tilde{X}_{\mu\nu}\,Y^{\mu\nu} \,(H^\dagger H)$\\

			\hline
			\hline
			\multicolumn{4}{||c||}{$\Psi^2 \Phi X$}\\
			\hline
			
			$\mathcal{O}_{X Q d H}$&
			$\color{blue}{(N_f^2)\,X_{\mu\nu}\,(\overline{d}_{p\alpha}\,\sigma^{\mu\nu} \,Q^{\alpha i}_{q}) \tilde{H}_i}$&
			$\mathcal{O}_{Y Q d H}$&
			$ \color{blue}{(N_f^2)\,Y_{\mu\nu} \,(\overline{d}_{p\alpha} \,\sigma^{\mu\nu} \,Q^{\alpha i}_{q}) \tilde{H}_i}$\\
			
			$\mathcal{O}_{X L e H}$&
			$ \color{blue}{(N_f^2)\,X_{\mu\nu} \,(\overline{e}_{p} \,\sigma^{\mu\nu} \,L^i_{q}) \tilde{H}_i}$&
			$\mathcal{O}_{Y L e H}$&
			$ \color{blue}{(N_f^2)\,Y_{\mu\nu} \,(\overline{e}_{p} \,\sigma^{\mu\nu} \,L^i_{q}) \tilde{H}_i}$\\
			
			$\mathcal{O}_{X Q u H}$&
			$ \color{blue}{(N_f^2)\,\epsilon_{ij} \,X_{\mu\nu}\,(\overline{u}_{p\alpha}\,\sigma^{\mu\nu}\,Q^{\alpha i}_{q}) H^j }$&
			$\mathcal{O}_{Y Q u H}$&
			$ \color{blue}{(N_f^2)\,\epsilon_{ij} \,Y_{\mu\nu}\,(\overline{u}_{p\alpha}\,\sigma^{\mu\nu}\,Q^{\alpha i}_{q}) \,H^j}$\\
			
			\hline
			\hline
			\multicolumn{4}{||c||}{$X^3$}\\
			\hline
			$\mathcal{O}_{BXY}$&
			$B_\mu^\rho X_\rho^\nu Y_\nu^\mu$ &
		 	$\mathcal{O}_{\tilde{B}XY}$&
		 	$\tilde{B}_\mu^\rho X_\rho^\nu Y_\nu^\mu$ \\

		\hline
	\end{tabular}}
\caption{SM extended by two abelian gauge groups: Additional operators of  dimension 6. Here $i, j$ and $\alpha$ are the $SU(2)$ and $SU(3)$ indices respectively. $\tau^I$ is the $SU(2)$ generator, $I=1,2,3$. $p,q=1,2,\cdots,N_f$ are the flavour indices.}
\label{table:abelian-gauge-extension}
\end{table}

%\clearpage

\subsection{Flavour ($N_f$) dependence and $B$, $L$, $CP$  violating operators}\label{subsec:flavour-dep-count}

In the SM, fermions appear in three flavours:
{\small\begin{eqnarray}
L_1 \equiv \begin{pmatrix}
\nu_{eL}\\ e_L
\end{pmatrix}, \,\, L_2 \equiv \begin{pmatrix}
\nu_{\mu L}\\ \mu_L
\end{pmatrix}, \,\, L_3 \equiv \begin{pmatrix}
\nu_{\tau L}\\ \tau_L
\end{pmatrix}, Q_1 \equiv \begin{pmatrix}
u_L\\ d_L
\end{pmatrix}, \,\, Q_2 \equiv \begin{pmatrix}
c_L\\ s_L
\end{pmatrix}, \,\, Q_3 \equiv \begin{pmatrix}
t_L\\ b_L
\end{pmatrix}, 
\end{eqnarray}}
and analogously for the right chiral singlets. In the unbroken SM, all flavours are in the same footing. The distinction is visible only after the symmetry breaking, once they acquire different masses. At the tree-level, there is a clear absence of lepton flavour violation while the same is induced in the quark sector through CKM mixing. But the insertion of effective operators certainly alters this observation. Here, we have presented our results in terms of $N_f$ flavour fermions.
The operators corresponding to different example BSMEFT scenarios are  classified into the following categories based on their fermion contents:
\begin{itemize}
	\item \underline{No fermion}: At dimensions 5 and 6 we have the $\Phi^5$ and  $\Phi^6$, $\Phi^4\mathcal{D}^2$, $X^3$, $\Phi^2X^2$ classes which do not contain any fermion fields. Thus the  number of operators belonging to these classes are independent of $N_f$ as expected.
	
	\item \underline{Bi-linear fermions}: We have found $\Psi^2 X$, $\Psi^2\Phi^2$  and $\Psi^2\Phi^2\mathcal{D}$, $\Psi^2\Phi X$, $\Psi^2\Phi^3$ classes at dimensions 5 and 6 respectively. The  number of operators belonging to these classes are of the following forms: 
	$\frac{1}{2}N_f(N_f-1), \frac{1}{2}N_f(N_f+1)$, and $N_f^2$ which correspond to overall anti-symmetric, symmetric and a combination of the two  respectively. The similar tensorial structures under internal and space-time symmetries also play pivotal roles to determine flavour ($N_f$) dependent coefficients.

	\item \underline{Quartic fermions}: There exists only  $\Psi^4$ class at dimension 6 which contains four fermion fields. Here, the number of operators is a function of the product of any two elements belonging to the set $\{\frac{1}{2}N_f(N_f-1),\frac{1}{2}N_f(N_f+1),N_f^2\}$.  But depending on the symmetry structure and fermion representation we may find more intricate combinations and those need to be analysed carefully, see  Ref.~\cite{Banerjee:2020bym} for a detailed discussion.
\end{itemize}
We have summarized the number of operators for each class for all the scenarios. We have listed  the number of additional dimension 5 operators  in Table \ref{table:number-of-ops-dim5}. The same information for dimension 6 operators has been collected in Tables \ref{table:number-of-ops-dim6} and \ref{table:number-of-ops-dim6-2}. We have also highlighted the number of $B$, $L$ and $CP$ violating operators for clarity.

\begin{table}[h]
	\centering
	\renewcommand{\arraystretch}{1.8}
	{\scriptsize\begin{tabular}{|c|c|c|c|}
			\hline
			\multirow{2}{*}{\textbf{BSM Field}}&
			\multirow{2}{*}{\textbf{Operator Class}}&
			\multicolumn{2}{c|}{\textbf{Number of Operators as $f(N_f)$}}\\
			\cline{3-4}
			
			&
			&
			Total Number (CPV Bosonic Ops.)&
			$B, \,L$ Violating Ops.\\
			\hline
			
			\multirow{6}{*}{$\delta^{+}$}&
			$\Phi^6$&
			3&
			0\\
			
			&
			$\Phi^4\mathcal{D}^2$&
			3&
			0\\
			
			&
			$\Phi^2X^2$&
			6 (3)&
			0\\
			
			&
			$\Psi^2\Phi^2\mathcal{D}$&
			$7N_f^2$&
			$2N_f^2$\\
			
			&
			$\Psi^2\Phi^3$&
			$9N_f^2-N_f$&
			$3N_f^2-N_f$\\
			
			&
			$\Psi^2\Phi X$&
			$2N_f^2$&
			$2N_f^2$\\
			\hline
			
			\multirow{6}{*}{$\rho^{++}$}&
			$\Phi^6$&
			3&
			0\\
			
			&
			$\Phi^4\mathcal{D}^2$&
			3&
			0\\
			
			&
			$\Phi^2X^2$&
			6 (3)&
			0\\
			
			&
			$\Psi^2\Phi^2\mathcal{D}$&
			$7N_f^2$&
			$2N_f^2$\\
			
			&
			$\Psi^2\Phi^3$&
			$9N_f^2+3N_f$&
			$3N_f^2+3N_f$\\
			
			&
			$\Psi^2\Phi X$&
			$N_f^2-N_f$&
			$N_f^2-N_f$\\
			\hline
			
			\multirow{6}{*}{$\Delta$}&
			$\Phi^6$&
			10&
			0\\
			
			&
			$\Phi^4\mathcal{D}^2$&
			7&
			0\\
			
			&
			$\Phi^2X^2$&
			10 (5)&
			0\\
			
			&
			$\Psi^2\Phi^2\mathcal{D}$&
			$9N_f^2$&
			$2N_f^2$\\
			
			&
			$\Psi^2\Phi^3$&
			$18N_f^2+4N_f$&
			$6N_f^2+4N_f$\\
			
			&
			$\Psi^2\Phi X$&
			$3N_f^2-N_f$&
			$3N_f^2-N_f$\\
			\hline

	\end{tabular}}
	\caption{Number of additional operators of different classes at dimension 6 with $N_f$ fermion flavours for each BSM model. The numbers in parentheses denote the counting for CP violating purely bosonic operators.}
	\label{table:number-of-ops-dim6}
\end{table}
\clearpage
\begin{table}[h]
	\centering
	\renewcommand{\arraystretch}{1.8}
	{\scriptsize\begin{tabular}{|c|c|c|c|}
			
			\hline
			\multirow{2}{*}{\textbf{BSM Field}}&
			\multirow{2}{*}{\textbf{Operator Class}}&
			\multicolumn{2}{c|}{\textbf{Number of Operators as $f(N_f)$}}\\
			\cline{3-4}
			
			&
			&
			Total Number (CPV Bosonic Ops.)&
			$B, \,L$ Violating Ops.\\
			\hline
			
			\multirow{4}{*}{$\Sigma$}&
			$\Psi^2\Phi^2\mathcal{D}$&
			$2N_f^2$&
			0\\
			
			&
			$\Psi^2\Phi^3$&
			$4N_f^2$&
			$4N_f^2$\\
			
			&
			$\Psi^2\Phi X$&
			$6N_f^2$&
			$6N_f^2$\\

			&
			$\Psi^4$&
			$\frac{69}{4}N_f^4-\frac{1}{2}N_f^3+\frac{9}{4}N_f^2$&
			$9N_f^4-N_f^3$\\
			\hline
			
			\multirow{4}{*}{$ V_{L,R};\, E_{L,R};\, N_{L,R}$}&
			$\Psi^2\Phi^3$&
			$20N_f^2$&
			$ 6N_f^2 $\\
			
			&
			$\Psi^2\Phi X$&
			$44N_f^2$&
			$ 12N_f^2 $\\
			
			&
			$\Psi^2\Phi^2\mathcal{D}$&
			$32N_f^2$&
			$ 12N_f^2 $\\
			
			&
			$\Psi^4$&
			$\frac{676}{3}N_f^4+6N_f^3+\frac{17}{3}N_f^2$ &
			$ \frac{805}{12}N_f^4+\frac{9}{2}N_f^3-\frac{7}{12}N_f^2 $\\
			
			\hline
			
			\multirow{6}{*}{$\chi_1$}&
			$\Phi^6$&
			7&
			0\\
			
			&
			$\Phi^4\mathcal{D}^2$&
			10&
			2\\
			
			&
			$\Phi^2X^2$&
			14 (7)&
			0\\
			
			&
			$\Psi^2\Phi^2\mathcal{D}$&
			$17N_f^2$&
			$6N_f^2$\\
			
			&
			$\Psi^2\Phi^3$&
			$38N_f^2$&
			$18N_f^2$\\
			
			&
			$\Psi^2\Phi X$&
			$6N_f^2$&
			$6N_f^2$\\
			\hline
			
			\multirow{6}{*}{$\varphi_1$}&
			$\Phi^6$&
			3&
			0\\
			
			&
			$\Phi^4\mathcal{D}^2$&
			4&
			0\\
			
			&
			$\Phi^2X^2$&
			10 (5)&
			0\\
			
			&
			$\Psi^2\Phi^2\mathcal{D}$&
			$12N_f^2$&
			$4N_f^2$\\
			
			&
			$\Psi^2\Phi^3$&
			$15N_f^2-3N_f$&
			$3N_f^2-3N_f$\\
			
			&
			$\Psi^2\Phi X$&
			$3N_f^2+N_f$&
			$3N_f^2+N_f$\\
			\hline
			
			\multirow{3}{*}{$X_{\mu},\,\,Y_{\mu}$}&
			$X^3$&
			2 (1)&
			0\\
			
			&
			$\Phi^2X^2$&
			14 (7)&
			0\\

			&
			$\Psi^2\Phi X$&
			12&
			0\\
			\hline
			
	\end{tabular}}
	\caption{Table \ref{table:number-of-ops-dim6} continued. The numbers in parentheses denote the counting for CP violating purely bosonic operators.}
	\label{table:number-of-ops-dim6-2}
\end{table}

\begin{table}[h]
	\centering
	\renewcommand{\arraystretch}{1.9}
	{\scriptsize\begin{tabular}{|c|c|c|c|}
			\hline
			\multirow{2}{*}{\textbf{BSM Field}}&
			\multirow{2}{*}{\textbf{Operator Class}}&
			\multicolumn{2}{c|}{\textbf{Number of Operators as $f(N_f)$}}\\
			\cline{3-4}
			
			&
			&
			Total Number&
			$B, \,L$ Violating Operators\\
			\hline
			
			$\delta^{+}$&
			$\Psi^2\Phi^2$&
			$7N_f^2+N_f$&
			$N_f^2+N_f$\\
			\hline
			
			\multirow{2}{*}{$\Delta$}&
			$\Psi^2\Phi^2$&
			$7N_f^2+N_f$&
			$N_f^2+N_f$\\
			
			&
			$\Phi^5$&
			6&
			0\\
			\hline
			
			\multirow{2}{*}{$\Sigma$}&
			$\Psi^2\Phi^2$&
			$4N_f^2$&
			$2N_f^2$\\
			
			&
			$\Psi^2X$&
			$2N_f^2$&
			0\\
			\hline
			
			\multirow{2}{*}{$ V_{L,R};\, E_{L,R};\, N_{L,R}$}&
			$\Psi^2\Phi^2$&
			$20N_f^2+4N_f$&
			$ 6N_f^2+4N_f $\\
			
			&
			$\Psi^2X$&
			$16N_f^2-2N_f$&
			$ 2N_f^2-2N_f $\\
			\hline
			
			$\chi_1$&
			$\Psi^2\Phi^2$&
			$19N_f^2 +N_f$&
			$13N_f^2+N_f$\\
			\hline
			
			$\varphi_1$&
			$\Psi^2\Phi^2$&
			$7N_f^2+N_f$&
			$7N_f^2+N_f$\\
			\hline
	\end{tabular}}
	\caption{Number of additional operators of different classes at dimension 5 with $N_f$ fermion flavours for each BSM model. There are no new dimension 5 operators for the models containing $\rho^{++}$ and $X_{\mu}, Y_{\mu}$.}
	\label{table:number-of-ops-dim5}
\end{table}

\section{Conclusions and Remarks}

In this paper, our chief objective has been to pave the way for BSMEFT. The UV model realised in nature, which is yet to be observed, may be residing over a range of energy scales containing a highly non-degenerate spectrum. Thus, it is very unlikely (unless it possesses a compressed spectrum) that all the non-SM particles are integrated out at the same scale leading to an effective theory described by the SMEFT Lagrangian. Instead, we expect to see a first glimpse of the full theory at ongoing high-energy experiments, where a new degree-of-freedom might appear on-shell. After obtaining the first hint of a new resonance, the imminent course of action will be to embed this new particle into an extension of SMEFT, where this particle acts as a new infrared degree of freedom in addition to all Standard Model particles. We denote this class of new effective theories as BSMEFT.

Already several rather minimal extensions of the SM exist, which attempt to solve or at least address its specific shortcomings. These extensions are therefore phenomenologically motivated and can be considered residual theories of multiple UV theories, e.g. a second scalar particle can arise from a plethora of very different UV models. Thus it would be wise to consider them as part of an effective theory, where the other heavy modes belonging to that unknown full theory have been integrated out. To capture their footprints we can include the lightest non-SM particle as the IR DOF along with the SM ones and construct the effective Lagrangian. This enlarges the operator set beyond the SMEFT and that is what we call BSMEFT. 

Looking into the possible well-motivated scenarios we have categorized the BSMEFT construction into three different classes: SM extended by additional uncolored and colored particles and gauge symmetries.  For each such class, we have adopted multiple example models. We have computed all non-redundant dimension 6 operators, extending SMEFT to BSMEFT. We have reached out to a variety of scenarios by adding color singlet scalars, fermions, colored Lepto-Quark scalars, vector-like fermions, and extending the gauge symmetry by two abelian groups. Many neutrino mass models contain complex $SU(2)_L$ singlets and (or) multiplets. All of them can be suitably described by a single effective theory containing a singly, or doubly charged scalar as the additional IR DOF. There are more complete theories, e.g., the parity conserving Pati-Salam, Left-Right Symmetric Model, etc. which contain all these DOFs in their minimal and (or) non-minimal versions. The suitable choices of heavy modes, consistent with phenomenological constraints,  will allow us to rewrite multiple theories in terms of an effective one. The future observation of the non-SM particle(s) will pinpoint the unique choice of additional IR DOF(s) and will open the gateway of BSMEFT.

\section{Acknowledgments}
 The work of JC, SP, UB, SUR is supported by
the Science and Engineering Research Board, Government of India, under the agreements SERB/PHY/2016348 (Early Career Research Award) and SERB/PHY/2019501
(MATRICS) and Initiation Research Grant, agreement number IITK/PHY/2015077, by
IIT Kanpur. M.S. is supported by the STFC under grant ST/P001246/1.

\appendix

\section{The SMEFT Effective Operator Basis}

For each BSMEFT scenario, only the additional effective operators in the presence of extra IR DOFs have been discussed. But while performing the complete analysis of these effective theories, one must not forget to add the SMEFT operators. For the sake of completeness we have provided the SM Lagrangian in Eqn.~\eqref{eq:SM-ren-lag} and the complete set of dimension 6 operators \cite{Grzadkowski:2010es} in Tables~\ref{table:dim6-ops-1} and \ref{table:dim6-ops-2}.
To avoid any ambiguity we have also listed the Standard Model degrees of freedom and their transformation properties under the gauge group $SU(3)_C \otimes SU(2)_L \otimes U(1)_Y$ in Table \ref{table:SM-fields}. 

\begin{table}[h!]
	\centering
	\renewcommand{\arraystretch}{1.7}
	{\scriptsize\begin{tabular}{|c|c|c|c|c|c|c|}
			\hline
			\textbf{Field} & \textbf{$SU(3)_C$} & \textbf{$SU(2)_{L}$}&\textbf{$U(1)_{Y}$}&Baryon No.&Lepton No.&Spin\\
			\hline
			$H$    &1&2&1/2&0&0&0\\
			$Q^p_L$       &3&2&1/6&1/3&0&1/2\\
			$u^p_R$     &3&1&2/3&1/3&0&1/2\\
			$d^p_R$     &3&1&-1/3&1/3&0&1/2\\
			$L^p_L$       &1&2&-1/2&0&-1&1/2\\
			$e^p_R$     &1&1&-1&0&-1&1/2\\
			\hline
			$G^A_{\mu}$ &8&1&0&0&0&1\\
			$W^I_{\mu}$ &1&3&0&0&0&1\\
			$B_{\mu}$   &1&1&0&0&0&1\\
			\hline
	\end{tabular}}
	\caption{\small Standard Model: Gauge and global quantum numbers and spins of the fields.} 
	\label{table:SM-fields}
\end{table}

{\small\begin{eqnarray}\label{eq:SM-ren-lag}
	\mathcal{L}^{(4)}_{SM}& = &-\frac{1}{4} G^{A}_{\mu\nu}\,G^{A\mu\nu} -\frac{1}{4} W^{I}_{\mu\nu}\,W^{I\mu\nu} -\frac{1}{4}B_{\mu\nu}\,B^{\mu\nu} \nonumber\\
	&&+\; i(\bar{L}^p_L\,\slashed{D}\, L^p_L + \bar{Q}^p_L\, \slashed{D}\, Q^p_L + \bar{e}^p_R\, \slashed{D}\, e^p_R + \bar{u}^p_R\, \slashed{D}\, u^p_R + \bar{d}^p_R\, \slashed{D}\, d^p_R) \nonumber \\
	&& - (y^{ps}_e\,\bar{L}^p_L\, e^s_R\, H +  y^{ps}_d\,\bar{Q}^p_L\, d^s_R\, H + y^{ps}_u\,\bar{Q}^p_L\, u^s_R\,\tilde{H}) + h.c. \nonumber \\
	&&+\; (D_{\mu}\,H)^{\dagger}\,(D^{\mu}\,H)\, - \,m^2\,(H^{\dagger}\,H)\, -\, \lambda\,(H^{\dagger}\,H)^2 .
\end{eqnarray}}
Here, $y_{e, d, u}$ are Yukawa matrices and $p, s$ are flavour indices, $\slashed{\mathcal{D}} = \gamma^{\mu}\mathcal{D}_{\mu}$ and the exact form of $\mathcal{D}_{\mu}$ for a specific field is determined based on its gauge quantum numbers. Also,
\newpage
{\small\begin{eqnarray}\label{eq:SM-field-tensor}
G^A_{\mu\nu} &=& \partial_\mu G^A_\nu - \partial_\nu G^A_\mu - g_3           f^{ABC}G^B_\mu G^C_\nu,\nonumber\\
W^I_{\mu\nu} &=& \partial_\mu W^I_\nu - \partial_\nu W^I_\mu - g \epsilon^{IJK}W^J_\mu W^K_\nu,\nonumber\\
B_{\mu\nu} &=& \partial_\mu B_\nu - \partial_\nu B_\mu .  
\end{eqnarray}}
are the field strength tensors corresponding to the $SU(3)_C$, $SU(2)_L$ and  $U(1)_Y$ gauge groups respectively with $A,B,C = 1,\cdots,8$ and $I,J,K = 1,2,3$. 

\begin{table}[h]
	\centering
	\renewcommand{\arraystretch}{2.0}
	{\small\begin{tabular}{||c|c||c|c||}
			\hline	
			\hline
			\multicolumn{4}{||c||}{$X^3$}\\
			\hline
			
			$\mathcal{O}_G$&
			$f^{ABC}\,G^{A\mu}_{\nu}\,G^{B\nu}_{\rho}\,G^{C\rho}_{\mu}$&
			$\mathcal{O}_{\tilde{G}}$&
			$f^{ABC}\,\tilde{G}^{A\mu}_{\nu}\,G^{B\nu}_{\rho}\,G^{C\rho}_{\mu}$\\
			
			$\mathcal{O}_W$&
			$\epsilon^{IJK}\,W^{I\mu}_{\nu}\,W^{J\nu}_{\rho}\,W^{K\rho}_{\mu}$&
			$\mathcal{O}_{\tilde{W}}$&
			$\epsilon^{IJK}\,\tilde{W}^{I\mu}_{\nu}\,W^{J\nu}_{\rho}\,W^{K\rho}_{\mu}$\\

			\hline	
			\hline
			\multicolumn{2}{||c||}{$\Psi^2\Phi^3$}&
			\multicolumn{2}{c||}{$\Phi^6,\hspace{0.2cm}\Phi^4\mathcal{D}^2$}\\
			\hline
			
			$\mathcal{O}_{eH}$&
			\textcolor{blue}{$(\overline{L}_{p}\,e_{q}\,H)\,(H^{\dagger}\,H)$}&
			$\mathcal{O}_{H}$&
			$(H^{\dagger}\,H)^6$\\
			
			$\mathcal{O}_{uH}$&
			\textcolor{blue}{$(\overline{Q}_{p}\,u_{q}\,\tilde{H})\,(H^{\dagger}\,H)$}&
			$\mathcal{O}_{H\square}$&
			$(H^{\dagger}\,H)\square(H^{\dagger}\,H)$\\
			
			$\mathcal{O}_{dH}$&
			\textcolor{blue}{$(\overline{Q}_{p}\,d_{q}\,H)\,(H^{\dagger}\,H)$}&
			$\mathcal{O}_{H\mathcal{D}}$&
			$(H^{\dagger}\,i\overleftrightarrow{\mathcal{D}}^{\mu}\,H)(H^{\dagger}\,i\overleftrightarrow{\mathcal{D}_{\mu}}\,H)$\\
			
			\hline
			\hline
			
			\multicolumn{2}{||c||}{$\Psi^2\Phi X$}&
			\multicolumn{2}{c||}{$\Psi^2\Phi^2\mathcal{D}$}\\
			\hline
			$\mathcal{O}_{e W}$&
			\textcolor{blue}{$(\overline{L}_{p}\,\sigma^{\mu\nu}\,e_{q})\,\tau^{I}\,H\,W^{I}_{\mu\nu}$}&
			$\mathcal{O}_{H L \mathcal{D}}^{(1)}$&
			$(H^{\dagger}\,i\overleftrightarrow{\mathcal{D}_{\mu}}\,H)\,(\overline{L}_{p}\,\gamma^{\mu}\,L_{q})$\\

			$\mathcal{O}_{e B}$&
			\textcolor{blue}{$(\overline{L}_{p}\,\sigma^{\mu\nu}\,e_{q})\,H\,B_{\mu\nu}$}&
			$\mathcal{O}_{H L \mathcal{D}}^{(3)}$&
			$(H^{\dagger}\,i\overleftrightarrow{\mathcal{D}^{I}_{\mu}}\,H)\,(\overline{L}_{p}\,\gamma^{\mu}\,\tau^{I}\,L_{q})$\\

			$\mathcal{O}_{u G}$&
			\textcolor{blue}{$(\overline{Q}_{p}\,\sigma^{\mu\nu}\,T^{A}\,u_{q})\,\tilde{H}\,G^{A}_{\mu\nu}$}&
			$\mathcal{O}_{H e \mathcal{D}}$&
			$(H^{\dagger}\,i\overleftrightarrow{\mathcal{D}_{\mu}}\,H)\,(\overline{e}_{p}\,\gamma^{\mu}\,e_{q})$\\

			$\mathcal{O}_{u W}$&
			\textcolor{blue}{$(\overline{Q}_{p}\,\sigma^{\mu\nu}\,u_{q})\,\tau^{I}\,\tilde{H}\,W^{I}_{\mu\nu}$}&
			$\mathcal{O}_{H Q \mathcal{D}}^{(1)}$&
			$(H^{\dagger}\,i\overleftrightarrow{\mathcal{D}_{\mu}}\,H)\,(\overline{Q}_{p}\,\gamma^{\mu}\,Q_{q})$\\
			
			$\mathcal{O}_{u B}$&
			\textcolor{blue}{$(\overline{Q}_{p}\,\sigma^{\mu\nu}\,u_{q})\,\tilde{H}\,B_{\mu\nu}$}&
			$\mathcal{O}_{H Q \mathcal{D}}^{(3)}$&
			$(H^{\dagger}\,i\overleftrightarrow{\mathcal{D}^{I}_{\mu}}\,H)\,(\overline{Q}_{p}\,\gamma^{\mu}\,\tau^{I}\,Q_{q})$\\

			$\mathcal{O}_{d G}$&
			\textcolor{blue}{$(\overline{Q}_{p}\,\sigma^{\mu\nu}\,T^{A}\,d_{q})\,H\,G^{A}_{\mu\nu}$}&
			$\mathcal{O}_{H u \mathcal{D}}$&
			$(H^{\dagger}\,i\overleftrightarrow{\mathcal{D}_{\mu}}\,H)\,(\overline{u}_{p}\,\gamma^{\mu}\,u_{q})$\\

			$\mathcal{O}_{d W}$&
			\textcolor{blue}{$(\overline{Q}_{p}\,\sigma^{\mu\nu}\,d_{q})\,\tau^{I}\,H\,W^{I}_{\mu\nu}$}&
			$\mathcal{O}_{H d \mathcal{D}}$&
			$(H^{\dagger}\,i\overleftrightarrow{\mathcal{D}_{\mu}}\,H)\,(\overline{d}_{p}\,\gamma^{\mu}\,d_{q})$\\

			$\mathcal{O}_{d B}$&
			\textcolor{blue}{$(\overline{Q}_{p}\,\sigma^{\mu\nu}\,d_{q})\,H\,B_{\mu\nu}$}&
			$\mathcal{O}_{H u d \mathcal{D}}$&
			\textcolor{blue}{$(\tilde{H}^{\dagger}\,i\mathcal{D}_{\mu}\,H)\,(\overline{u}_{p}\,\gamma^{\mu}\,d_{q})$}\\
			
			\hline
	\end{tabular}}
	\caption{SMEFT dimension 6 operators. Here, $T^A$ and $\tau^I$ are $SU(3)$ and $SU(2)$ generators respectively. $A,B,C=1,2,\cdots,8$ and $I,J,K=1,2,3$. $p, q=1,2,\cdots,N_f$ are flavour indices. Operator naming convention has been adopted from \cite{Grzadkowski:2010es}.}
	\label{table:dim6-ops-1}
\end{table}
\clearpage
\begin{table}[h]
	\centering
	\renewcommand{\arraystretch}{2.0}
	{\small\begin{tabular}{||c|c||c|c||}
			\hline
			\hline
			\multicolumn{4}{||c||}{$\Phi^2X^2$}\\
			\hline
			$\mathcal{O}_{H G}$&
			$(H^{\dagger}\,H)\,(G^{A}_{\mu\nu}\,G^{A\mu\nu})$&
			$\mathcal{O}_{H \tilde{G}}$&
			$(H^{\dagger}\,H)\,(\tilde{G}^{A}_{\mu\nu}\,G^{A\mu\nu})$\\
			
			$\mathcal{O}_{H W}$&
			$(H^{\dagger}\,H)\,(W^{I}_{\mu\nu}\,W^{I\mu\nu})$&
			$\mathcal{O}_{H \tilde{W}}$&
			$(H^{\dagger}\,H)\,(\tilde{W}^{I}_{\mu\nu}\,W^{I\mu\nu})$\\
			
			$\mathcal{O}_{H B}$&
			$(H^{\dagger}\,H)\,(B_{\mu\nu}\,B^{\mu\nu})$&
			$\mathcal{O}_{H \tilde{B}}$&
			$(H^{\dagger}\,H)\,(\tilde{B}_{\mu\nu}\,B^{\mu\nu})$\\
			
			$\mathcal{O}_{H W B}$&
			$(H^{\dagger}\,\tau^{I}\,H)\,(W^{I}_{\mu\nu}\,B^{\mu\nu})$&
			$\mathcal{O}_{H \tilde{W} B}$&
			$(H^{\dagger}\,\tau^{I}\,H)\,(\tilde{W}^{I}_{\mu\nu}\,B^{\mu\nu})$\\
			
			\hline
			\hline
			\multicolumn{4}{||c||}{$\Psi^4$}\\
			\hline
			$\mathcal{O}_{LL}$&
			$(\overline{L}_{p}\,\gamma_{\mu}\,L_{q})\,(\overline{L}_{r}\,\gamma^{\mu}\,L_{s})$&
			$\mathcal{O}_{ee}$&
			$(\overline{e}_{p}\,\gamma_{\mu}\,e_{q})\,(\overline{e}_{r}\,\gamma^{\mu}\,e_{s}) $\\
			
			$\mathcal{O}^{(1)}_{QQ}$&
			$(\overline{Q}_{p}\,\gamma_{\mu}\,Q_{q})\,(\overline{Q}_{r}\,\gamma^{\mu}\,Q_{s}) $&
			$\mathcal{O}_{uu}$&
			$(\overline{u}_{p}\,\gamma_{\mu}\,u_{q})\,(\overline{u}_{r}\,\gamma^{\mu}\,u_{s}) $\\
			
			$\mathcal{O}^{(3)}_{QQ} $&
			$(\overline{Q}_{p}\,\gamma_{\mu}\,\tau^{I}\,Q_{q})\,(\overline{Q}_{r}\,\gamma^{\mu}\,\tau^{I}\,Q_{s}) $&
			$\mathcal{O}_{dd}$&
			$(\overline{d}_{p}\,\gamma_{\mu}\,d_{q})\,(\overline{d}_{r}\,\gamma^{\mu}\,d_{s}) $\\
			
			$\mathcal{O}^{(1)}_{LQ} $&
			$(\overline{L}_{p}\,\gamma_{\mu}\,L_{q})\,(\overline{Q}_{r}\,\gamma^{\mu}\,Q_{s}) $&
			$\mathcal{O}_{Le}$&
			$(\overline{L}_{p}\,\gamma_{\mu}\,L_{q})\,(\overline{e}_{r}\,\gamma^{\mu}\,e_{s}) $\\
			
			$\mathcal{O}^{(3)}_{LQ} $&
			$(\overline{L}_{p}\,\gamma_{\mu}\,\tau^{I}\,L_{q})\,(\overline{Q}_{r}\,\gamma^{\mu}\,\tau^{I}\,Q_{s}) $&
			$\mathcal{O}_{Lu} $&
			$(\overline{L}_{p}\,\gamma_{\mu}\,L_{q})\,(\overline{u}_{r}\,\gamma^{\mu}\,u_{s}) $\\
			
			$\mathcal{O}_{eu} $&
			$(\overline{e}_{p}\,\gamma_{\mu}\,e_{q})\,(\overline{u}_{r}\,\gamma^{\mu}\,u_{s}) $&
			$\mathcal{O}_{Ld} $&
			$(\overline{L}_{p}\,\gamma_{\mu}\,L_{q})\,(\overline{d}_{r}\,\gamma^{\mu}\,d_{s}) $\\
			
			$\mathcal{O}_{ed} $&
			$(\overline{e}_{p}\,\gamma_{\mu}\,e_{q})\,(\overline{d}_{r}\,\gamma^{\mu}\,d_{s}) $&
			$\mathcal{O}_{Qe} $&
			$(\overline{Q}_{p}\,\gamma_{\mu}\,Q_{q})\,(\overline{e}_{r}\,\gamma^{\mu}\,e_{s}) $\\
			
			$\mathcal{O}^{(1)}_{ud} $&
			$(\overline{u}_{p}\,\gamma_{\mu}\,u_{q})\,(\overline{d}_{r}\,\gamma^{\mu}\,d_{s}) $&
			$\mathcal{O}^{(1)}_{Qu} $&
			$(\overline{Q}_{p}\,\gamma_{\mu}\,Q_{q})\,(\overline{u}_{r}\,\gamma^{\mu}\,u_{s}) $\\
			
			$\mathcal{O}^{(8)}_{ud} $&
			$(\overline{u}_{p}\,\gamma_{\mu}\,T^{A}\,u_{q})\,(\overline{d}_{r}\,\gamma^{\mu}\,T^{A}\,d_{s}) $&
			$\mathcal{O}^{(8)}_{Qu} $&
			$(\overline{Q}_{p}\,\gamma_{\mu}\,T^{A}\,Q_{q})\,(\overline{u}_{r}\,\gamma^{\mu}\,T^{A}\,u_{s}) $\\
			
			$\mathcal{O}^{(1)}_{Qd} $&
			$(\overline{Q}_{p}\,\gamma_{\mu}\,Q_{q})\,(\overline{d}_{r}\,\gamma^{\mu}\,d_{s}) $&
			$\mathcal{O}^{(8)}_{Qd} $&
			$(\overline{Q}_{p}\,\gamma_{\mu}\,T^{A}\,Q_{q})\,(\overline{d}_{r}\,\gamma^{\mu}\,T^{A}\,d_{s}) $\\

			$\mathcal{O}_{LedQ}$&
			\textcolor{blue}{$(\overline{L}_{pj}\,e_{q})\,(\overline{d}_{r}\,Q^{j}_{s})$}&
			$\mathcal{O}_{duQ}$&
			\textcolor{purple}{$\epsilon^{\alpha\beta\gamma}\,\epsilon_{jk}\,[(d^{\alpha}_{p})^T\,C\,u^{\beta}_{q}]\,[(Q^{j\gamma}_{r})^T\,C\,L^{k}_{s}]$}\\
			
			$\mathcal{O}^{(1)}_{LeQu}$&
			\textcolor{blue}{$\epsilon_{jk}\,(\overline{L}_{pj}\,e_{q})\,(\overline{Q}_{rk}\,u_{s})$}&
			$\mathcal{O}_{duu}$&
			\textcolor{purple}{$\epsilon^{\alpha\beta\gamma}\,[(d^{\alpha}_{p})^T\,C\,u^{\beta}_{q}]\,[(u^{\gamma}_{r})^T\,C\,e_{s}]$}\\
			
			$\mathcal{O}^{(3)}_{LeQu}$&
			\textcolor{blue}{$\epsilon_{jk}\,(\overline{L}_{pj}\,\sigma_{\mu\nu}\,e_{q})\,(\overline{Q}_{rk}\,\sigma^{\mu\nu}\,u_{s})$}&
			$\mathcal{O}_{QQQ}$&
			\textcolor{purple}{$\epsilon^{\alpha\beta\gamma}\,\epsilon_{jn}\,\epsilon_{km}\,[(Q^{j\alpha}_{p})^T\,C\,Q^{k\beta}_{q}]\,[(Q^{m\gamma}_{r})^T\,C\,L^{n}_{s}]$}\\
			
			$\mathcal{O}^{(1)}_{QuQd}$&
			\textcolor{blue}{$\epsilon_{jk}\,(\overline{Q}_{pj}\,u_{q})\,(\overline{Q}_{rk}\,d_{s})$}&
			$\mathcal{O}_{QQu}$&
			\textcolor{purple}{$\epsilon^{\alpha\beta\gamma}\,\epsilon_{jk}\,[(Q^{j\alpha}_{p})^T\,C\,Q^{k\beta}_{q}]\,[(u^{\gamma}_{r})^T\,C\,e_{s}]$}\\
			
			$\mathcal{O}^{(8)}_{QuQd}$&
			\textcolor{blue}{$\epsilon_{jk}\,(\overline{Q}_{pj}\,T^{A}\,u_{q})\,(\overline{Q}_{rk}\,T^{A}\,d_{s})$}&
			&
			\\
			\hline
	\end{tabular}}
	\caption{Table~\ref{table:dim6-ops-1} continued. Here $j, k, m, n$ and $\alpha,\beta,\gamma$ are the $SU(2)$ and $SU(3)$ indices respectively and $p, q, r, s=1,2,\cdots,N_f$ are the flavour indices. Operators in {\it red} violate lepton and baryon numbers.}
	\label{table:dim6-ops-2}
\end{table}
\clearpage
\section{ BSMEFT: a few more popular scenarios}
\subsection{The Operator Bases}

\subsubsection*{\underline{SM + $SU(2)$ Quadruplet Scalar ($\Theta$)}}

The SM can be extended by an $ SU(2)_L $   quadruplet scalar $(\Theta)$ with hypercharge $3/2$. After the breaking of electroweak symmetry, the components of the multiplet emerge as electromagnetically charged fields\footnote{An $SU(2)$ quadruplet has $T_3$ values (+3/2, +1/2, -1/2, -3/2). So, using $Q = T_3 + Y$, we get the electromagnetic charges (+3, +2, +1, 0) since $Y = 3/2$.} and we can write them as $\Theta$ = $(\Theta^{+++},\,\,\Theta^{++},\,\,\Theta^{+},\,\,\Theta^{0})$. Since the quadruplet contains charged scalars they offer very interesting phenomenology, e.g., neutrino mass generation, lepton number and flavour violations \cite{Babu:2009aq,Kumericki:2012bh,McDonald:2013hsa, Chakrabortty:2015zpm,delAguila:2013hla,delAguila:2013yaa,Bambhaniya:2013yca} in the presence of additional particles which can be heavy enough to be integrated out. This would lead to an effective Lagrangian described by the SM DOFs along with the quadruplet scalar. The operators of mass dimensions 5 and 6 involving $\Theta$ have been catalogued in Tables~\ref{table:QuadrupletScalar-dim5-ops-1} and \ref{table:QuadrupletScalar-dim6-ops-1} respectively. 
\noindent
While writing the operators in their covariant forms, we have to be careful with  the quadruplet $\Theta$. That is why we have worked with its component form $\Theta_{ijk}$ with $i,j,k$ = $1,2$ and we identify the components as $\Theta_{111} = \Theta^{+++}$, $\Theta_{112} = \Theta^{++}$, $\Theta_{122} = \Theta^{+}$ and $\Theta_{222} = \Theta^{0}$.  To compute the higher tensor products of $\Theta$ with the $SU(2)_L$ doublets, i.e., $L, Q, H$ and with the triplet $W_{\mu\nu}$, we have introduced the $4\times 4$ generators of $SU(2)$ and we denote them as $\tau^{I}_{(4)}$:

{\small\begin{eqnarray}
	\tau^{1}_{(4)} = \begin{pmatrix}
	0\ \ \frac{\sqrt{3}}{2} \ \  0\ \ 0\\
	\frac{\sqrt{3}}{2} \ \ 0\ \ 1  \ \ 0\\
	0 \ \ 1 \ \  0 \ \ \frac{\sqrt{3}}{2}\\
	0 \ \ 0\ \  \frac{\sqrt{3}}{2} \ \ 0
	\end{pmatrix}, \hspace{0.5cm} \tau^{2}_{(4)} = \begin{pmatrix}
	0\ \ \text{-}\frac{\sqrt{3}}{2}i \ \  0\ \ 0\\
	\frac{\sqrt{3}}{2}i \ \ 0\ \ \text{-}i  \ \ 0\\
	0 \ \ i \ \  0 \ \ \text{-}\frac{\sqrt{3}}{2}i\\
	0 \ \ 0\ \  \frac{\sqrt{3}}{2}i \ \ 0
	\end{pmatrix}, \hspace{0.5cm} \tau^{3}_{(4)} = \begin{pmatrix}
	\frac{3}{2} \ \ 0 \ \  0 \ \ 0\\
	0 \ \ \frac{1}{2} \ \  0 \ \ 0\\
	0 \ \ 0 \ \  \text{-}\frac{1}{2} \ \ 0\\
	0 \ \ 0 \ \  0 \ \ \text{-}\frac{3}{2}
	\end{pmatrix}.
	\end{eqnarray}}
To avoid confusion, for this model we have denoted the $2\times 2$ Pauli matrices as $\tau^{I}_{(2)}$.
\\
\\
\noindent\textbf{Features of the additional operators:}
\begin{itemize}	
	\item At dimension 5, we have a lepton number violating operator $\tilde{\mathcal{O}}_{LH\Theta}$.
	
	\item At dimension 6, most of the operators possess similar structures as found in the case of the complex triplet scalar ($\Delta$).

	\item This scenario does not offer any baryon and lepton number violation at dimension 6.
\end{itemize}

\begin{table}[h]
	\centering
	\renewcommand{\arraystretch}{1.9}
	{\scriptsize\begin{tabular}{||c|c||}
			\hline
			\hline
			\multicolumn{2}{||c||}{$\Psi^2\Phi^2$}\\
			\hline
			$\tilde{\mathcal{O}}_{LH\Theta}$&
			$\color{purple}{\frac{1}{2} (N_f^2+N_f) \,L^i_{p} \,L^j_{q} \,\tilde{H}_k \,\Theta_{ijk}}$
			\\
			\hline
	\end{tabular}}
	\caption{SM extended by $SU(2)$ Quadruplet Scalar ($\Theta$): Additional operators of dimension 5. Here $i, j, k$ are the $SU(2)$ indices and $p,q=1,2,\cdots,N_f$ are the flavour indices. The operator violates lepton number.}
	\label{table:QuadrupletScalar-dim5-ops-1}
\end{table}

\begin{table}[h]
	\centering
	\renewcommand{\arraystretch}{1.9}
	{\scriptsize\begin{tabular}{||c|c||c|c||}
			
			\hline
			\multicolumn{4}{||c||}{$\Phi^4\mathcal{D}^2$}\\
			\hline
			
			$\mathcal{O}_{H\Theta\square}$&
			$(\Theta^{\dagger} \,\Theta) \,\square \,(H^{\dagger} \,H)$&
			$\mathcal{O}_{H\Theta\mathcal{D}}^{(1)}$&
			$(\Theta^{\dagger} \,i\overleftrightarrow{\mathcal{D}}_{\mu} \,\Theta) \,(H^{\dagger} \,i\overleftrightarrow{\mathcal{D}}^{\mu} \,H)$\\
			
			$\mathcal{O}_{H\Theta\mathcal{D}}^{(2)}$&
			$(H^{\dagger} \,H) \,[(\mathcal{D}^{\mu} \,\Theta)^{\dagger} \,(\mathcal{D}_{\mu} \,\Theta)]$&
			$\mathcal{O}_{H\Theta\mathcal{D}}^{(3)}$&
			$(\Theta^{\dagger} \,\Theta) \,[(\mathcal{D}^{\mu} \,H)^{\dagger} \,(\mathcal{D}_{\mu} \,H)]$\\
			
			$\mathcal{O}_{\Theta\square}^{(1)}$&
			$(\Theta^{\dagger} \,\Theta) \,\square \,(\Theta^{\dagger} \,\Theta)$&
			$\mathcal{O}_{\Theta\mathcal{D}}^{(1)}$&
			$(\Theta^{\dagger} \,\Theta) \,[(\mathcal{D}^{\mu} \,\Theta)^{\dagger} \,(\mathcal{D}_{\mu} \,\Theta)]$\\
			
			$\mathcal{O}_{\Theta\mathcal{D}}^{(2)}$&
			$(\Theta^{\dagger} \,i\overleftrightarrow{\mathcal{D}}_{\mu} \,\Theta) \,(\Theta^{\dagger} \,i\overleftrightarrow{\mathcal{D}}^{\mu} \,\Theta)$&
			$\mathcal{O}_{\Theta\mathcal{D}}^{(3)}$&
			$[(\mathcal{D}^{\mu} \,\Theta_{ijk})^{\dagger} \,\Theta^{ilm} \,\Theta^{\dagger}_{lmn}\,(\mathcal{D}_{\mu} \,\Theta^{jkn})]$\\
			
			\hline
			\hline
			\multicolumn{4}{||c||}{$\Psi^2\Phi^2\mathcal{D}$}\\
			\hline
			
			$\mathcal{O}^{(1)}_{Q \Theta \mathcal{D}}$
			&$(N_f^2) \,(\overline{Q}_{p\alpha i} \,\gamma^{\mu} \,Q^{\alpha i}_{q}) \,(\Theta^{\dagger} \,i\overleftrightarrow{\mathcal{D}}_{\mu} \,\Theta)$&
			$\mathcal{O}^{(2)}_{Q \Theta \mathcal{D}}$&
			$(N_f^2) \,(\overline{Q}_{p\alpha i} \,\tau^I \,\gamma^{\mu} \,Q^{\alpha i}_{q}) \,(\Theta^{\dagger} \,i\overleftrightarrow{\mathcal{D}}_{\mu}^I \,\Theta)$
			\\
			
			$\mathcal{O}^{(1)}_{L \Theta \mathcal{D}}$&
			$(N_f^2) \,(\overline{L}_{pi} \,\gamma^{\mu} \,L^{i}_{q}) \,(\Theta^{\dagger} \,i\overleftrightarrow{\mathcal{D}}_{\mu} \,\Theta)$&
			$\mathcal{O}^{(2)}_{L \Theta \mathcal{D}}$&
			$(N_f^2) \,(\overline{L}_{pi} \,\tau^I \,\gamma^{\mu} \,L^{i}_{q}) \,(\Theta^{\dagger} \,i\overleftrightarrow{\mathcal{D}}_{\mu}^I \,\Theta)$\\
			
			$\mathcal{O}_{u \Theta \mathcal{D}}$&
			$(N_f^2) \,(\overline{u}_{p\alpha} \,\gamma^{\mu} \,u^{\alpha}_{q}) \,(\Theta^{\dagger} \,i\overleftrightarrow{\mathcal{D}}_{\mu} \,\Theta)$
			&$\mathcal{O}_{d \Theta \mathcal{D}}$&
			$(N_f^2) \,(\overline{d}_{p\alpha} \,\gamma^{\mu} \,d^{\alpha}_{q}) \,(\Theta^{\dagger} \,i\overleftrightarrow{\mathcal{D}}_{\mu} \,\Theta)$
			\\
			
			$\mathcal{O}_{e \Theta \mathcal{D}}$&
			$(N_f^2) \,(\overline{e}_{p} \,\gamma^{\mu} \,e_{q}) \,(\Theta^{\dagger} \,i\overleftrightarrow{\mathcal{D}}_{\mu} \,\Theta)$&
			&
			\\
			\hline
			\hline
			\multicolumn{4}{||c||}{$\Phi^6$}\\
			\hline
			$\mathcal{O}_{\Theta}^{(1)}$&
			$(\Theta^{\dagger} \,\Theta)^3$&
			$\mathcal{O}_{\Theta}^{(2)}$&
			$(\Theta^{\dagger}_{ijk} \,\Theta^{ilm} \,\Theta^{\dagger}_{lmn} \,\Theta^{jkn})\,(\Theta^{\dagger} \,\Theta)$\\
			
			$\mathcal{O}_{\Theta}^{(3)}$&
			$(\Theta^{\dagger}_{ijk} \,\Theta^{ilm} \,\Theta^{\dagger}_{lmn} \,\Theta^{nrp} \,\Theta^{\dagger}_{rpq} \,\Theta^{jkq})$&
			$\mathcal{O}_{H^2\Theta^4}^{(1)}$&
			$(\Theta^{\dagger} \,\Theta)^2 \,(H^{\dagger} \,H)$
			\\
			
			$\mathcal{O}_{H^2\Theta^4}^{(2)}$&
			$(\Theta^{\dagger}_{ijk} \,\Theta^{ilm} \,\Theta^{\dagger}_{lmn} \,\Theta^{jkn}) \,(H^{\dagger} \,H)$&
			$\mathcal{O}_{H^2\Theta^4}^{(3)}$&
			$(\Theta^{\dagger} \,\Theta) \,(\tilde{H}^i \,\Theta^{\dagger}_{ijk} \,\Theta^{jkl} \,H_l)$
			\\
			
			$\mathcal{O}_{H^2\Theta^4}^{(4)}$&
			$(\tilde{H}^i \,\Theta^{\dagger}_{ijk} \,\Theta^{jkl} \,\Theta^{\dagger}_{lmn} \,\Theta^{mnr} \,H_{r})$&$\mathcal{O}_{H^4\Theta^2}^{(1)}$&
			$(\Theta^{\dagger} \,\Theta) \,(H^{\dagger} \,H)^2$
			\\
			
			$\mathcal{O}_{H^4\Theta^2}^{(2)}$&
			$(\tilde{H}^i \,\Theta^{\dagger}_{ijk} \,\Theta^{jkl} \,H_l) \,(H^{\dagger} \,H)$&$\mathcal{O}_{H^4\Theta^2}^{(3)}$&
			$(\tilde{H}^i \,\tilde{H}^j \,\Theta^{\dagger}_{ijk} \,\Theta^{klm}\,H_l\,H_m)$
			\\
			
			$\mathcal{O}_{H^5 \Theta}$&
			$\color{blue}{(H^i \,H^j \,H^k \,\Theta^{\dagger}_{ijk}) \,(H^{\dagger} \,H)}$
			&$\mathcal{O}_{H^3\Theta^3}^{(1)}$&
			$\color{blue}{(H^i \,H^j \,H^k \,\Theta^{\dagger}_{ijk}) \,(\Theta^{\dagger} \,\Theta)}$
			\\

			&
			&$\mathcal{O}_{H^3\Theta^3}^{(2)}$&
			$\color{blue}{(H^i \,H^j \,H^k \,\Theta^{\dagger}_{ijn}) \,(\Theta^{\dagger}_{lmk} \,\Theta^{lmn})}$
			\\
			
			\hline
			\hline
			\multicolumn{4}{||c||}{$\Phi^2X^2$}\\
			\hline
			
			$\mathcal{O}_{B\Theta}$&
			$B_{\mu\nu} \,B^{\mu\nu} \,(\Theta^{\dagger} \,\Theta)$&
			$\mathcal{O}_{\tilde{B}\Theta}$&
			$\tilde{B}_{\mu\nu} \,B^{\mu\nu} \,(\Theta^{\dagger} \,\Theta)$
			\\
			
			$\mathcal{O}_{G\Theta}$&
			$G^{A}_{\mu\nu}\,G^{A\mu\nu}\,(\Theta^{\dagger} \,\Theta)$&$\mathcal{O}_{\tilde{G}\Theta}$&
			$\tilde{G}^{A}_{\mu\nu}\,G^{A\mu\nu}\,(\Theta^{\dagger} \,\Theta)$\\
			
			$\mathcal{O}_{W\Theta}^{(1)}$&
			$W^{I}_{\mu\nu} \,W^{I\mu\nu} \,(\Theta^{\dagger} \,\Theta)$&$\mathcal{O}_{\tilde{W}\Theta}^{(1)}$&
			$\tilde{W}^{I}_{\mu\nu} \,W^{I\mu\nu} \,(\Theta^{\dagger} \,\Theta)$\\
			
			$\mathcal{O}_{W\Theta}^{(2)}$&
			$\epsilon_{IJK} \,W^I_{\mu\nu} \,W^{J\mu\nu} \,(\Theta^{\dagger} \,\tau^K_{(4)} \,\Theta)$&
			$\mathcal{O}_{\tilde{W}\Theta}^{(2)}$&
			$\epsilon_{IJK} \,\tilde{W}^I_{\mu\nu} \,W^{J\mu\nu} \,(\Theta^{\dagger} \,\tau^K_{(4)} \,\Theta)$\\
			
			$\mathcal{O}_{BW\Theta}$&
			$B_{\mu\nu} \,W^{I\mu\nu} \,(\Theta^{\dagger} \,\tau^I_{(4)} \,\Theta)$
			&
			$\mathcal{O}_{B\tilde{W}\Theta}$&
			$B_{\mu\nu} \,\tilde{W}^{I\mu\nu} \,(\Theta^{\dagger} \,\tau^I_{(4)} \,\Theta)$
			\\

			\hline
			\hline
			\multicolumn{4}{||c||}{$\Psi^2\Phi^3$}\\
			\hline
			
			$\mathcal{O}_{QdH\Theta}^{(1)}$&
			$\textcolor{blue}{(N_f^2) \,(\overline{Q}_{p\alpha i} \,d^{\alpha}_{q}) \,H^i \,(\Theta^{\dagger} \,\Theta)}$&
			$\mathcal{O}_{QdH\Theta}^{(2)}$&
			$\color{blue}{(N_f^2) \,(\overline{Q}_{p\alpha i} \,d^{\alpha}_{q})\,\tau^I_{(2)} \,H^i \,(\Theta^{\dagger} \,\tau^I_{(4)} \,\Theta)}$
			\\
			
			$\mathcal{O}_{QuH\Theta}^{(1)}$&
			$\textcolor{blue}{(N_f^2) \,\epsilon_{ij} \,(\overline{Q}_{p\alpha i} \,u^{\alpha}_{q}) \,\tilde{H}^j \,(\Theta^{\dagger} \,\Theta)}$&
			$\mathcal{O}_{QdH\Theta}^{(2)}$&
			$\textcolor{blue}{(N_f^2) \,\epsilon_{ij}  \,(\overline{Q}_{p\alpha i} \,u^{\alpha}_{q}) \,\tau^I_{(2)} \,\tilde{H}^j \,(\Theta^{\dagger} \,\tau^I_{(4)} \,\Theta)}$
			\\
			
			$\mathcal{O}_{eLH\Theta}^{(1)}$
			&$\color{blue}{(N_f^2) \,(\overline{L}_{pi} \,e_{q}) \,H^i \,(\Theta^{\dagger} \,\Theta)}$&
			$\mathcal{O}_{eLH\Theta}^{(2)}$&
			$\color{blue}{(N_f^2) \,(\overline{L}_{pi} \,e_{q}) \,\tau^I_{(2)} \,H^i \,(\Theta^{\dagger} \,\tau^I_{(4)} \,\Theta)}$
			\\
			
			$\mathcal{O}_{eLH^2\Theta}$&
			$\textcolor{blue}{(N_f^2) \,(\overline{L}_{pi} \,e_{q}) \,\tilde{H}^{j} \,\tilde{H}^k \,\Theta_{ijk}}$
			&$\mathcal{O}_{dQH^2\Theta}$&
			$\textcolor{blue}{(N_f^2) \,(\overline{Q}_{p\alpha i}  \,d^{\alpha}_{q}) \,(\tilde{H}^j \,\tilde{H}^k \,\Theta_{ijk})}$
			\\
			
			$\mathcal{O}_{uQH^2\Theta}$&
			$\textcolor{blue}{(N_f^2) \,(\overline{Q}_{p\alpha i} \,u^{\alpha}_{q}) \,(H^j \,H^k \,\Theta^{\dagger}_{ijk})}$&
			&\\
			\hline
	\end{tabular}}
	\caption{SM extended by  $SU(2)$ Quadruplet Scalar ($\Theta$): Additional operators of dimension 6. Here $i, j, k, l ,m, n$ and $\alpha$ are the $SU(2)$ and $SU(3)$ indices respectively. $\tau^I_{(2)}$ and  $\tau^{I}_{(4)}$ are $SU(2)$ generators in $2\times 2$ and $4\times 4$ representations respectively. $A=1,2,\cdots,8$ and $I,J,K=1,2,3$. $p, q = 1,2,\cdots,N_f$ are the flavour indices.}
	\label{table:QuadrupletScalar-dim6-ops-1}
\end{table}

\clearpage
\subsubsection*{\underline{SM + Lepto-Quarks ($\chi_2$, $\varphi_2$)}}

We have catalogued the effective operators for two other cases where the SM is extended by additional Lepto-Quarks. These are similar to the previous Lepto-Quark scenarios but having different hypercharges, see Table~\ref{table:Lepto-Quark-quantum-no-2}.
The effective operators of dimensions 5 and 6 containing $\chi_{2}$ have been collected in Table~\ref{table:Lepto-QuarkModel3-dim5-ops-1} and Tables~\ref{table:Lepto-QuarkModel3-dim6-ops-1}, \ref{table:Lepto-QuarkModel3-dim6-ops-1-contd} respectively. We have listed the same for  $\varphi_2$ as well in Table~\ref{table:Lepto-QuarkModel4-dim5-ops-1} and Tables~\ref{table:Lepto-QuarkModel4-dim6-ops-1}, \ref{table:Lepto-QuarkModel4-dim6-ops-1-contd}. The operators with distinct hermitian conjugates have been coloured blue.\\

\begin{table}[h]
	\centering
	\renewcommand{\arraystretch}{1.6}
	{\scriptsize\begin{tabular}{|c|c|c|c|c|c|c|}
			\hline
			\textbf{Non-SM IR DOFs}&
			\multirow{2}{*}{$SU(3)_C$}&
			\multirow{2}{*}{$SU(2)_L$}&
			\multirow{2}{*}{$U(1)_Y$}&
			\multirow{2}{*}{\textbf{Spin}}&
			\multirow{2}{*}{\textbf{Baryon No.}}&
			\multirow{2}{*}{\textbf{Lepton No.}}\\
			
			\textbf{(Lepto-Quarks)}&
			&
			&
			&
			&
			&
			\\
			\hline

			$\chi_2$&
			3&
			2&
			7/6&
			0&
			1/3&
			-1\\
			\hline
						
			$\varphi_2$&
			3&
			1&
			-1/3&
			0&
			1/3&
			-1\\
			\hline
	\end{tabular}}
	\caption{\small Additional IR DOFs (Lepto-Quarks) as representations of the SM gauge groups along with their spin, baryon and lepton numbers.}
	\label{table:Lepto-Quark-quantum-no-2}
\end{table} 

\noindent\textbf{Features of the additional operators:}
\begin{itemize}
	\item For $\chi_2$, we obtain a single operator at mass dimension 5 which violates baryon and lepton numbers, whereas for $\varphi_2$ we obtain 2 operators at mass dimension 5 and both of them violate only lepton number.

	\item We have noted  the lepton and baryon number violations, signifying the mixing between quark and lepton sectors within the $\Psi^2\Phi^2\mathcal{D}$, $\Psi^2\Phi^3$ and $\Psi^2\Phi X$ classes.
	
	\item We have also observed the  mixing between $B_{\mu\nu}$, $W^{I}_{\mu\nu}$ and $G^{A}_{\mu\nu}$ within the $\Phi^2X^2$ class similar to the case of $\chi_1$ and $\varphi_1$.
\end{itemize}

\begin{table}[h]
	\centering
	\renewcommand{\arraystretch}{1.9}
	{\scriptsize
\begin{tabular}{||c|c||}
			\hline
			\hline
			\multicolumn{2}{||c||}{$\Psi^2\Phi^2$}\\
			\hline
			$\tilde{\mathcal{O}}_{d H \chi_2}$&
			$\boxed{\color{purple}{\frac{1}{2}(N_f^2-N_f) \,\epsilon_{\alpha\beta\gamma} \,((d^{\alpha}_{p})^T \,C \,d^{\beta}_{q}) \,(\tilde{H}_i \,\chi_2^{\gamma i})}}$
			\\
			\hline
	\end{tabular}}
	\caption{SM extended by Lepto-Quark ($\chi_2$): Additional operators of dimension 5. Here $i$ and $\alpha,\beta,\gamma$ are the $SU(2)$ and $SU(3)$ indices respectively. $p, q=1,2,\cdots,N_f$ are the flavour indices. This operator violates baryon and lepton number. }
	\label{table:Lepto-QuarkModel3-dim5-ops-1}
\end{table}

\begin{table}[h]
	\centering
	\renewcommand{\arraystretch}{1.9}
	{\scriptsize\begin{tabular}{||c|c||c|c||}
			\hline
			\hline
			\multicolumn{4}{||c||}{$\Psi^2\Phi^2$}\\
			\hline
			$\tilde{\mathcal{O}}_{d \varphi_2}$&
			$\color{purple}{\frac{1}{2}(N_f^2+N_f) \,((d^{\alpha}_{p})^T \,C \,d^{\beta}_{q}) \,(\varphi^{\dagger}_{2,\alpha} \,\varphi^{\dagger}_{2,\beta})}$
			&$\tilde{\mathcal{O}}_{L d H \varphi_2}$&
			$\color{purple}{(N_f^2) \,\epsilon_{ij} \,(\overline{d}_{p\alpha} \,L^i_{q}) \,(H^j \,\varphi^{\alpha}_2)}$
			\\
			\hline
	\end{tabular}}
	\caption{SM extended by Lepto Quark ($\varphi_2$): Additional operators of dimension 5. Here $i, j$ and $\alpha,\beta $ are the $SU(2)$ and $SU(3)$ indices respectively. $p,q=1,2,\cdots,N_f$ are the flavour indices. Both operators violate lepton number.}
	\label{table:Lepto-QuarkModel4-dim5-ops-1}
\end{table}

\begin{table}[h]
	\centering
	\renewcommand{\arraystretch}{1.9}
	{\scriptsize\begin{tabular}{||c|c||c|c||}
			\hline
			\hline
			\multicolumn{2}{||c||}{$\Phi^4\mathcal{D}^2$}&
			\multicolumn{2}{c||}{$\Phi^6$}\\
			\hline
			$\mathcal{O}_{\chi_2 \square}^{(1)}$&
			$(\chi^{\dagger}_2 \,\chi_2) \,\square \,(\chi^{\dagger}_2 \,\chi_2)$&
			$\mathcal{O}_{\chi_2}^{(1)}$&
			$(\chi_2^{\dagger} \,\chi_2)^3$ \\
			
			$\mathcal{O}_{\chi_2 \square}^{(2)}$&
			$(\chi^{\dagger}_2 \,T^A \,\chi_2) \,\square \,(\chi^{\dagger}_2 \,T^A \,\chi_2)$&
			$\mathcal{O}_{\chi_2}^{(2)}$&
			$(\chi_2^{\dagger} \,T^A \,\chi_2) \,(\chi_2^{\dagger} \,T^A \,\chi_2) \,(\chi_2^{\dagger} \,\chi_2)$
			\\
			
			$\mathcal{O}_{\chi_2 \mathcal{D}}^{(1)}$&
			$(\chi^{\dagger}_2 \,\overleftrightarrow{\mathcal{D}}_{\mu} \,\chi_2)\,(\chi^{\dagger}_2 \,\overleftrightarrow{\mathcal{D}}^{\mu} \,\chi_2)$&
			$\mathcal{O}_{H^2 \chi_2^4}^{(1)}$&
			$(\chi^{\dagger}_2 \,\chi_2)^2 \,(H^{\dagger} \,H)$
			\\
			
			$\mathcal{O}_{\chi_2 \mathcal{D}}^{(2)}$&
			$(\chi^{\dagger}_2 \,\overleftrightarrow{\mathcal{D}}_{\mu}^{A} \,\chi_2)\,(\chi^{\dagger}_2 \,\overleftrightarrow{\mathcal{D}}^{\mu A} \,\chi_2)$&
			$\mathcal{O}_{H^2 \chi_2^4}^{(2)}$&
			$(\chi^{\dagger}_2 \,T^A \,\chi_2) \,(\chi^{\dagger}_2 \,T^A \,\chi_2) \,(H^{\dagger} \,H)$
			\\
			
			$\mathcal{O}_{ H \chi_2 \square}$&
			$(\chi^{\dagger}_2 \,\chi_2) \,\square \,(H^{\dagger} \,H)$&
			$\mathcal{O}_{H^2 \chi_2^4}^{(3)}$&
			$(\chi^{\dagger}_2 \,\chi_2) \,(\chi^{\dagger}_2 \,\tau^I \,\chi_2) \,(H^{\dagger} \,\tau^I\,H)$\\
			
			$\mathcal{O}_{ H \chi_2 \mathcal{D}}^{(2)}$&
			$(\chi^{\dagger}_2 \,\overleftrightarrow{\mathcal{D}}_{\mu}^I \,\chi_2)  \,(H^{\dagger} \,\overleftrightarrow{\mathcal{D}}^{\mu I} \,H)$&
			$\mathcal{O}_{H^4 \chi_2^2}^{(1)}$&
			$(\chi^{\dagger}_2 \,\chi_2) \,(H^{\dagger} \,H)^2$\\
\hline
	\end{tabular}}
	\caption{SM extended by Lepto-Quark ($\chi_2$): Additional operators of dimension 6. Here $i, j$ and $\alpha$ are the $SU(2)$ and $SU(3)$ indices respectively. $T^A$ and $\tau^I$ are $SU(3)$ and $SU(2)$ generators respectively. $A,B,C=1,2,\cdots,8$ and $I=1,2,3$. $p, q=1,2,\cdots,N_f$ are the flavour indices. Operators in {\it red} violate lepton number.}
	\label{table:Lepto-QuarkModel3-dim6-ops-1}
\end{table}
\clearpage
\begin{table}[h]
	\centering
	\renewcommand{\arraystretch}{1.9}
	{\scriptsize\begin{tabular}{||c|c||c|c||}

			\hline
			\hline
			\multicolumn{4}{||c||}{$\Psi^2\Phi^3$}\\
			\hline
			
			$\mathcal{O}_{Q d H \chi_2}^{(1)}$&
			$\textcolor{blue}{(N_f^2) \,(\overline{Q}_{p\alpha i} \,d^{\alpha}_{q}) \,H^i \,(\chi^{\dagger}_2 \,\chi_2)}$&
			$\mathcal{O}_{Q d H \chi_2}^{(2)}$&
			$\color{blue}{(N_f^2) \,(\overline{Q}_{p\alpha i} \,T^A \,d^{\alpha}_{q}) \,H^i \,(\chi^{\dagger}_2 \,T^A \,\chi_2)}$\\
			
			$\mathcal{O}_{Q d H \chi_2}^{(3)}$&
			$\color{blue}{(N_f^2) \,(\overline{Q}_{p\alpha i} \,d^{\alpha}_{q}) \,\tau^I \,H^i \,(\chi^{\dagger}_2 \,\tau^I \,\chi_2)}$&
			$\mathcal{O}_{Q d H \chi_2}^{(4)}$&
			$\color{blue}{(N_f^2) \,(\overline{Q}_{p\alpha i} \,T^A \,d^{\alpha}_{q}) \,\tau^I \,H^i \,(\chi^{\dagger}_2 \,T^A \,\tau^I \,\chi_2)}$
			\\
			
			$\mathcal{O}_{Q u H \chi_2}^{(1)}$&
			$\textcolor{blue}{(N_f^2) \,\epsilon_{ij} \,(\overline{Q}_{p\alpha i} \,u^{\alpha}_{q}) \,\tilde{H}_j \,(\chi^{\dagger}_2 \,\chi_2)}$
			&$\mathcal{O}_{Q u H \chi_2}^{(2)}$&
			$\textcolor{blue}{(N_f^2) \,\epsilon_{ij}  \,(\overline{Q}_{p\alpha i} \,T^A \,u^{\alpha}_{q}) \,\tilde{H}_j \,(\chi^{\dagger}_2 \,T^A \,\chi_2)}$
			\\
			
			$\mathcal{O}_{Q u H \chi_2}^{(3)}$&
			$\textcolor{blue}{(N_f^2) \,\epsilon_{ij}  \,(\overline{Q}_{p\alpha i} \,u^{\alpha}_{q}) \,\tau^I \,\tilde{H}_j \,(\chi^{\dagger}_2 \,\tau^I \,\chi_2)}$&
			$\mathcal{O}_{Q u H \chi_2}^{(4)}$&
			$\textcolor{blue}{(N_f^2) \,\epsilon_{ij}  \,(\overline{Q}_{p\alpha i} \,T^A \,u^{\alpha}_{q}) \,\tau^I \,\tilde{H}_j \,(\chi^{\dagger}_2 \,T^A \,\tau^I \,\chi_2)}$
			\\ 
			
			$\mathcal{O}_{L e H \chi_2}^{(1)}$&
			$\textcolor{blue}{(N_f^2) \,(\overline{L}_{pi} \,e_{q}) \,H_i \,(\chi_2^{\dagger} \,\chi_2)}$&
			$\mathcal{O}_{L e H \chi_2}^{(2)}$&
			$\textcolor{blue}{(N_f^2) \,(\overline{L}_{pi} \,e_{q}) \,\tau^I \,H_i \,(\chi_2^{\dagger} \,\tau^I \,\chi_2)}$
			\\
\hline
	\end{tabular}}
	\caption{Table \ref{table:Lepto-QuarkModel3-dim6-ops-1} continued. Operators in {\it red} violate lepton number.}
	\label{table:Lepto-QuarkModel3-dim6-ops-1-contd}
\end{table}

\begin{table}[h]
	\centering
	\renewcommand{\arraystretch}{1.9}
	{\scriptsize\begin{tabular}{||c|c||c|c||}
			
			\hline
			\hline
			\multicolumn{4}{||c||}{$\Phi^2 X^2$}\\
			\hline
			
			$\mathcal{O}_{B \varphi_2}$&
			$B_{\mu\nu} \,B^{\mu\nu} \,(\varphi^{\dagger}_2 \,\varphi_2)$&
			$\mathcal{O}_{\tilde{B} \varphi_2}$&
			$\tilde{B}_{\mu\nu} \,B^{\mu\nu} \,(\varphi^{\dagger}_2 \,\varphi_2)$\\
			
			$\mathcal{O}_{G \varphi_2}^{(1)}$&
			$G^{A}_{\mu\nu}\,G^{A\mu\nu}\,(\varphi^{\dagger}_2 \,\varphi_2)$&
			$\mathcal{O}_{G \varphi_2}^{(2)}$&
			$d_{ABC} \,G^{A}_{\mu\nu} \,G^{B\mu\nu} \,(\varphi^{\dagger}_2 \,T^C \,\varphi_2) $\\
			
			$\mathcal{O}_{\tilde{G} \varphi_2}^{(1)}$&
			$\tilde{G}^{A}_{\mu\nu}\,G^{A\mu\nu}\,(\varphi^{\dagger}_2 \,\varphi_2)$&
			$\mathcal{O}_{\tilde{G} \varphi_2}^{(2)}$&
			$d_{ABC} \,\tilde{G}^{A}_{\mu\nu} \,G^{B\mu\nu} \,(\varphi^{\dagger}_2 \,T^C \,\varphi_2) $\\

			$\mathcal{O}_{W \varphi_2}$&
			
			$W^{I}_{\mu\nu} \,W^{I\mu\nu} \,(\varphi^{\dagger}_2 \,\varphi_2)$&
			
			$\mathcal{O}_{\tilde{W} \varphi_2}$&
			$\tilde{W}^{I}_{\mu\nu} \,W^{I\mu\nu} \,(\varphi^{\dagger}_2 \,\varphi_2)$\\

			$\mathcal{O}_{B G \varphi_2}$&
			$B_{\mu\nu} \,G^{A\mu\nu} \,(\varphi^{\dagger}_2 \,T^A \,\varphi_2)$&
			$\mathcal{O}_{B \tilde{G}\varphi_2}$&
			$B_{\mu\nu} \,\tilde{G}^{A\mu\nu} \,(\varphi^{\dagger}_2 \,T^A \,\varphi_2)$\\  
			\hline
			\hline
			\multicolumn{4}{||c||}{$\Psi^2\Phi^2\mathcal{D}$}\\
			\hline
			
			$\mathcal{O}_{Q \varphi_2 \mathcal{D}}^{(1)}$
			&$(N_f^2) \,(\overline{Q}_{p\alpha i} \,\gamma^{\mu} \,Q^{\alpha i}_{q}) \,(\varphi^{\dagger}_2 \,i\overleftrightarrow{\mathcal{D}}_{\mu} \,\varphi_2)$
			&$\mathcal{O}_{Q \varphi_2 \mathcal{D}}^{(2)}$&
			$(N_f^2) \,(\overline{Q}_{p\alpha i} \,T^A \,\gamma^{\mu} \,Q^{\alpha i}_{q}) \,(\varphi^{\dagger}_2 \,i\overleftrightarrow{\mathcal{D}}_{\mu}^A \,\varphi_2)$
			\\
\hline
	\end{tabular}}
	\caption{SM extended by Lepto Quark ($\varphi_2$): Additional operators of dimension 6. Here $i, j$ and $\alpha,\beta,\gamma$ are the $SU(2)$ and $SU(3)$ indices respectively. $T^A$ are the $SU(3)$ generators. $A,B,C=1,2,\cdots,8$ and $I=1,2,3$. $p, q=1,2,\cdots,N_f$ are the flavour indices. Operators in {\it red} violate lepton and baryon numbers.}
	\label{table:Lepto-QuarkModel4-dim6-ops-1}
\end{table}
\clearpage
\begin{table}[h]
	\centering
	\renewcommand{\arraystretch}{1.9}
	{\scriptsize\begin{tabular}{||c|c||c|c||}
			\hline
			\hline
			\multicolumn{2}{||c||}{$\Phi^4\mathcal{D}^2$}&
			\multicolumn{2}{c||}{$\Phi^6$}\\
			\hline
			
			$\mathcal{O}_{\varphi_2 \mathcal{D}}^{(1)}$&
			$(\varphi^{\dagger}_2 \,\varphi_2) \,\square \,(\varphi^{\dagger}_2 \,\varphi_2)$&$\mathcal{O}_{\varphi_2}$&
			$(\varphi_2^{\dagger} \,\varphi_2)^3$\\
			
			$\mathcal{O}_{\varphi_2 \mathcal{D}}^{(2)}$&
			$(\varphi^{\dagger}_2 \,\overleftrightarrow{\mathcal{D}}_{\mu} \,\varphi_2)\,(\varphi^{\dagger}_2 \,\overleftrightarrow{\mathcal{D}}^{\mu} \,\varphi_2)$&
			$\mathcal{O}_{H^2 \varphi_2^4}$&
			$(\varphi^{\dagger}_2 \,\varphi_2)^2 \,(H^{\dagger} \,H)$
			\\
			
			$\mathcal{O}_{H \varphi_2 \mathcal{D}}^{(1)}$&
			$(\varphi^{\dagger}_2 \,\varphi_2) \,\left[(\mathcal{D}^{\mu}\,H)^{\dagger}(\mathcal{D}_{\mu}\,H)\right]$&
			$\mathcal{O}_{H^4 \varphi_2^2}$&
			$ (\varphi^{\dagger}_2 \,\varphi_2) \,(H^{\dagger} \,H)^2$
			\\
			
			$\mathcal{O}_{H \varphi_2 \mathcal{D}}^{(2)}$&
			$(H^{\dagger}\,H)\,\left[(\mathcal{D}^{\mu}\,\varphi_2)^{\dagger}(\mathcal{D}_{\mu}\,\varphi_2)\right]$
			&
			&\\
			
			\hline
			\hline
			\multicolumn{4}{||c||}{$\Psi^2\Phi^3$}\\
			\hline
			
			$\mathcal{O}_{Q d H \varphi_2}^{(1)}$&
			$\textcolor{blue}{(N_f^2) \,(\overline{Q}_{p\alpha i} \,d^{\alpha}_{q}) \,H^i \,(\varphi^{\dagger}_2 \,\varphi_2)}$&
			$\mathcal{O}_{Q d H \varphi_2}^{(2)}$&
			$\color{blue}{(N_f^2) \,(\overline{Q}_{p\alpha i} \,T^A \,d^{\alpha}_{q}) \,H^i \,(\varphi^{\dagger}_2 \,T^A \,\varphi_2)}$\\
			
			$\mathcal{O}_{Q H \varphi_2}^{(1)}$&
			$\textcolor{blue}{(N_f^2) \,\epsilon_{ij} \,(\overline{Q}_{p\alpha i} \,u^{\alpha}_{q}) \,\tilde{H}^j \,(\varphi^{\dagger}_2 \varphi_2)}$&
			$\mathcal{O}_{Q H \varphi_2}^{(2)}$&
			$\textcolor{blue}{(N_f^2) \,\epsilon_{ij}  \,(\overline{Q}_{p\alpha i} \,T^A \,u^{\alpha}_{q}) \,\tilde{H}^j \,(\varphi^{\dagger}_2 \,T^A \,\varphi_2)}$
			\\
\hline
	\end{tabular}}
	\caption{Table \ref{table:Lepto-QuarkModel4-dim6-ops-1} continued. Operators in {\it red} violate lepton and baryon numbers.}
	\label{table:Lepto-QuarkModel4-dim6-ops-1-contd}
\end{table}

\subsection{Flavour ($N_f$) dependence and $B$, $L$, $CP$  violating operators}
Based on the ideas discussed in subsection \ref{subsec:flavour-dep-count},  we have tabulated the total number of operators of each class for the additional scenarios discussed in the previous subsections. We have displayed the results for dimension 5 in Table \ref{table:number-of-ops-dim5-app} and dimension 6 in Table \ref{table:number-of-ops-dim6-app}.  We have highlighted the number of $B$, $L$ and $CP$ violating operators wherever needed. 

\begin{table}[h]
	\centering
	\renewcommand{\arraystretch}{1.7}
	{\scriptsize\begin{tabular}{|c|c|c|c|}
			\hline
			\multirow{2}{*}{\textbf{BSM Field}}&
			\multirow{2}{*}{\textbf{Operator Class}}&
			\multicolumn{2}{c|}{\textbf{Number of Operators as $f(N_f)$}}\\
			\cline{3-4}
			
			&
			&
			Total Number&
			$B, \,L$ Violating Operators\\
			\hline
			
			$\Theta$&
			$\Psi^2\Phi^2$&
			$N_f^2+N_f$&
			$N_f^2+N_f$\\
			\hline
			
			$\chi_2$&
			$\Psi^2\Phi^2$&
			$N_f^2-N_f$&
			$N_f^2-N_f$\\
			\hline
			
			$\varphi_2$&
			$\Psi^2\Phi^2$&
			$3N_f^2+N_f$&
			$3N_f^2+N_f$\\
			\hline
			
	\end{tabular}}
	\caption{Number of additional operators of different classes at dimension 5 with $N_f$ fermion flavours, for the models containing $\chi_2$ and $\varphi_2$. }
	\label{table:number-of-ops-dim5-app}
\end{table}

\begin{table}[h]
	\centering
	\renewcommand{\arraystretch}{1.7}
	{\scriptsize\begin{tabular}{|c|c|c|c|}
			\hline
			\multirow{2}{*}{\textbf{BSM Field}}&
			\multirow{2}{*}{\textbf{Operator Class}}&
			\multicolumn{2}{c|}{\textbf{Number of Operators as $f(N_f)$}}\\
			\cline{3-4}
			
			&
			&
			Total Number (CPV Bosonic Ops.)&
			$B, \,L$ Violating Ops.\\
			\hline
			
			\multirow{5}{*}{$\Theta$}&
			$\Phi^6$&
			16&
			0\\
			
			&
			$\Phi^4\mathcal{D}^2$&
			8&
			0\\
			
			&
			$\Phi^2X^2$&
			10 (5)&
			0\\
			
			&
			$\Psi^2\Phi^2\mathcal{D}$&
			$7N_f^2$&
			0\\
			
			&
			$\Psi^2\Phi^3$&
			$18N_f^2$&
			0\\
			\hline

			\multirow{6}{*}{$\chi_2$}&
			$\Phi^6$&
			7&
			0\\
			
			&
			$\Phi^4\mathcal{D}^2$&
			8&
			0\\
			
			&
			$\Phi^2X^2$&
			14 (7)&
			0\\
			
			&
			$\Psi^2\Phi^2\mathcal{D}$&
			$19N_f^2$&
			$8N_f^2$\\
			
			&
			$\Psi^2\Phi^3$&
			$38N_f^2$&
			$18N_f^2$\\
			
			&
			$\Psi^2\Phi X$&
			$12N_f^2$&
			$12N_f^2$\\
			\hline
			
			\multirow{6}{*}{$\varphi_2$}&
			$\Phi^6$&
			3&
			0\\
			
			&
			$\Phi^4\mathcal{D}^2$&
			4&
			0\\
			
			&
			$\Phi^2X^2$&
			10 (5)&
			0\\
			
			&
			$\Psi^2\Phi^2\mathcal{D}$&
			$14N_f^2$&
			$8N_f^2$\\
			
			&
			$\Psi^2\Phi^3$&
			$27N_f^2+N_f$&
			$7N_f^2+N_f$\\
			
			&
			$\Psi^2\Phi X$&
			$20N_f^2$&
			$10N_f^2$\\
			\hline			
	\end{tabular}}
	\caption{Number of additional operators of different classes at dimension 6 with $N_f$ fermion flavours, for models containing $\Theta$, $\chi_2$ and $\varphi_2$. The numbers in parentheses denote the counting for CP violating purely bosonic operators. }
	\label{table:number-of-ops-dim6-app}
\end{table}

\section*{}
\providecommand{\href}[2]{#2}
\addcontentsline*{toc}{section}{}
\bibliographystyle{JHEP}
\bibliography{BSMEFT}

\end{document}